%% file: main.tex
\documentclass[preprint,journal,svgnames]{vgtc}            

\onlineid{1214}

\vgtccategory{Research}

\vgtcpapertype{evaluation}

\newcommand{\revised}[1]{{\color{purple}#1}}
\newcommand{\removed}[1]{{\color{gray}\st{#1}}}

\renewcommand{\removed}[1]{}
\renewcommand{\revised}[1]{{\color{black}#1}}

\title{Improved Visual Saliency of Graph Clusters with Orderable Node-Link Layouts}

\author{%
  \authororcid{Nora Al-Naami}{0009-0005-5077-2062},
  \authororcid{Nicolas M\'edoc}{0000-0002-7419-8748},
  \authororcid{Matteo Magnani}{0000-0002-3437-9018},
  \authororcid{Mohammad Ghoniem}{0000-0001-6745-3651}
}

\authorfooter{
  \item
  	Nora Al-Naami, Nicolas M\'edoc and Mohammad Ghoniem are with Luxembourg Institute of Science and Technology.\\
  	E-mail: \{nora.alnaami\,$|$\,nicolas.medoc\,$|$\,mohammad.ghoniem\}@list.lu\,.

  \item 
        Matteo Magnani is with Uppsala University.\\
  	E-mail: matteo.magnani@it.uu.se\,.

}

\abstract{%
Graphs are often used to model relationships between entities. 
The identification and visualization of clusters in graphs enable insight discovery in many application areas, such as life sciences and social sciences.
Force-directed graph layouts promote the visual saliency of clusters, as they bring adjacent nodes closer together, and push non-adjacent nodes apart.
\revised{At the same time, matrices can effectively show clusters when a suitable row/column ordering is applied, but are less appealing to untrained users not providing an intuitive node-link metaphor. It is thus worth exploring layouts combining the strengths of the node-link metaphor and node ordering.}
In this work, we study the impact of node ordering on the visual saliency of clusters in orderable node-link diagrams, namely radial diagrams, arc diagrams and symmetric arc diagrams. 
Through a crowdsourced controlled experiment, we show that users can count clusters consistently more accurately, and to a large extent faster, with orderable node-link diagrams than with three state-of-the art force-directed layout algorithms, i.e., \emph{`Linlog'}, \emph{`Backbone'} and \emph{`sfdp'}. 
The measured advantage is greater in case of low cluster separability and/or low compactness. 
  A free copy of this paper and all supplemental materials are available at \url{https://osf.io/kc3dg/}.
}


\teaser{
    $\hfill\vcenter{\hbox{\includegraphics[width=0.23\textwidth]{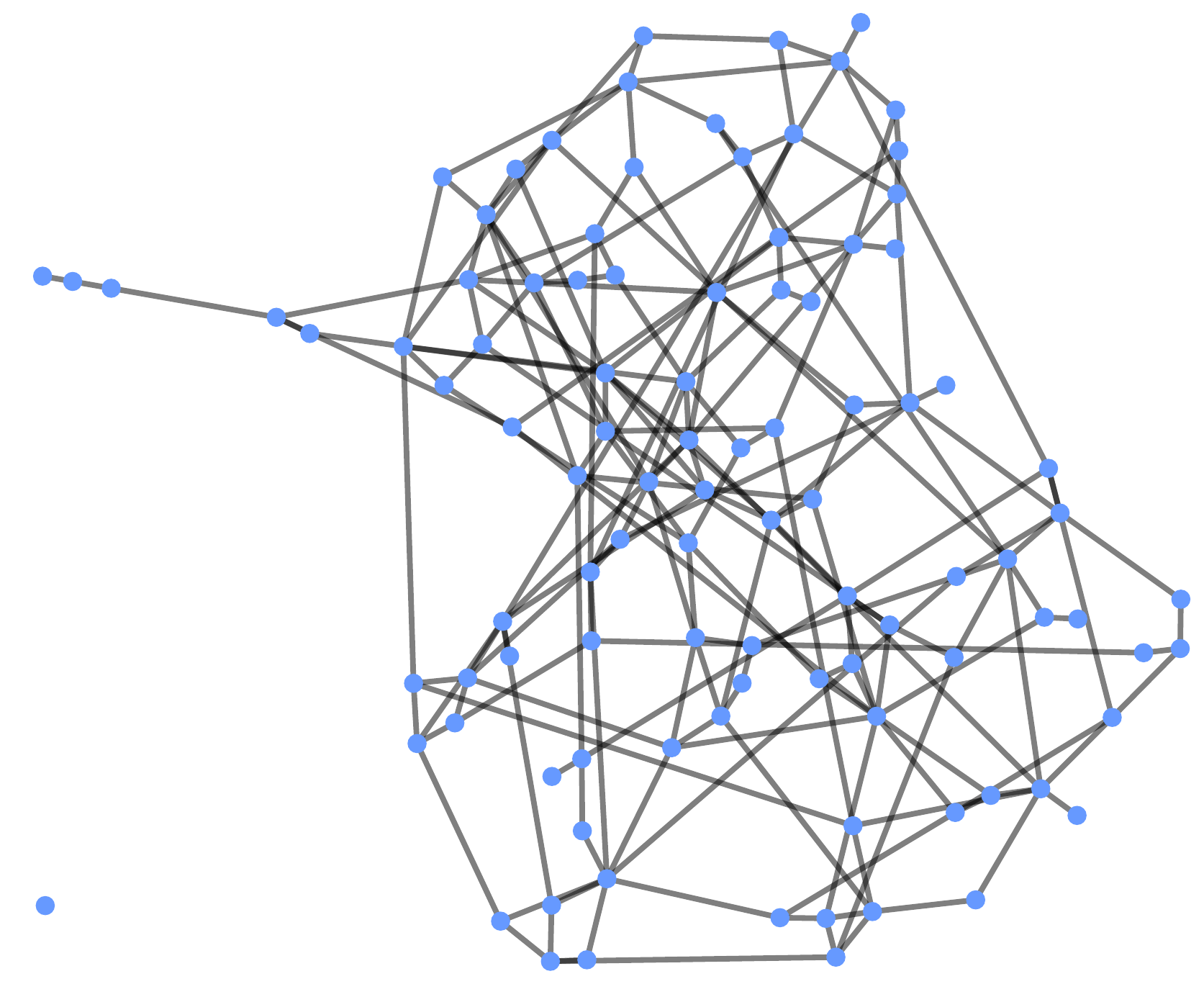}}}
    \hfill        \vcenter{\hbox{\includegraphics[width=0.23\textwidth]{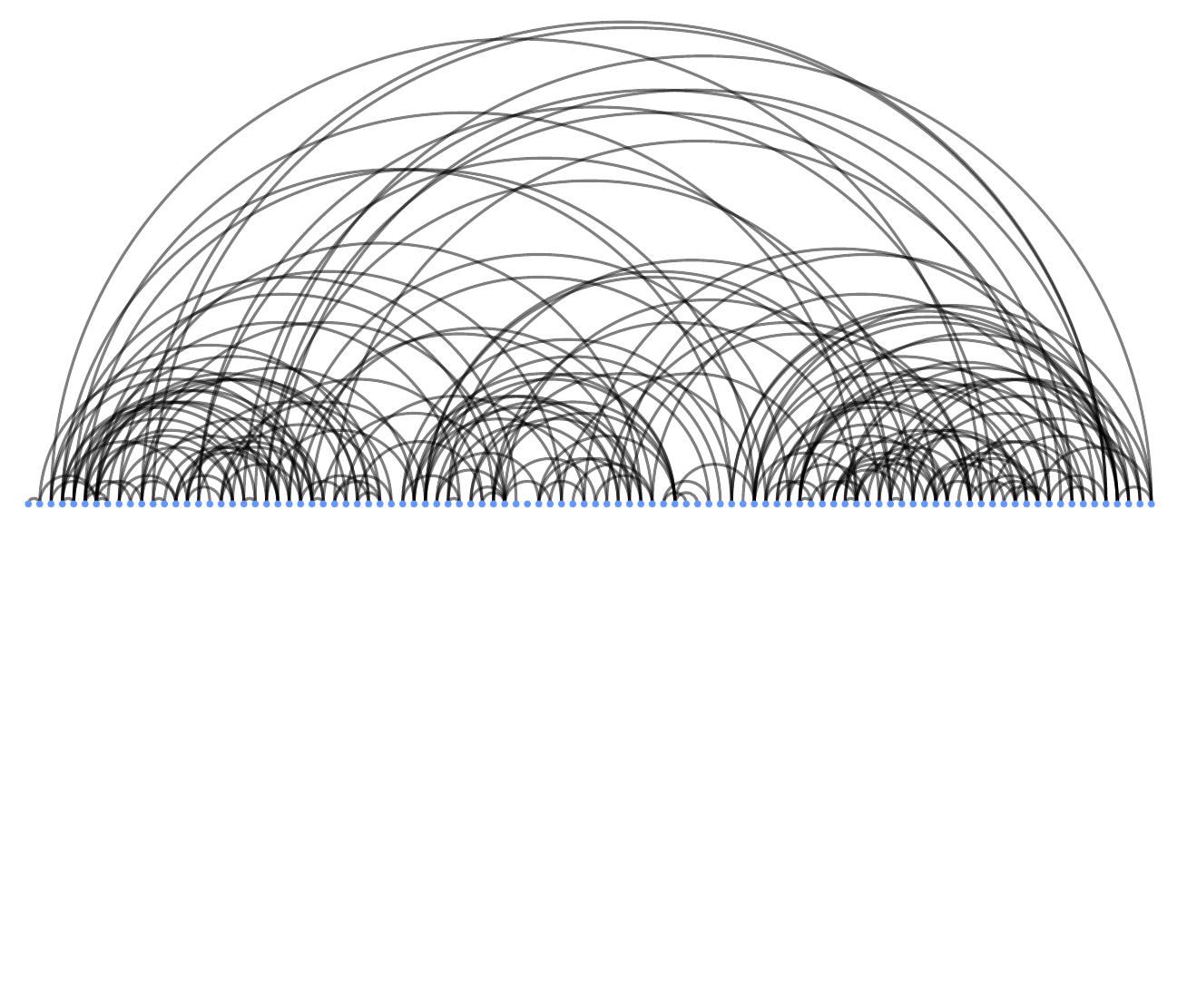}}}
    \hfill        \vcenter{\hbox{\includegraphics[width=0.23\textwidth]{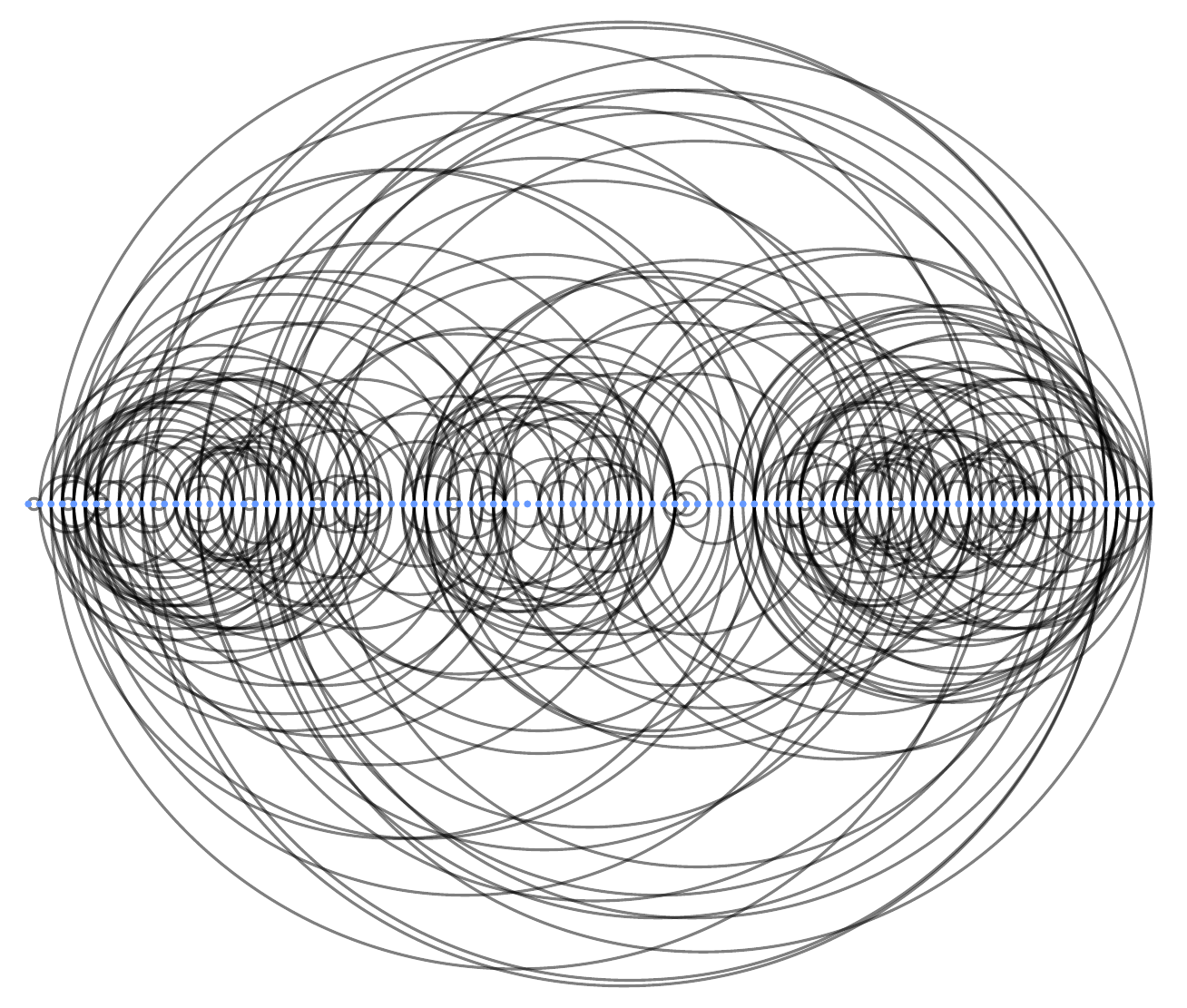}}}
    \hfill
    \vcenter{\hbox{\includegraphics[width=0.23\textwidth]{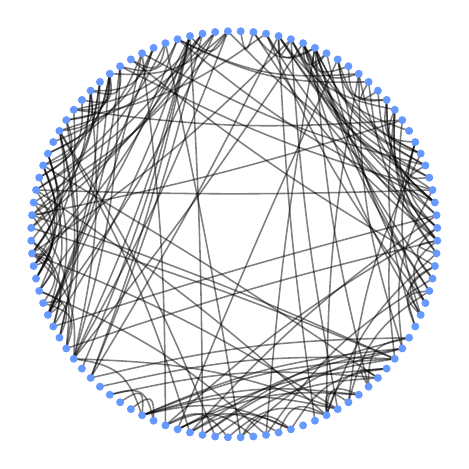}}}\hfill$
    \caption{A 100-node synthetic network having three clusters with `\emph{low}' within-cluster and `\emph{low}' between-cluster link probability, laid out from left to right according to the `\emph{Linlog}' layout, vanilla `\emph{arc diagram}', symmetric `\emph{arc diagram}' and `\emph{radial diagram}'. The clusters are indiscernible in the Linlog diagram, and clearly visible in the orderable layouts, where \revised{nodes} are in the original graph generation order.}
  \label{fig:teaser}
}

\graphicspath{{figs/}{figures/}{pictures/}{images/}{./}} 

\usepackage{mathptmx}                  

\usepackage{svg}
\usepackage[export]{adjustbox}
\usepackage{makecell}

\usepackage{multirow}
\usepackage{pifont}
\newcommand{\cmark}{\ding{51}}%
\newcommand{\xmark}{\ding{55}}%

\usepackage{mathtools}
\DeclarePairedDelimiter\norm{\lVert}{\rVert}

\definecolor{linlog}{HTML}{1B9E77}
\definecolor{backbone}{HTML}{D95F02}
\definecolor{sfdp}{HTML}{7570B3}

\definecolor{gen}{HTML}{1F78B4}
\definecolor{cr}{HTML}{B2DF8A}
\definecolor{olo}{HTML}{A6CEE3}

\usepackage{booktabs}

\usepackage{tikz}

\newcommand{\square}[1]{
\begin{tikzpicture}[scale=1]  
\fill [#1] (0,0) rectangle (0.2,0.2);
\end{tikzpicture}%
}

\usepackage{soul}

\newcommand{\nestedIcon}[1]{
\begingroup
\setbox0=\hbox{\includegraphics[height=9pt]{#1}}%
\hspace{-1pt}\parbox{\wd0}{\box0}\hspace{-2pt}
\endgroup
}

\begin{document}



\maketitle

\input{inc-body}

\section*{Supplemental Materials}
\label{sec:supplemental_materials}


All supplemental materials are available on OSF at \url{https://osf.io/kc3dg/}, released under a CC BY 4.0 license.
In particular, they include 
\begin{enumerate}[itemsep=0pt]
    \item A zip file containing \textbf{I) the material used to train study participants} (video tutorials and training data sets). \textbf{II) The stimuli} used in the user study in various formats (SVG, PDF), and \textbf{III) All graph data sets} in GraphML format to support study replication. 
    \item Additional statistics including I) a comparison of cluster count completion time between the Linlog layout and orderable layouts (accuracy is already analyzed above), II) a comparison of completion time between all orderable layouts across all cluster types, III) a comparison of the accuracy and completion time of backbone, respectively sfdp, to orderable layouts. 
\end{enumerate}

\acknowledgments{%
The authors wish to thank Helen Purchase for her valuable advice about the experimental setup.
This work was supported in part by a grant from Luxembourg FNR (\# INTER/ANR/20/14956115).%
}

\bibliographystyle{abbrv-doi-hyperref-narrow}

\bibliography{template}

\appendix 

\input{inc-appendices}

\end{document}

%% file: inc-body.tex
\section{Introduction}

Graph visualization is widely used to support network analysis tasks in various areas of science and engineering~\cite{herman-survey}. 
One popular network analysis task consists in identifying locally dense subgraphs~\cite{lee06taxonomy}, often called `\emph{clusters}', or `\emph{communities}' in social network analysis~\cite{bedi2016community}, or `\emph{modules}' in biological network analysis~\cite{bonneau2008modules}. Indeed, graph clusters correspond to meaningful sub-structures in many applications~\cite{landesberger-survey}. Hence, when asked to lay out a graph interactively, users tend to prioritize the formation of clusters over other aesthetic considerations~\cite{rogowitz08cluster}.

Graph clusters are often visually identified by drawing the graph using a force-directed node-link layout \nestedIcon{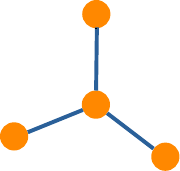} algorithm. This category of algorithms uses physics-based models, whereby attraction forces place adjacent nodes (see Section~\ref{sec:related}) closer to each other, and repulsive forces place non-adjacent nodes farther apart~\cite{landesberger-survey,gestalt-graphs,gibson2Dsurvey}. 
Hence, when the graph contains clusters, nodes belonging to the same cluster tend to clump together in the 2D plane. 
This makes clusters visually salient in virtue of the Gestalt rule of proximity~\cite{gestalt-graphs}, without using any additional visual cues to denote cluster membership, such as node color. 
In a recent comparison of many force-directed layouts, the LinLog layout~\cite{linlog2004,linlog2007} has consistently outperformed other alternatives~\cite{Meidiana2019AQM}.
The best cluster separation is generally achieved when internal link density within clusters is high, and external link density is low, i.e., relatively fewer links exist between nodes sitting in distinct clusters~\cite{linlog2007}.

Another approach uses adjacency matrices as a visual metaphor to analyze graphs.
Nodes are mapped to the rows and columns of a matrix, while links appear as colored rectangles at the intersection of adjacent (rows/columns) nodes. 
Besides being more readable for many low-level perceptual tasks, for large and dense graphs~\cite{Ghoniem2005OnTR}, matrices are effective at showing graph clusters when a suitable row/column ordering is applied~\cite{Riche2006MatrixExplorerAD}. 
Following Bertin's work on table seriation~\cite{bertin1983semiology}, much work compared various methods for reordering matrix visualizations to elicit structural properties~\cite{Fekete2015ReorderjsAJ,behrisch2016ordering}.
Matrices are yet a less popular graph visualization, and are less appealing to untrained users~\cite{Ghoniem2005OnTR}. 

In this work, we retain the node-link metaphor owing to its popularity and intuitiveness and, we apply node ordering to elicit clusters, inspired by the proven effectiveness of node ordering in matrix visualizations. 
The node-link layouts where nodes can be ordered include the ones that place nodes on a continuous non-self-intersecting topological curve, e.g., a circle, an ellipse, a straight line or a space-filling curve.
Nodes can be ordered along such a curve, e.g., based on topological or attribute-based criteria. 
In this paper, we will refer to such layouts, as `\emph{orderable node-link layouts}'. 
This category includes `\emph{arc diagrams}'\nestedIcon{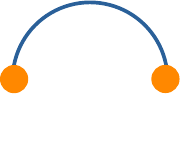}~\cite{wattenberg-arc} and `\emph{radial diagrams}'\nestedIcon{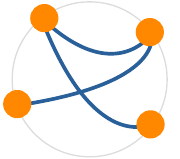}~\cite{Krzywinski18062009}, which place the nodes along a straight line and along a circle, respectively. Edges are drawn as half circles in arc layouts \nestedIcon{img/icons/Arc.pdf}, and as B\'ezier curves or straight lines in radial layouts \nestedIcon{img/icons/radial.pdf}.
Although radial and arc layouts~\nestedIcon{img/icons/radial.pdf}\nestedIcon{img/icons/Arc.pdf} are not new representations, we lack user studies regarding their readability for perceptual tasks. 

This paper raises the following research question: how do orderable node-link layouts \nestedIcon{img/icons/Arc.pdf} \nestedIcon{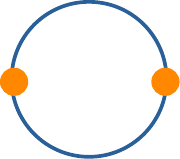} \nestedIcon{img/icons/radial.pdf} compare to force-directed layouts \nestedIcon{img/icons/NL.pdf} and to one another regarding their ability to reveal node clusters in graphs? To do so, we contribute a crowdsourced empirical study whose dependent variables are the accuracy and completion time of cluster count judgement, under varying values of independent variables which are cluster compactness, cluster separability, node ordering method, graph size and graph layout.
We elicit new insights regarding how three popular orderable node-link diagrams (arc diagrams \nestedIcon{img/icons/Arc.pdf}, symmetric arc diagrams \nestedIcon{img/icons/ArcSym.pdf}, radial diagrams \nestedIcon{img/icons/radial.pdf}) compare to three competitive force-directed layouts~\nestedIcon{img/icons/NL.pdf} (Linlog, Backbone and sfdp). We also assess the impact of two different node ordering methods (the optimal leaf ordering and the cross reduction algorithms) on the effectiveness of orderable layouts, in addition to the original node order provided by the graph generator.
Finally, we derive guidelines for the use of these layouts based on graph characteristics and type of clusters.
\vspace{0.5em}

\section{Related Work}
\label{sec:related}

Our work builds on previous research in the areas of graph visualization, and graph seriation, as well as user studies related to these areas.

\subsection{Graph Visualization}

Graphs are commonly used to model relationships between entities. 
They have been used to analyze complex phenomena in many application fields, such as social sciences, biology, and transportation~\cite{landesberger-survey}.
Formally, a graph $G$ is defined as $G=(V,E)$ where $V$ is a set of vertices and $E\subseteq V \times V$ is the set of edges connecting pairs of vertices in $G$.
Two vertices $u\in V$ and $v\in V$ are said to be \emph{adjacent}, if $\exists e\in E : e=(u,v)$. 

The popularity of graphs has motivated much research on graph drawing and network visualization~\cite{graph-drawing-book,diaz:2002:survey}. In
\emph{node-link diagrams}, vertices are displayed as nodes, e.g., circles, and edges as links, e.g., curves. 
Many algorithms produce such diagrams in 2D and 3D~\cite{gibson2Dsurvey,herman-survey}. 
In their survey, von Landesberger et al.~\cite{landesberger-survey} categorize node-link layouts into \emph{force-directed}, \emph{constraint-based}~\cite{dwyer09constraint}, \emph{multiscale}~\cite{koren02ace}, \emph{layered}~\cite{dwyer05cola} and \emph{non-standard} layouts, e.g., projection-based layouts~\cite{harel02embedding}.

Force-directed layouts \nestedIcon{img/icons/NL.pdf} adopt a physics-based model, e.g., springs and masses, or electric particles, whose equilibrium determines the position of nodes such that adjacent nodes are placed closer to each other~\cite{Heer:2010:ATTTVZ,McGuffin:2012:SAFNVAT}. They are hence well-suited for cluster visualization~\cite{gibson2Dsurvey}, in virtue of the Gestalt proximity law~\cite{gestalt-graphs}. The `\emph{Linlog}' energy models improve the visual separation of clusters compared to prior force-directed layouts~\cite{linlog2004,linlog2007}. Meidiana et al.~\cite{Meidiana2019AQM} used a quality metric score to compare force-directed layouts (e.g., Fruchter-Reingold \cite{Fruchterman}), multi-level layouts (e.g., FM3 \cite{hachulFM3}, sfdp \cite{graphviz, huSFDP}), multi-dimensional scaling methods (e.g., MDS \cite{torgerson1952multidimensional}, pivot MDS \cite{brandes2006eigensolver}), stress-based layouts (e.g., Stress Majorization \cite{stressmajorization}), spectral layouts (e.g., with graph Laplacian) with other layouts designed for cluster discovery, such as Backbone \cite{nocaj2015untangling}, tsNET~\cite{KruigerTSNE} and Linlog~\cite{linlog2004}. They found that the Linlog layout performed well across many graph data sets, and generally better than competing algorithms~\cite{gibson2Dsurvey}, followed by Backbone and tsNET.

Other node-link layouts place nodes along a geometric locus. 
For instance, radial layouts \nestedIcon{img/icons/radial.pdf} place nodes around a circle~\cite{draper:2009:radialsurvey}, whereas arc diagrams \nestedIcon{img/icons/Arc.pdf}~\cite{wattenberg-arc,Nagel:2013:AVSOAD} and BioFabric~\cite{BioFabric} place them along a straight line. 
These layouts are special cases of constraint-based layouts~\cite{dwyer09constraint}, where graph nodes are placed along the given locus, be it a circle, an ellipse, a straight-line, or a space-filling curve, e.g., Hilbert curves~\cite{spacefilling, spacefilling2}. 
Such layouts are orderable, since nodes can be sorted along the underlying geometric locus.
User studies on the impact of node ordering on the readability of radial \nestedIcon{img/icons/radial.pdf} or arc \nestedIcon{img/icons/Arc.pdf} layouts are yet lacking. 

When the vertices and edges of the graph have attributes attached to them, distinct types of graphs and graph visualization problems arise. 
This has given birth to specific lines of graph visualization research including, for example, multivariate graphs~\cite{nobre:2019:TSOTAIVMN}, multipartite graphs~\cite{asratian:1998:bipartite}, dynamic graphs~\cite{beck:2017:dynamic} and multilayer graphs~\cite{mcgee:2019:multilayer}. Many attribute-based techniques take a faceted layout approach to show subgraphs, e.g., parallel 1-D axes in PivotGraph~\cite{pivotgraph}, parallel 2-D planes (showing edges on demand) as in Semantic Substrates~\cite{semanticSubstrates} or NodeXL~\cite{nodeXL}.

In this work, we focus on the perception of clusters in single-mode unweighted unattributed graphs, without prior ground truth about clusters. We compare three orderable node-link layouts \nestedIcon{img/icons/Arc.pdf}\nestedIcon{img/icons/ArcSym.pdf}\nestedIcon{img/icons/radial.pdf} to three state-of-the-art force-directed layouts~\nestedIcon{img/icons/NL.pdf}, i.e., Linlog, Backbone and sfdp, in terms of visual cluster perception.

\subsection{Graph Seriation}

Graph seriation~\cite{liivseriation} is the arrangement of nodes and their edges in a visualization to elicit hidden graph patterns, which reveal key properties of the underlying network~\cite{behrisch2016ordering, muellerNatrixOrderingcomparison}. Seriation was mainly applied to tables and matrices~\cite{perin14bertifier,liivseriation,rao1994table}.
In a network, node ordering algorithms aim to reveal latent network patterns to the user~\cite{McGuffin:2012:SAFNVAT}. They can be divided into \emph{attribute-based} algorithms, e.g., ordering nodes based on their label and \emph{topology-based} or \emph{seriation} algorithms~\cite{liivseriation}.

Focusing on adjacency matrices, Behrisch et al.~\cite{behrisch2016ordering} distinguish seven families of ordering methods, based on common underlying reordering concepts~\cite{behrisch2016ordering}:
\begin{enumerate*}[label=\arabic*)]
 \item Robinsonian methods, e.g., clustering-based algorithms, including the \emph{Optimal Leaf Ordering}~\cite{BarJoseph2001FastOL} (OLO);
 \item spectral methods, e.g., rank-two ellipse seriation~\cite{chen2002generalized};
 \item dimensionality reduction methods, e.g., multi-dimensional scaling~\cite{borg2012multidimensional};
 \item heuristic methods, e.g., the crossing reduction heuristic (CR), also known as the barycenter heuristic~\cite{Mkinen2005TheBH};
 \item graph-theoretic methods, e.g., based on the Traveling Salesman Problem (TSP)~\cite{lenstra1974clustering};
 \item biclustering methods~\cite{cheng2000biclustering}; 
 \item interactive reordering methods, e.g., the Bertifier~\cite{perin14bertifier}.
\end{enumerate*}
They benchmark many seriation methods against many real and synthetic graphs, in terms of execution time and quality of the resulting matrix visualization using the \emph{minimum linear arrangement} score~\cite{jordi04minla}.
They conclude that cluster identification is better supported by clustering-based techniques. They recommend the use of OLO, as it optimizes for cluster patterns at the global and local levels, e.g., placing nodes at the boundary of their cluster closer to their neighbors in other clusters. 

In this empirical study, we take a user perception perspective on node ordering, with three understudied orderable node-link layouts \nestedIcon{img/icons/Arc.pdf}\nestedIcon{img/icons/ArcSym.pdf}\nestedIcon{img/icons/radial.pdf}. Our study considers the OLO order, as recommended by Berisch et~al.~\cite{behrisch2016ordering}, 
and the CR order, which was also used by Oeke et al.~\cite{Okoe2019NodeLinkOA} in their cluster perception task, and in PivotGraph~\cite{pivotgraph} to minimize edge crossings between parallel 1-D axes. In this work, OLO is based on the hierarchical clustering algorithm with average linkage. This puts this work close to other studies using hierarchical clustering to order graph nodes, e.g., by Abdelaal et al.~\cite{Abdelaal2022ComparativeEO} who do not mention using leaf order optimization explicitly. With this empirical study, we compare the two ordering methods systematically to one another when applied to \nestedIcon{img/icons/Arc.pdf} \nestedIcon{img/icons/ArcSym.pdf} \nestedIcon{img/icons/radial.pdf} and also to three state-of-the-art force-directed layouts\mbox{\nestedIcon{img/icons/NL.pdf}}, i.e., Linlog, Backbone and sfdp, in terms of cluster perception. 

\subsection{User Studies on Cluster Perception in Graphs}

Lee et al. categorize graph visualization tasks into \emph{topology-based}, \emph{attribute-based}, \emph{browsing} and \emph{overview} tasks~\cite{lee06taxonomy}. 
They consider cluster identification as a topology-based task, and underscore that an overview of the graph will usually help to find patterns, such as clusters, if a suitable graph layout is used. 
This is typically what force-directed node-link layouts aim to do, as well as matrix seriation methods, as pointed out earlier. 
Vehlow et al. survey visual encodings of group structure information in graph visualizations~\cite{vehlow2016survey}. 
They deliberately exclude the use of layout for this purpose from their scope, but mention in passing the possibility to use 1-D layouts to implicitly encode group information, where nodes are placed along a circle or a linear axis. They focus on other visual encodings, like the use of node color or hulls and contours for cluster membership.
They categorize groups into flat versus hierarchically structured, and disjoint versus overlapping. 
They also position previous work based on the type of visual encoding used to convey group membership information.

Many user studies on cluster perception in graph visualizations compare force-directed layouts to adjacency matrices and cover various graph visualization tasks such as path-related, attribute-based, overview and memorability tasks. Okoe et al.~\cite{Okoe2019NodeLinkOA} compare node-link layouts generated using \emph{neato}~\cite{graphviz} and matrix visualizations ordered by the crossing reduction method from \emph{Reorderjs}~\cite{Fekete2015ReorderjsAJ}. Based on two real graphs having 258 and 332 nodes, they show that matrices outperform force-directed layouts for the perception of colored clusters. Nobre et~al. also confirmed the superiority of matrices for cluster perception in multivariate networks, based on two real 75- and 25-node graphs~\cite{Nobre2020EvaluatingMN}. Chen et al. also studied cluster perception using an interactive toroidal layout, with automatic panning~\cite{chen2021sa}. They used synthetic graphs of variable density comprising 68 to 134 nodes and 3 to 8 clusters. They show that toroidal wrap layouts support faster and more accurate cluster perception than `standard' force-directed layouts. 
Several studies on clustered graphs have used much larger graphs, though~\cite{graphEvalMethods}.

Other studies go beyond force-directed node-link layouts and adjacency matrices. For example, to solve path-related problems typical of adjacency matrices, Henry-Riche et al.~\cite{Riche2007MatLinkEM} compare the Linlog layout to MatLink, a hybrid visualization composed of an arc diagram drawn on the side of a matrix ordered by a TSP heuristic. Based on six open data sets with less than 100 nodes with varying densities, they find that MatLink outperforms force-directed layouts for the identification of the largest cluster in the graph. 
Also, Abdelaal et al.~\cite{Abdelaal2022ComparativeEO} compare force-directed layouts generated by \emph{d3-force}, to a bipartite layout placing nodes along two (orderable) parallel axes with links in between them and a matrix visualization. They assess user performance for five tasks, including a cluster perception task, on graphs as large as 500 nodes. The bipartite layouts and the matrices were ordered using an agglomerative hierarchical clustering algorithm. For the cluster perception task, they report no significant difference between matrices and force-directed layouts, whereas bipartite layouts are much worse. 

While in prior studies the overall graph density is considered as a factor affecting user performance, this study aims to determine to what extent cluster perception is affected by link probability between clusters and within clusters, across various network sizes and visualization types, including three state-of-the-art force-directed layouts~\mbox{\nestedIcon{img/icons/NL.pdf}}, i.e., Linlog, Backbone and sfdp and three orderable layouts\nestedIcon{img/icons/Arc.pdf}\nestedIcon{img/icons/ArcSym.pdf}\nestedIcon{img/icons/radial.pdf}. To this end, we generated 60 different graph structures with a controlled within- and between-cluster probability (two levels: `\emph{low}' and `\emph{high}' for both probabilities), a controlled number of clusters $k \in [3..7]$ of variable sizes, and a controlled network size in terms of number of nodes (50, 100 and 300 nodes). Also, we assess the impact of the node ordering method (CR and OLO) on the performance of orderable layouts compared to the three force-directed node-link layouts. 
While Linlog is close to the optimum in translating mathematical clusters into a visual clustering~\cite{noack2009}, the interplay between within- and between-cluster probability seems underexplored in network visualization. This focus on graph cluster compactness and separability is inspired by cluster validity indices~\cite{cluster_sep,cluster_silhouettes,cluster_indices} used in pattern mining research.
Finally, our study is purely perceptual to evaluate the layouts in their own right, as navigation interactions may add more factors to control.

\section{Methods}
In this section, we describe the key parameters we control to generate the set of stimuli used in this user study. We also present our hypotheses and their motivation, as well as the study design and procedure.

\subsection{The Visual Stimuli}
\label{sec:visual_stimuli}

\paragraph{Network Size} In this study, we generated synthetic data to better control key network parameters. Compared to previous work surveyed by Yoghourdjian et al.~\cite{Yoghourdjian2018ExploringTL} where most studies use graphs with less than 100 nodes, our study considers 50-, 100- and 300-node graphs. By doing so, we aim to study both small and larger networks, without the need to use interaction techniques. We also use large networks to reduce the impact of random fluctuations during data generation.

\paragraph{Number of clusters and cluster size} Since in real networks clusters may vary in size, we use the Gaussian random partition graph generator~\cite{Brandes2003ExperimentsOG} of NetworkX~\cite{Hagberg2008ExploringNS} to create networks with $k\in[3..7]$ clusters, each drawn from a normal distribution with a mean cluster size and variance of cluster size distribution. We did not go beyond seven clusters because counting the clusters might become tedious without adding insights concerning cluster perception.

\begin{figure}[t]
\centering
$\vcenter{\hbox{\includegraphics[width=0.4\columnwidth]{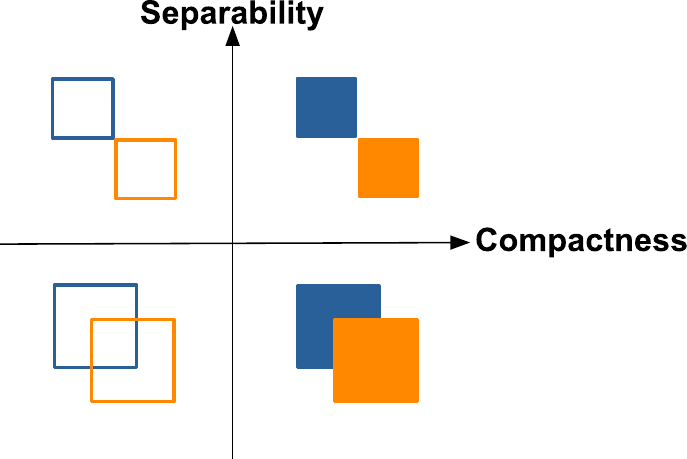}}}$
 \caption{The four quandrants of cluster types.}

 \label{fig:cluster_settings}
\end{figure}

\paragraph{Cluster compactness and separability} We set the within- and between-cluster link probability, $p_{in}$ and $p_{out}$ respectively, to control cluster compactness and separability. A high value of $p_{in}$ corresponds to \emph{compact} clusters, while a low value leads to \emph{loose} clusters. Likewise, a high $p_{out}$ creates relatively many links between clusters, i.e., poor cluster separation, whereas a low $p_{out}$ leads to better cluster separation. To keep the duration of user sessions under one hour, we consider two cluster compactness profiles $p_{in}\in\{low=0.3, high=0.6\}$ and two cluster separability profiles $p_{out}\in\{low=0.3, high=0.6\}$ as shown in \autoref{fig:cluster_settings}. In the rest of this paper, we will refer to clusters as \emph{separable} in the low $p_{out}$ condition, and as \emph{inseparable} for high $p_{out}$. Also, we describe clusters as \emph{compact} if they result from high $p_{in}$, and \emph{loose} otherwise. For each network size, we generate four graph structures, one in each quadrant. We use the following icons to denote these quadrants: \nestedIcon{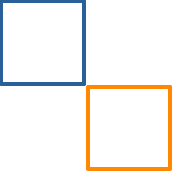} (loose, separable) clusters, \nestedIcon{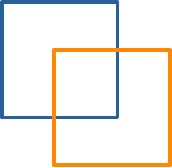} (loose, inseparable) clusters, \nestedIcon{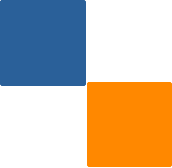}~(compact, separable) clusters, and \nestedIcon{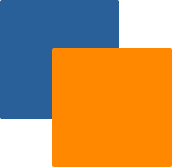} (compact, inseparable) clusters. An important concern regarding cluster compactness and separability is to choose the so-called `low' and `high' values for $p_{in}$ and $p_{out}$ such that the obtained graph clusters are meaningful. Determining the absolute upper and lower bounds beyond which clusters cease to exist is beyond the scope of this study. 

\begin{figure*}[t]
 \centering
 \begingroup
 \setlength{\tabcolsep}{1pt}

 \begin{tabular}{cccccc}
 ~&$k=3$&$k=4$&$k=5$&$k=6$&$k=7$\\
 \vspace{-1pt}
 \makecell{\includegraphics[width=11pt]{img/icons/loose-inseparable}}&
 \makecell{\includegraphics[width=0.14\textwidth]{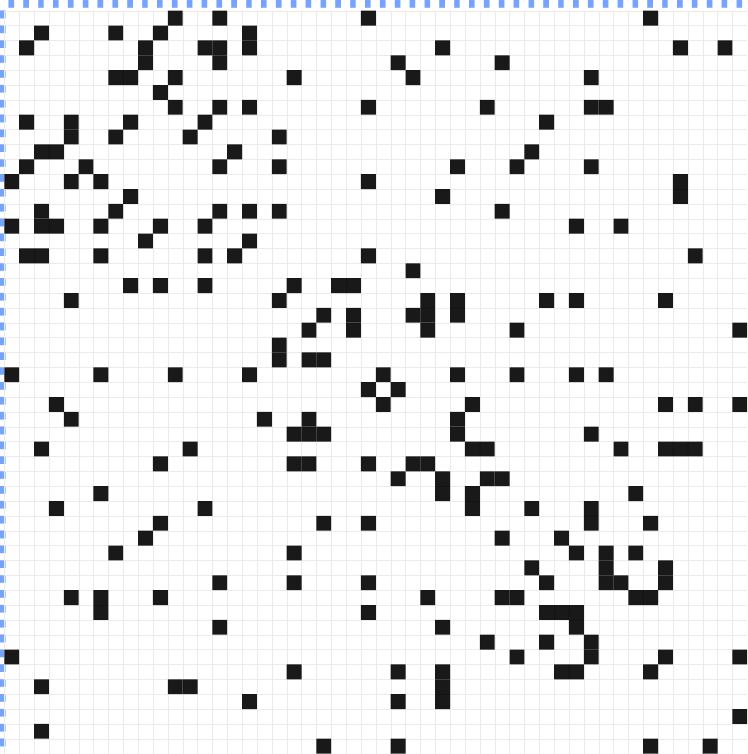}}&
 \makecell{\includegraphics[width=0.14\textwidth]{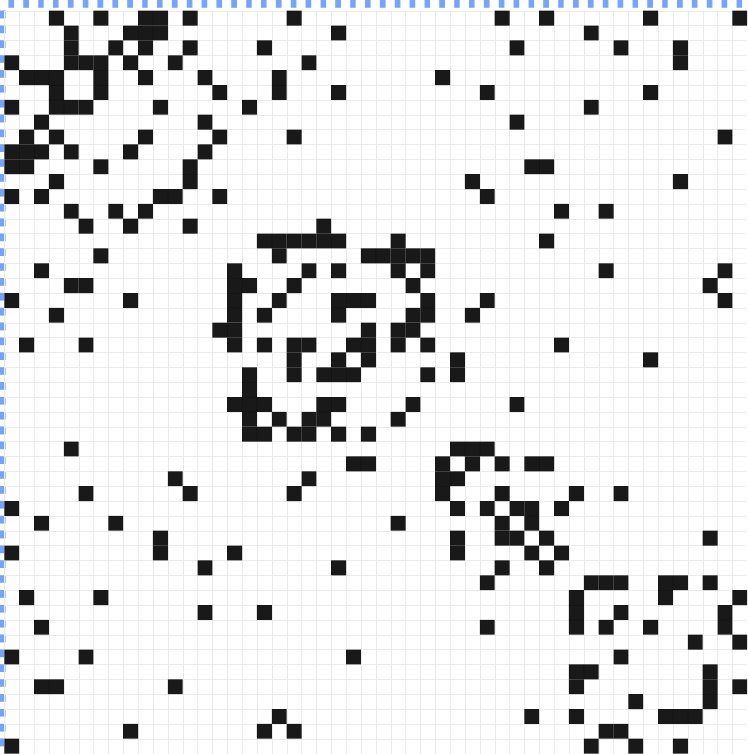}}&
 \makecell{\includegraphics[width=0.14\textwidth]{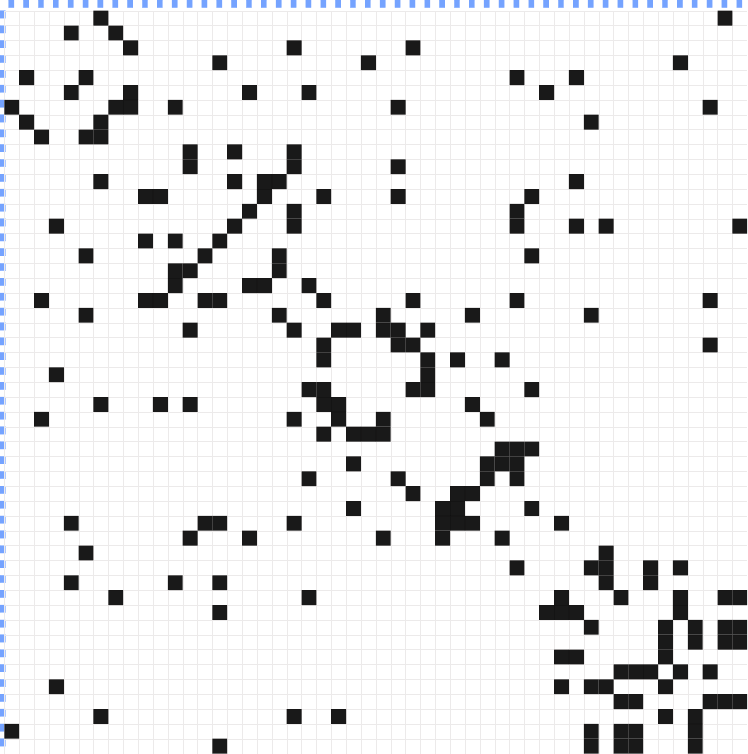}}&
 \makecell{\includegraphics[width=0.14\textwidth]{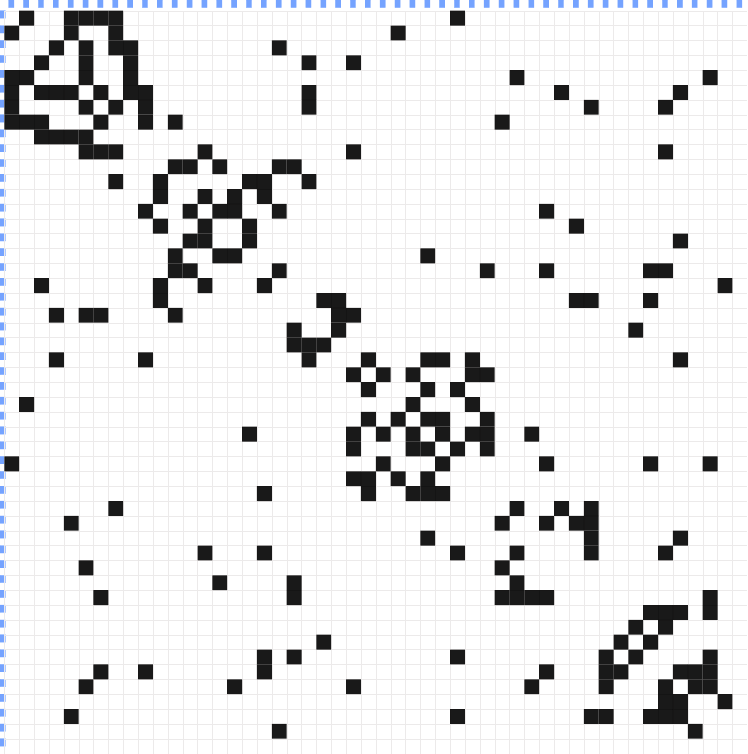}}&\makecell{\includegraphics[width=0.14\textwidth]{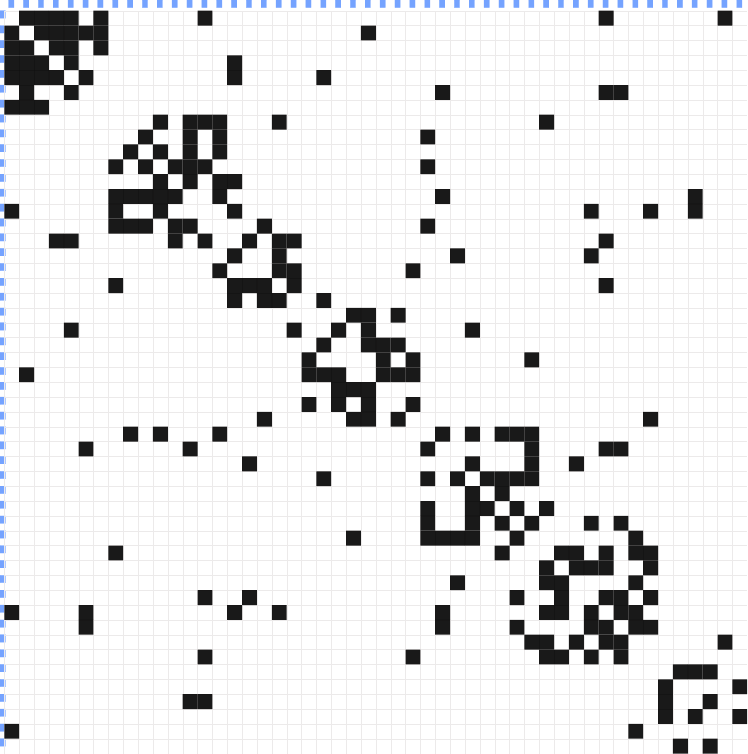}}
 \\ \vspace{-1pt} 
 \makecell{\includegraphics[width=11pt]{img/icons/loose-separable}}&
 \makecell{\includegraphics[width=0.14\textwidth]{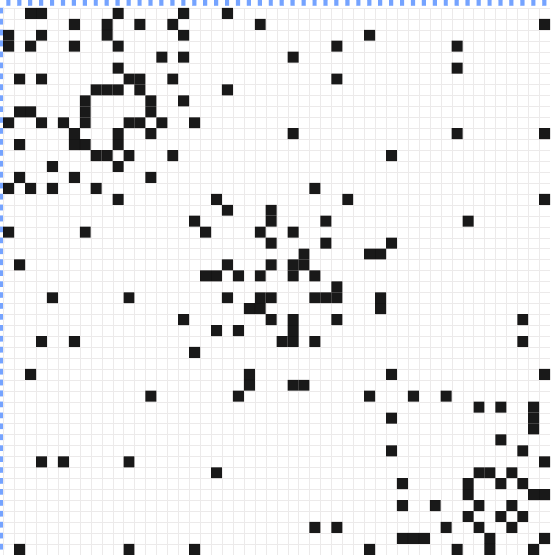}}&
 \makecell{\includegraphics[width=0.14\textwidth]{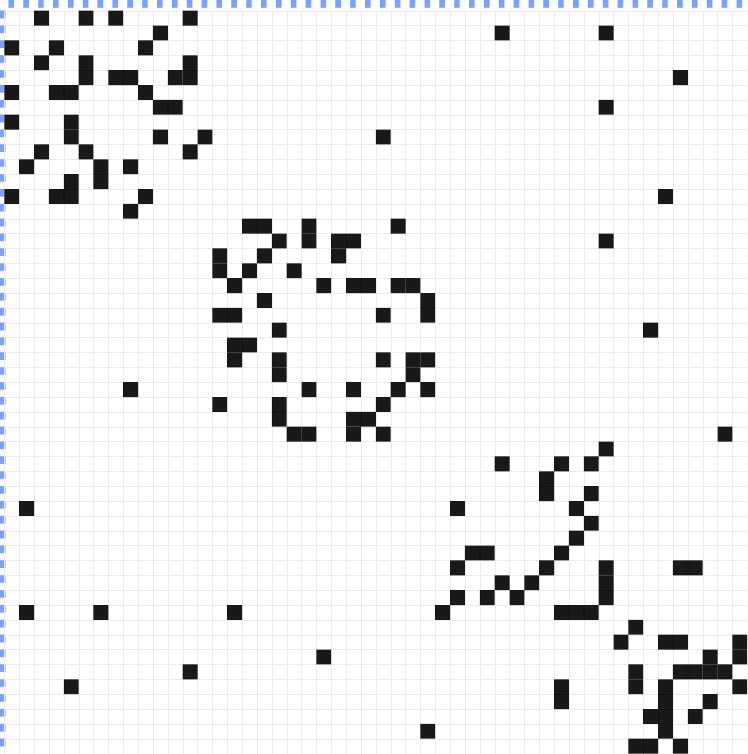}}&
 \makecell{\includegraphics[width=0.14\textwidth]{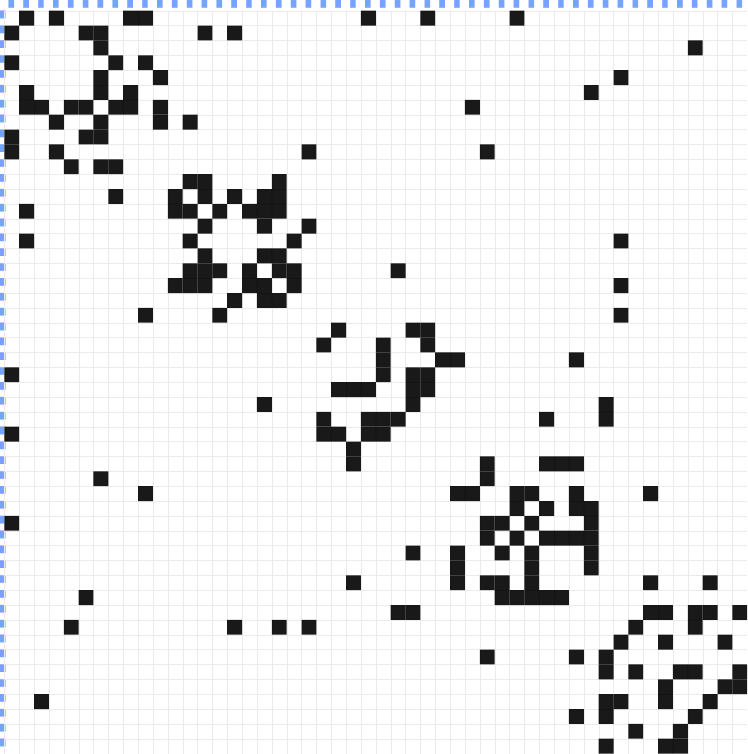}}&
 \makecell{\includegraphics[width=0.14\textwidth]{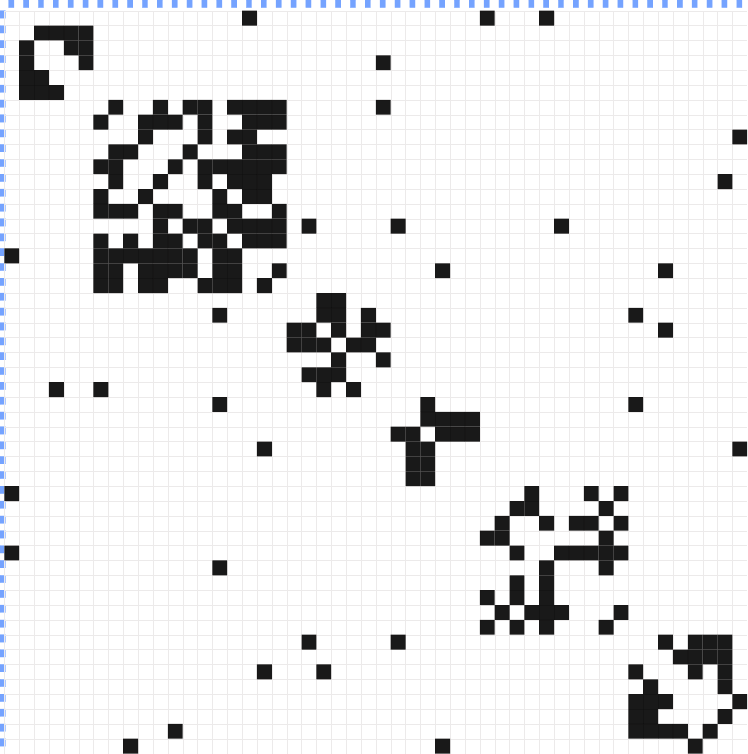}}&
 \makecell{\includegraphics[width=0.14\textwidth]{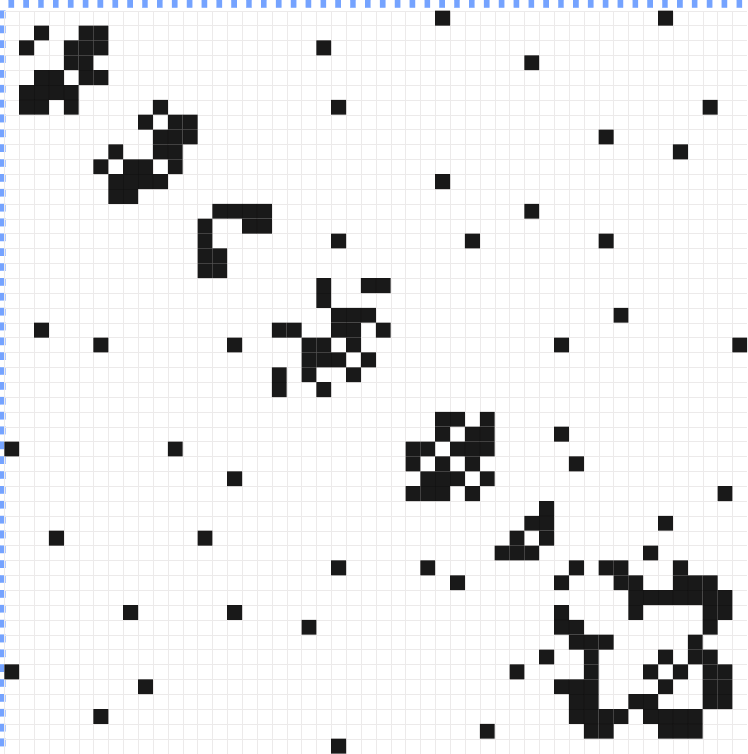}}
 \\ \vspace{-1pt}
 \makecell{\includegraphics[width=11pt]{img/icons/compact-inseparable}}&
 \makecell{\includegraphics[width=0.14\textwidth]{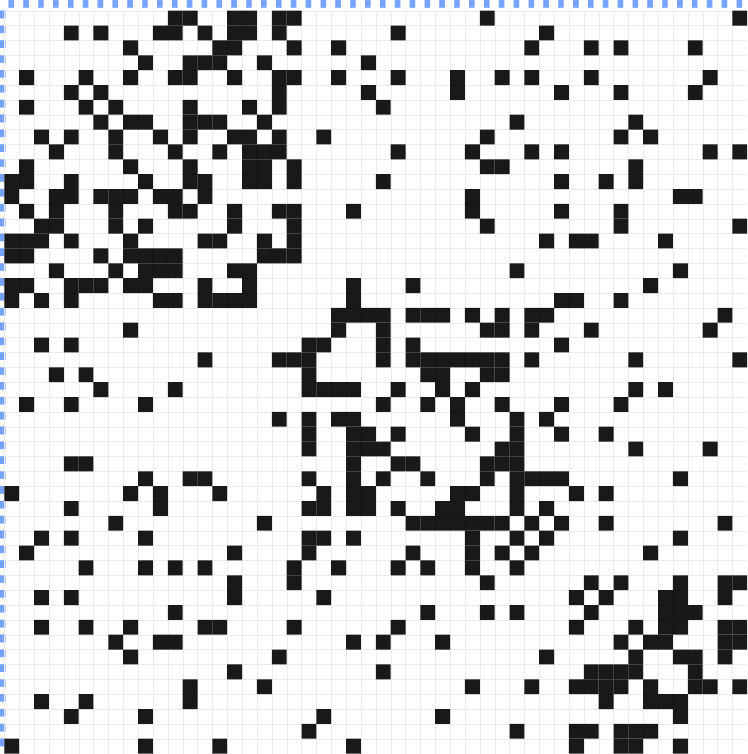}}&
 \makecell{\includegraphics[width=0.14\textwidth]{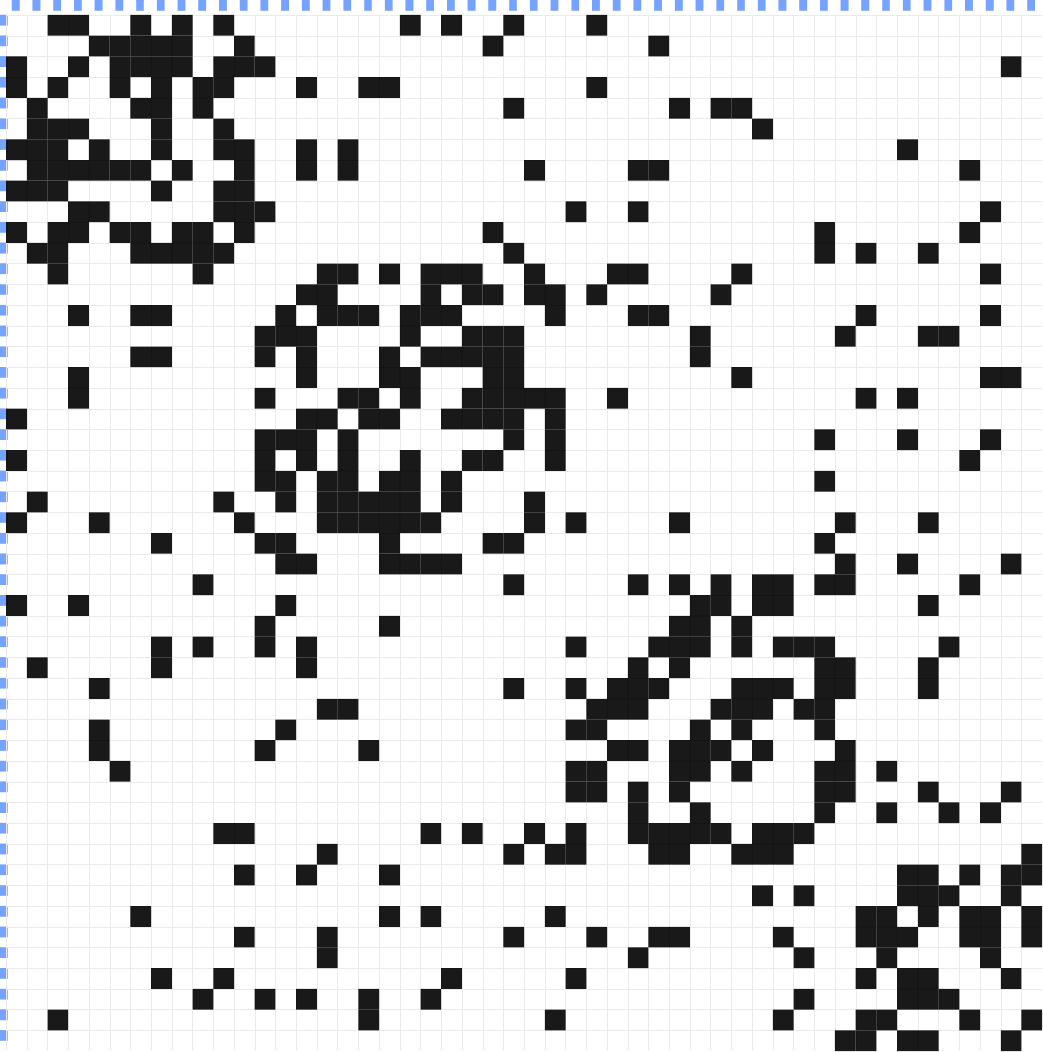}}&
 \makecell{\includegraphics[width=0.14\textwidth]{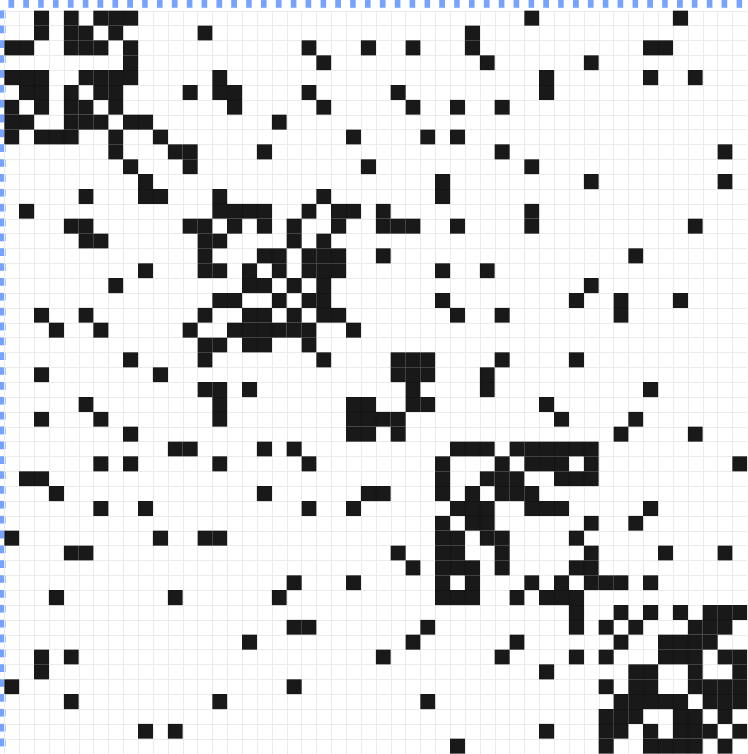}}&
 \makecell{\includegraphics[width=0.14\textwidth]{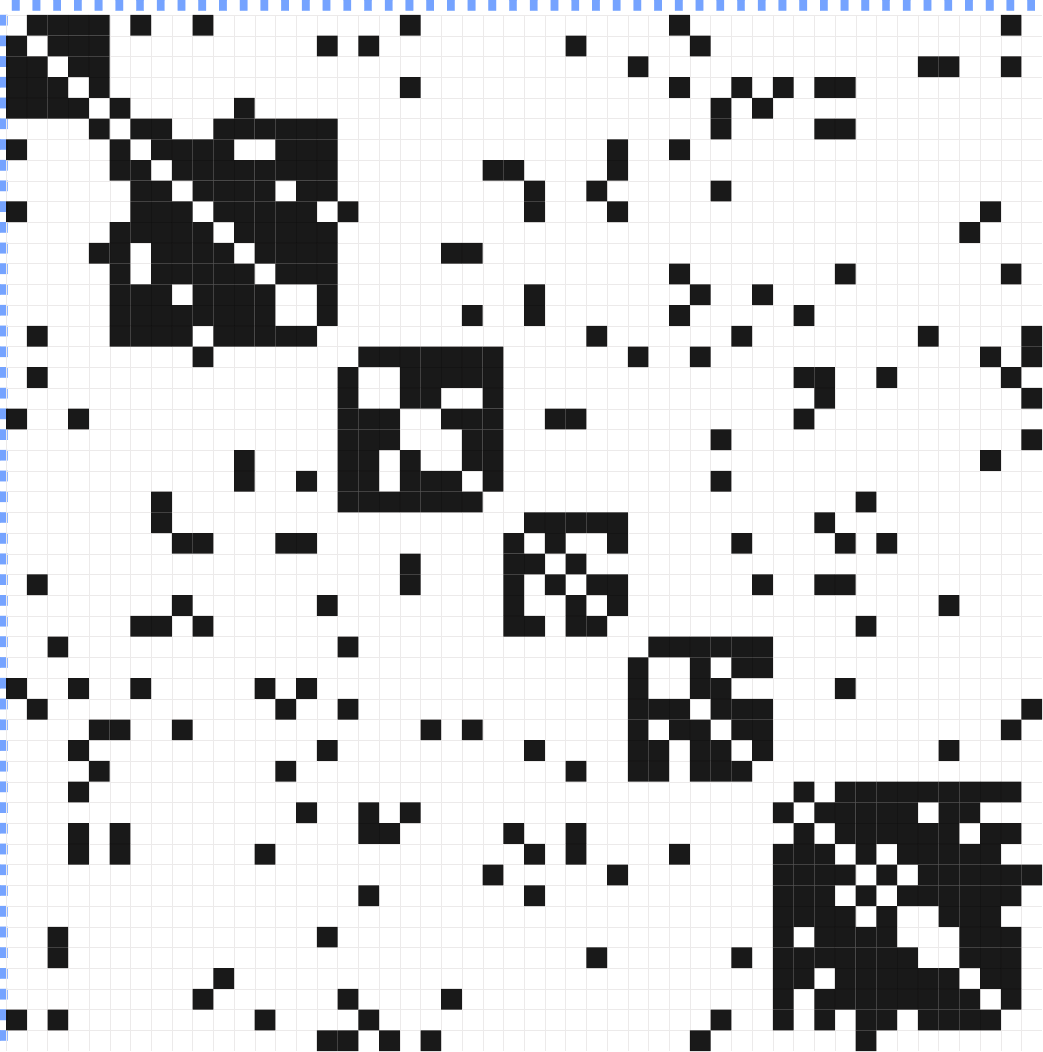}}&\makecell{\includegraphics[width=0.14\textwidth]{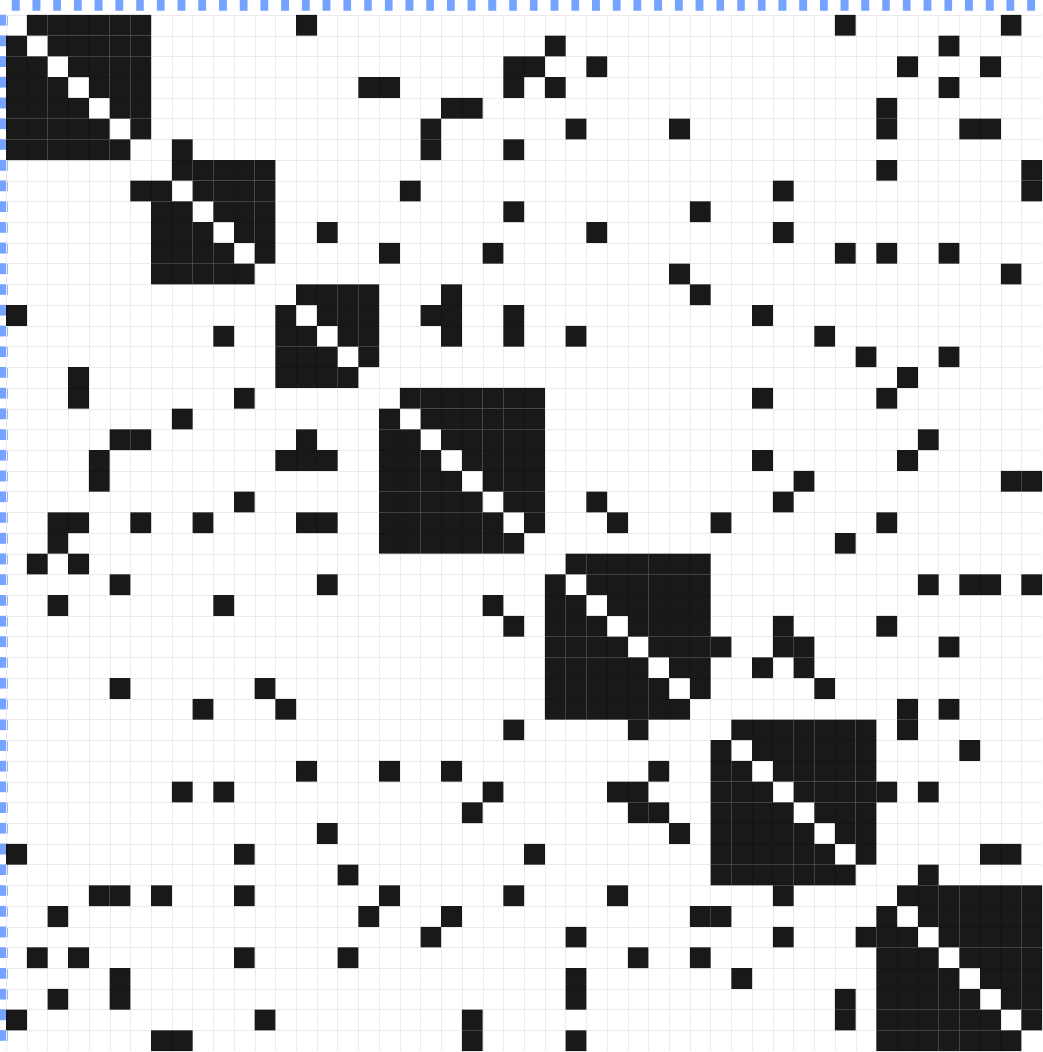}}
 \\
 \makecell{\includegraphics[width=11pt]{img/icons/compact-separable}}&
 \makecell{\includegraphics[width=0.14\textwidth]{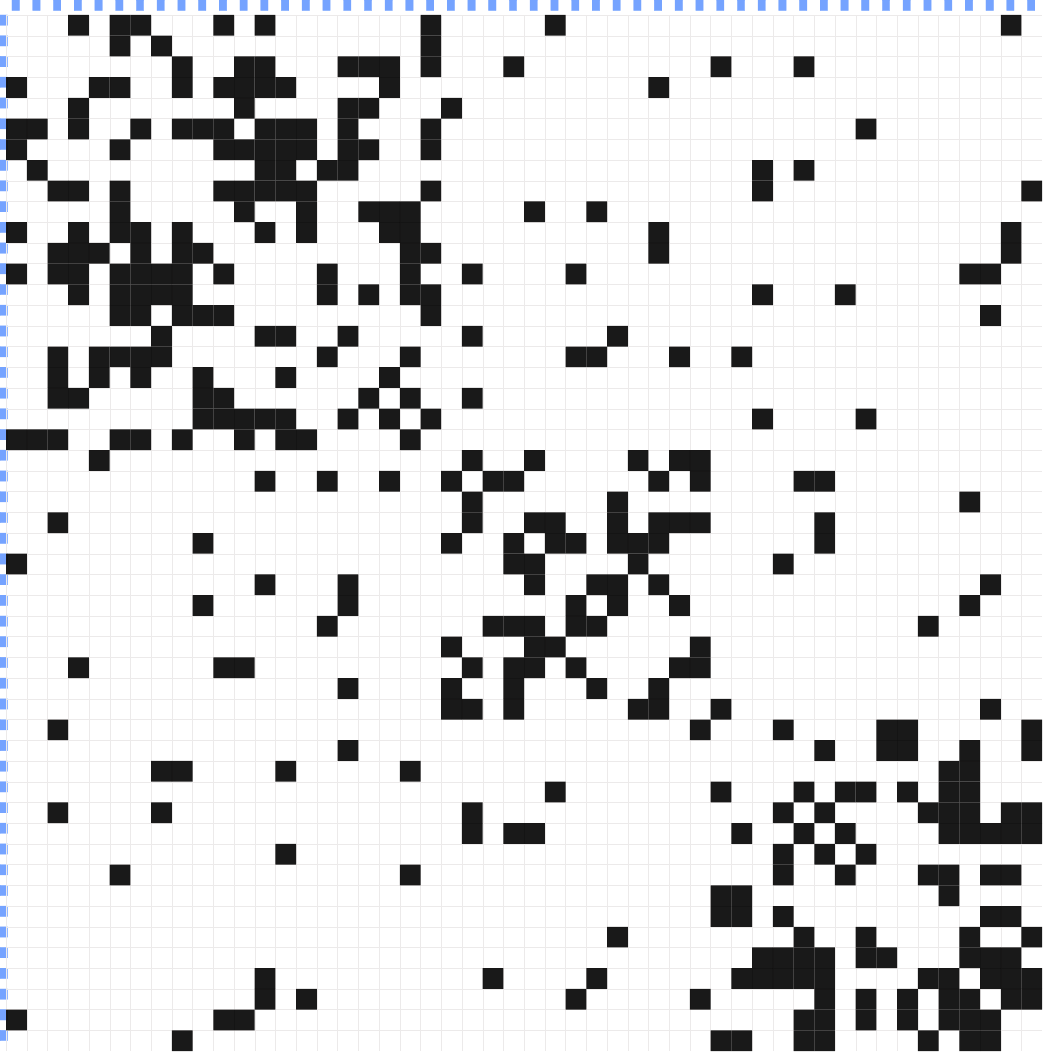}}&
 \makecell{\includegraphics[width=0.14\textwidth]{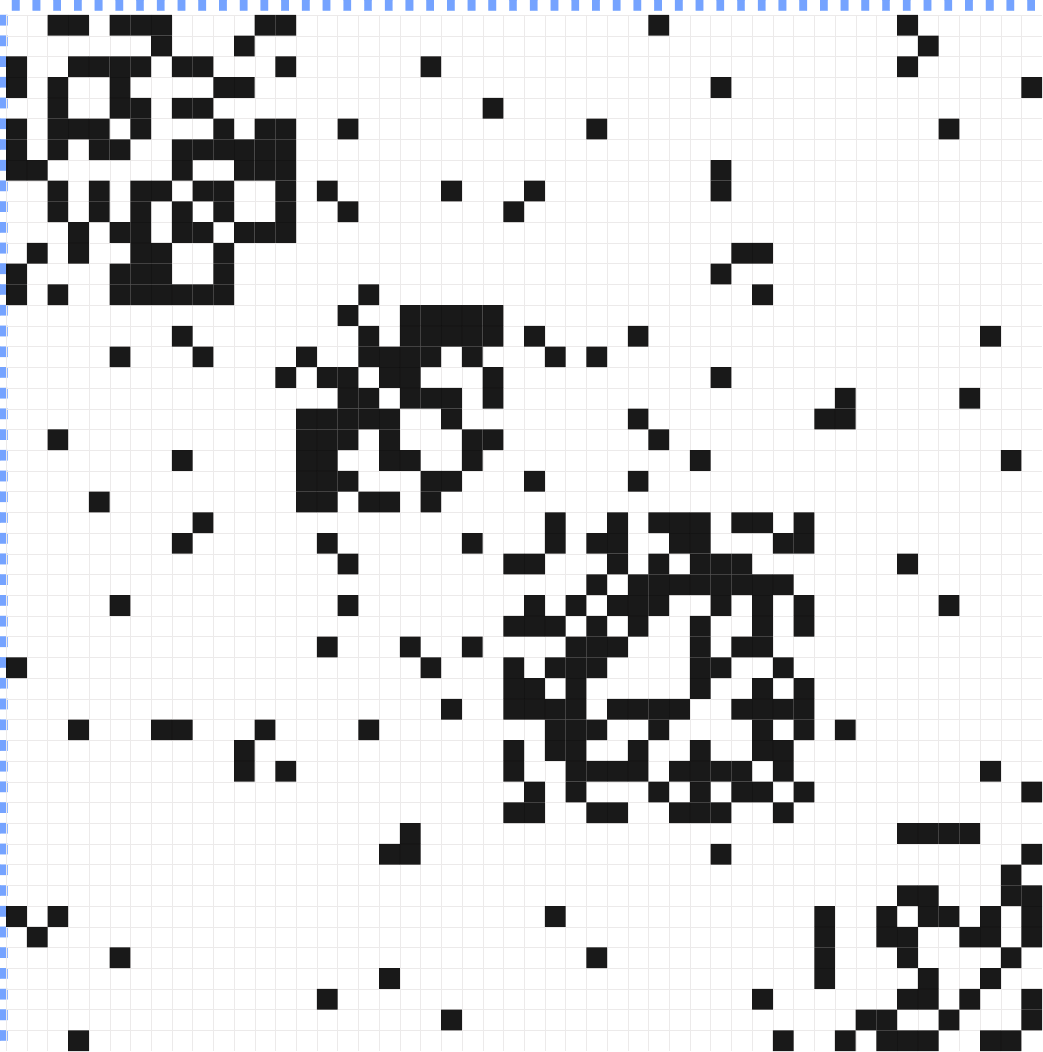}}&
 \makecell{\includegraphics[width=0.14\textwidth]{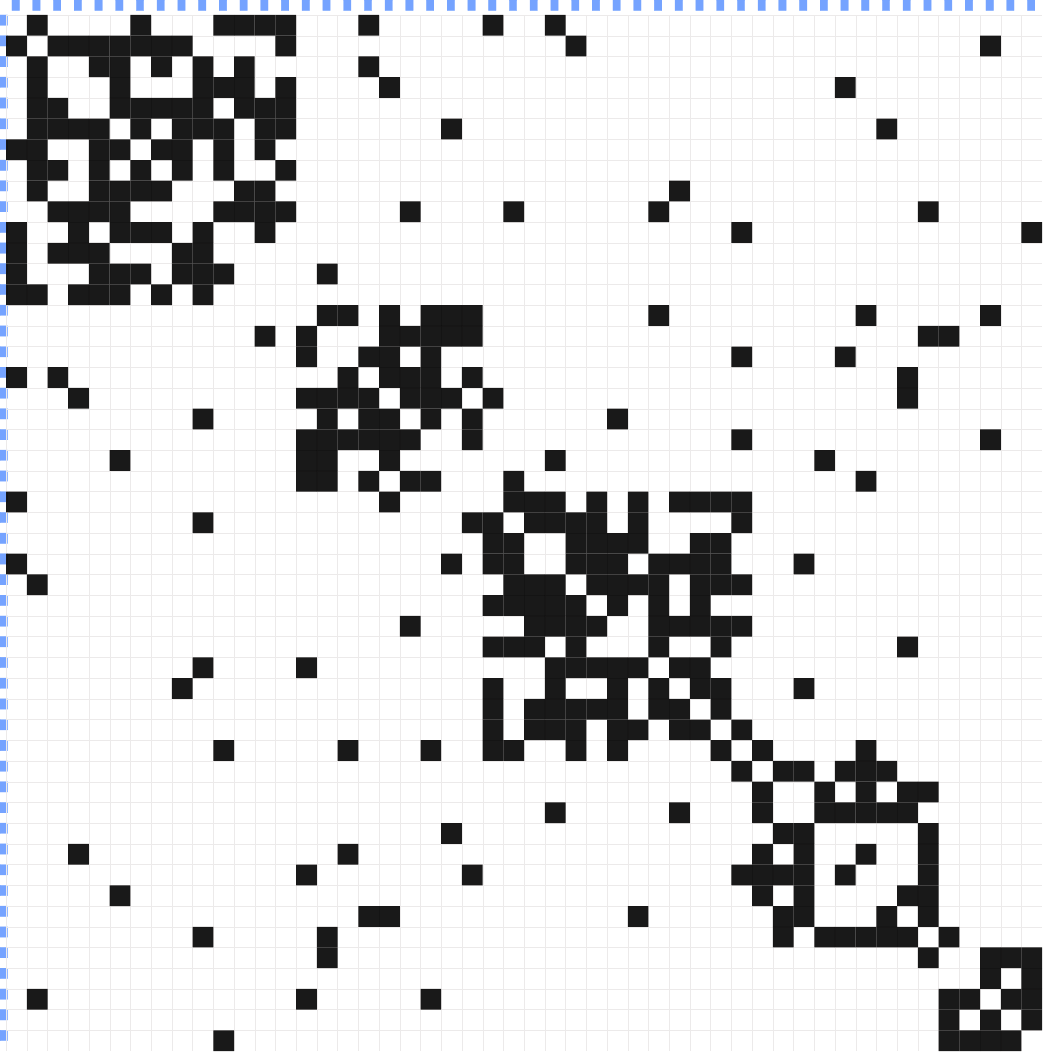}}&
 \makecell{\includegraphics[width=0.14\textwidth]{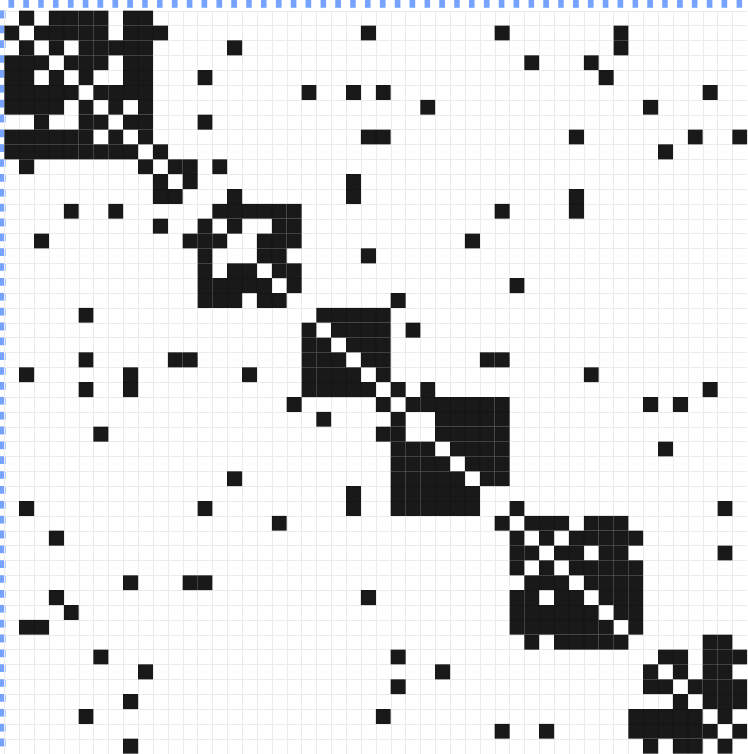}}&\makecell{\includegraphics[width=0.14\textwidth]{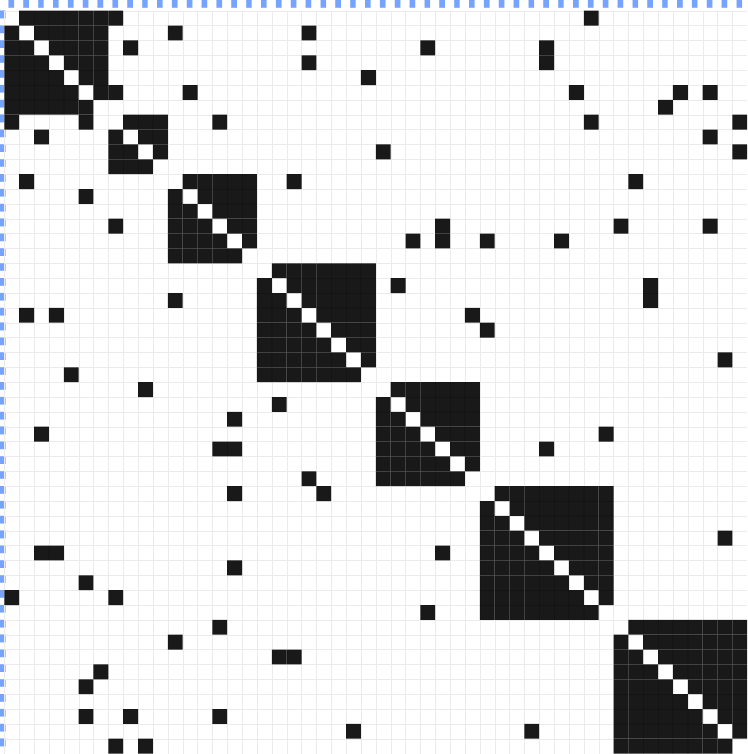}}
 \end{tabular}
 \endgroup
 \caption{Visual assessment of clustering tendency based on the adjacency matrix of the 50-node graphs in their original node order.}

 \label{fig:VAT}
\end{figure*}

A pragmatic validation measure to ensure the clustering tendency of the generated graphs consisted in visually inspecting the corresponding adjacency matrices to verify the presence of characteristic diagonal blocks. The rows and columns of the matrix were ordered according the original node order provided by the graph generator. In essence, such a visual approach to assess clustering tendency is tantamount to the VAT method and its many variants~\cite{VATsurvey} which also look for diagonal blocks in matrices, albeit originally for multivariate data in general. This approach also leverages prior knowledge about the effectiveness of matrix visualizations in showing clusters, subject to vertex seriation. \autoref{fig:VAT} shows such matrices for the 50 node-graphs for all four cluster type conditions, retaining the original node order provided by the graph generator. Clustering tendency is at stake in the top three rows, where clusters are loose and/or inseparable. Yet, we can see the right number of diagonal blocks in all cases, the signal being admittedly faint in the top left case.

\paragraph{Visualization types and settings}
Meidiana et al.~\cite{Meidiana2019AQM} showed that Linlog is the best for cluster detection. Also, Jacomy et al.~\cite{ForceAtlas2} argue that Force Atlas2 with Linlog energy is faster and leads to a better cluster definition than other layouts like Fruchterman-Reingold \cite{Fruchterman}, Yifan-Hu and Force Atlas2 without Linlog energy. We study Linlog, Backbone and sfdp, three performant layouts from the study of Meidiana et al.
We generated the Linlog layouts using Gephi~\cite{ICWSM09154} v.~0.10.1 based on its implementation of the Forceatlas2 algorithm with Linlog~\cite{linlog2004} mode on, and with the node overlap prevention mode on. We increased the edge width in these diagrams to improve link visibility. The Backbone layout was generated using VisOne v.~2.26, with backbone type = `Quadrilateral Simmelian' (recommended in~\cite{nocaj2015untangling}), backbone strength type = `redundancy' and compared edge = `true'. The sfdp layouts came from Graphviz v.~2.43.0 with the default parameters. We also implemented three types of orderable node-link layouts in D3, namely radial \nestedIcon{img/icons/radial.pdf}, arc\nestedIcon{img/icons/Arc.pdf} and symmetric arc\nestedIcon{img/icons/ArcSym.pdf} layouts. For each graph, we applied two node ordering methods: the crossing reduction method from Reorderjs~\cite{Fekete2015ReorderjsAJ} v.~2.2.4, i.e., the barycenter heuristic~\cite{Mkinen2005TheBH}, and the optimal leaf ordering method~\cite{BarJoseph2001FastOL} with average linkage and Jaccard distance from scipy v.~1.11.4. We also generated the stimuli corresponding to the original node order provided by the graph generator (denoted as GEN hereafter). Finally, each stimulus is saved as an SVG image which is scaled to occupy all the available canvas, with no possible user interaction. To reduce digital eyestrain due to screen brightness~\cite{eyestrain}, the stimuli were rendered with white semi-transparent links (and blue nodes) on a black background (as in \autoref{fig:drawing}), whereas most images in this paper use the reverse encoding to save ink. All stimuli are provided in supplemental material. Digital eyestrain is yet a complex phenomenon for which we still lack definitive guidelines.

\subsection{Hypotheses}

Below, we list our hypotheses about the visual saliency of clusters in orderable and force-directed node-link layouts across the cluster compactness and separability quadrants. Given the known ground truth of the graph generation process, we use the users' ability to count clusters as a proxy of cluster visual saliency. We use the same accuracy score as Okoe et al.~\cite{Okoe2019NodeLinkOA}, defined as:
$Acc = max(0,1-\frac{\norm{PA - TA}}{\norm{TA}})$, where PA stands for participant answer and TA for the true answer.\vspace{0.3em} 

 \noindent\ding{110}\,\textbf{H1} Graph clusters are more salient visually in orderable layouts \nestedIcon{img/icons/Arc.pdf} \nestedIcon{img/icons/ArcSym.pdf} \nestedIcon{img/icons/radial.pdf} compared to force-directed layouts \nestedIcon{img/icons/NL.pdf}, for all combinations of compactness and separability except for \nestedIcon{img/icons/compact-separable.pdf} (compact, well-separated) clusters, which is the ideal scenario for force-directed layouts.\vspace{0.3em} 
 
 \noindent\emph{Rationale:} Prior studies~\cite{Riche2007MatLinkEM, Nobre2020EvaluatingMN, Okoe2019NodeLinkOA} on various cluster perception tasks suggest that ordered matrices improve the visual saliency of clusters (which appear as dense blocks of links). We expect that node ordering in arc \nestedIcon{img/icons/Arc.pdf}, symmetric arc \nestedIcon{img/icons/ArcSym.pdf} and radial \nestedIcon{img/icons/radial.pdf} layouts will also create locally high link concentrations easing cluster perception, as in \autoref{fig:teaser}.\vspace{0.3em}

 \noindent\ding{110}\,\textbf{H2} Graph clusters are more salient visually in orderable node-link diagrams ordered by the optimal leaf ordering method (OLO) than the layouts ordered by the crossing reduction method (CR).\vspace{0.3em} 

 \noindent\emph{Rationale:} Based on the study by Behrisch et al.~\cite{behrisch2016ordering}, OLO~\cite{BarJoseph2001FastOL} which is a clustering based method is preferable for cluster perception tasks to other types of ordering methods, in this case CR~\cite{Mkinen2005TheBH}. In a sense, the visual saliency of clusters resulting from CR is an indirect product of reducing link crossings in a network. A good example of CR's ability to reveal clusters is provided in \autoref{fig:miserables}. Yet, OLO is designed to reveal clusters, and should consistently surpass CR for cluster perception.\vspace{0.3em} 

 \noindent\ding{110}\,\textbf{H3} Graph clusters are more salient visually in symmetric arc diagrams \nestedIcon{img/icons/ArcSym.pdf} than in plain arc diagrams \nestedIcon{img/icons/Arc.pdf}.\vspace{0.3em} 

 \noindent\emph{Rationale:} According to the Gestalt principles, the elements of symmetric components tend to be seen as belonging together~\cite{gestalt-graphs}. Also, symmetry preservation is a known
 aesthetic rule in graph drawing~\cite{graph-drawing-book, aestheticsSurvey}. Hence, the round disc-shaped clusters in symmetric arc layouts might improve cluster perception as opposed to half discs in plain arc layouts.

\begin{figure*}[h]
 \centering

 \includegraphics[width=0.75\textwidth]{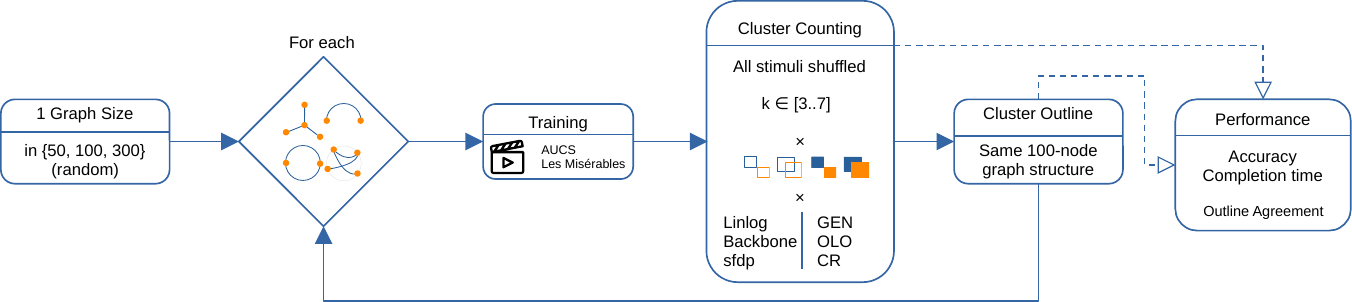}
 \caption{Flowchart of the study procedure. Credit: video icon by Ilham Fitrotul Hayat from Noun Project (CC BY 3.0).}
 \label{fig:study_procedure}
\end{figure*}

\subsection{Study Design}

This user study combines between-subject and within-subject designs. We adopted a between-subject approach at the level of network size. That is, participants were exposed to a single network size only throughout their session. Within a session, we adopted a within subject approach where we varied all other factors as detailed in Section~\ref{sec:proc}.

We ran three pilot tests with a few users each to troubleshoot any usability issues and validate our data collection procedure, our estimation of task completion time and the total duration of each participant session. We aimed to keep the total session duration under one hour.
We used Prolific to recruit English-speaking online participants, 

equipped with a laptop or desktop computer. 
Like in similar user studies~\cite{Abdelaal2022ComparativeEO,Okoe2019NodeLinkOA}, we aimed for 40 to 50 participants for each condition. In total, we recruited $n=139$ participants, mainly between the ages of 18 and 34 (64\%); the remaining 36\% were above 35. Among them, 57.24\% declared to be males, 44.27\% females and the rest (0.76)\% did not declare their gender. 
Most participants were geographically located in the UK(27\%), South Africa (17\%), the USA (9\%) and Portugal (9\%).

\subsection{Study Procedure}
\label{sec:proc}
\begin{figure*}[h]
 \centering
 \begin{tabular}{cccc}
 $\vcenter{\hbox{\includegraphics[width=0.2\textwidth]{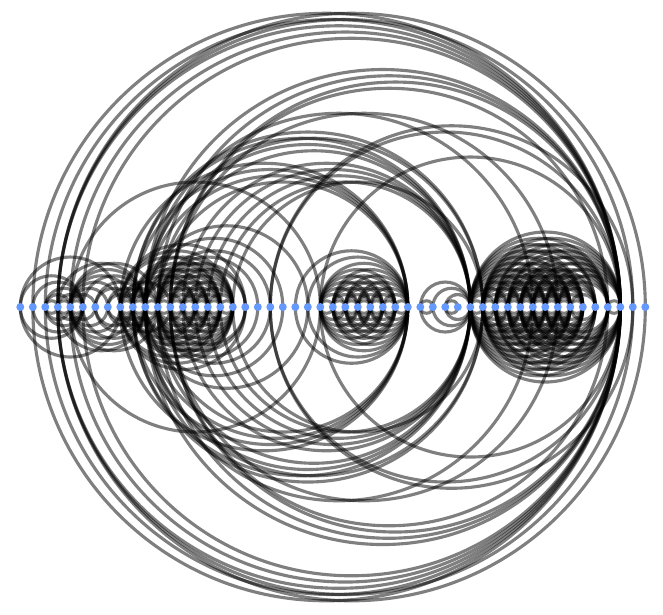}}}$
 &
 $\vcenter{\hbox{\includegraphics[width=0.2\textwidth]{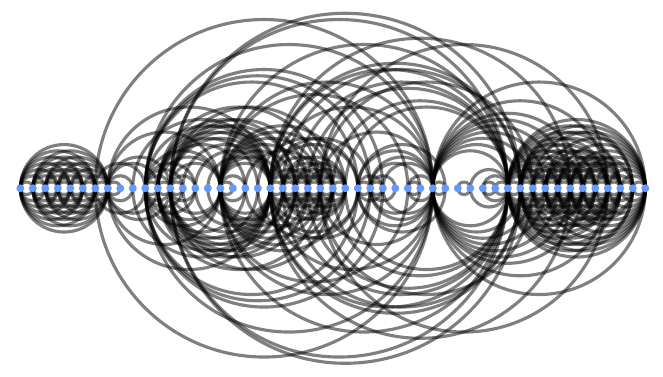}}}$
 &
 $\vcenter{\hbox{\includegraphics[width=0.2\textwidth]{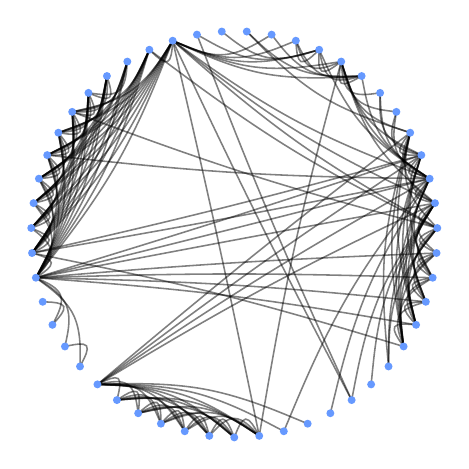}}}$
 &
 $\vcenter{\hbox{\includegraphics[width=0.2\textwidth]{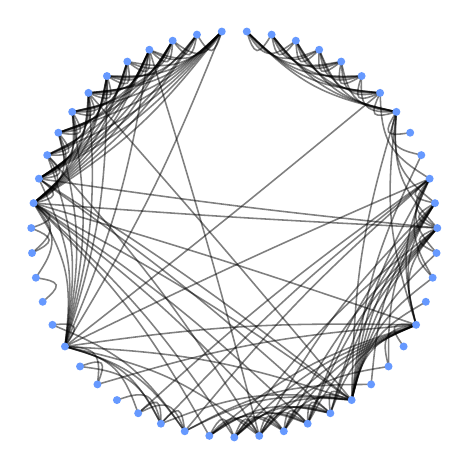}}}$ \\
 \small (a) Symmetric arc layout, OLO & \small (b) Symmetric arc layout, CR & \small (c) Radial layout, OLO & \small (d) Radial layout, CR\\
 \end{tabular}
 \begin{tabular}{ccc}
 $\vcenter{\hbox{\includegraphics[width=0.2\textwidth]{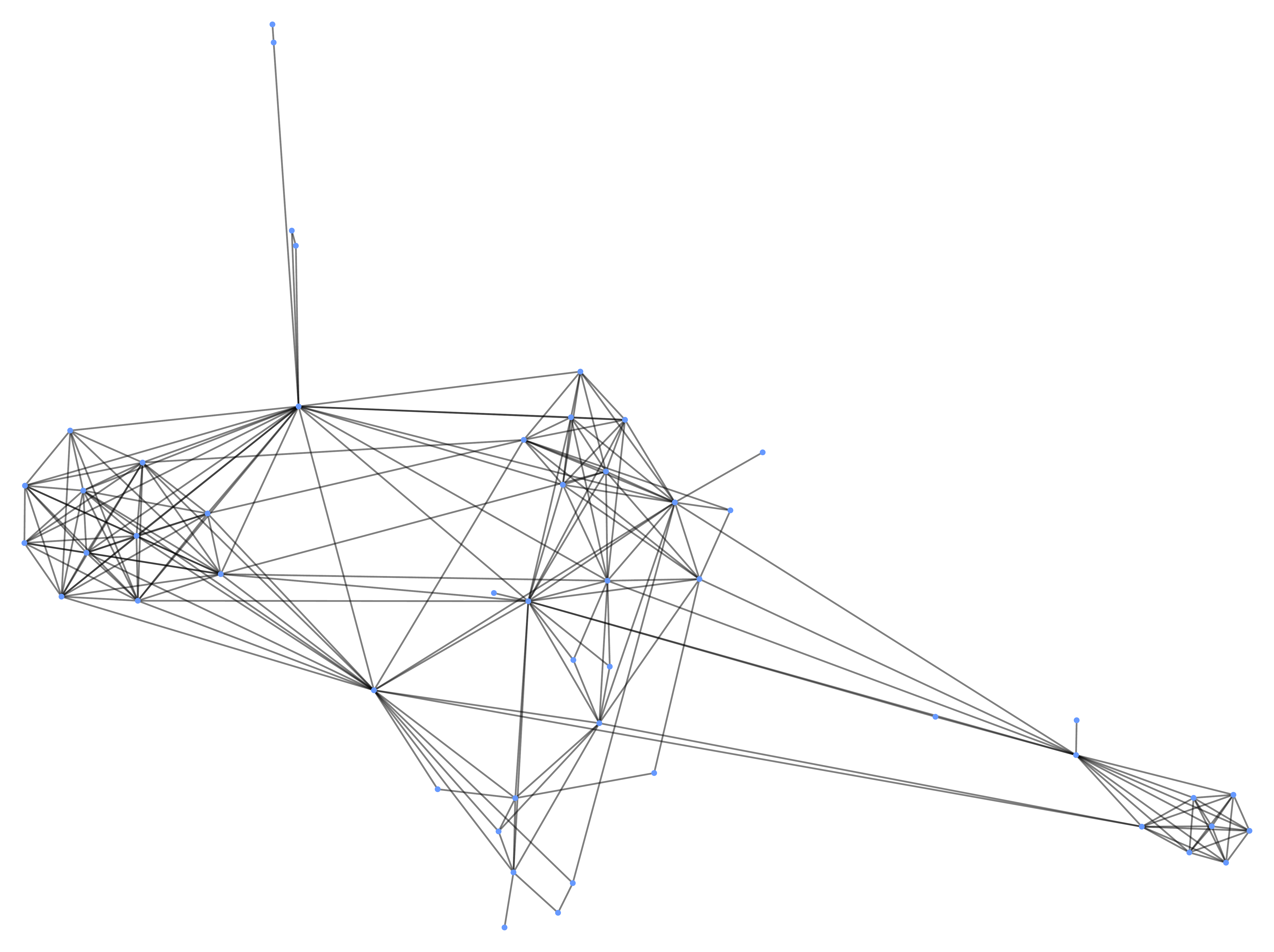}}}$
 &
 $\vcenter{\hbox{\includegraphics[width=0.2\textwidth]{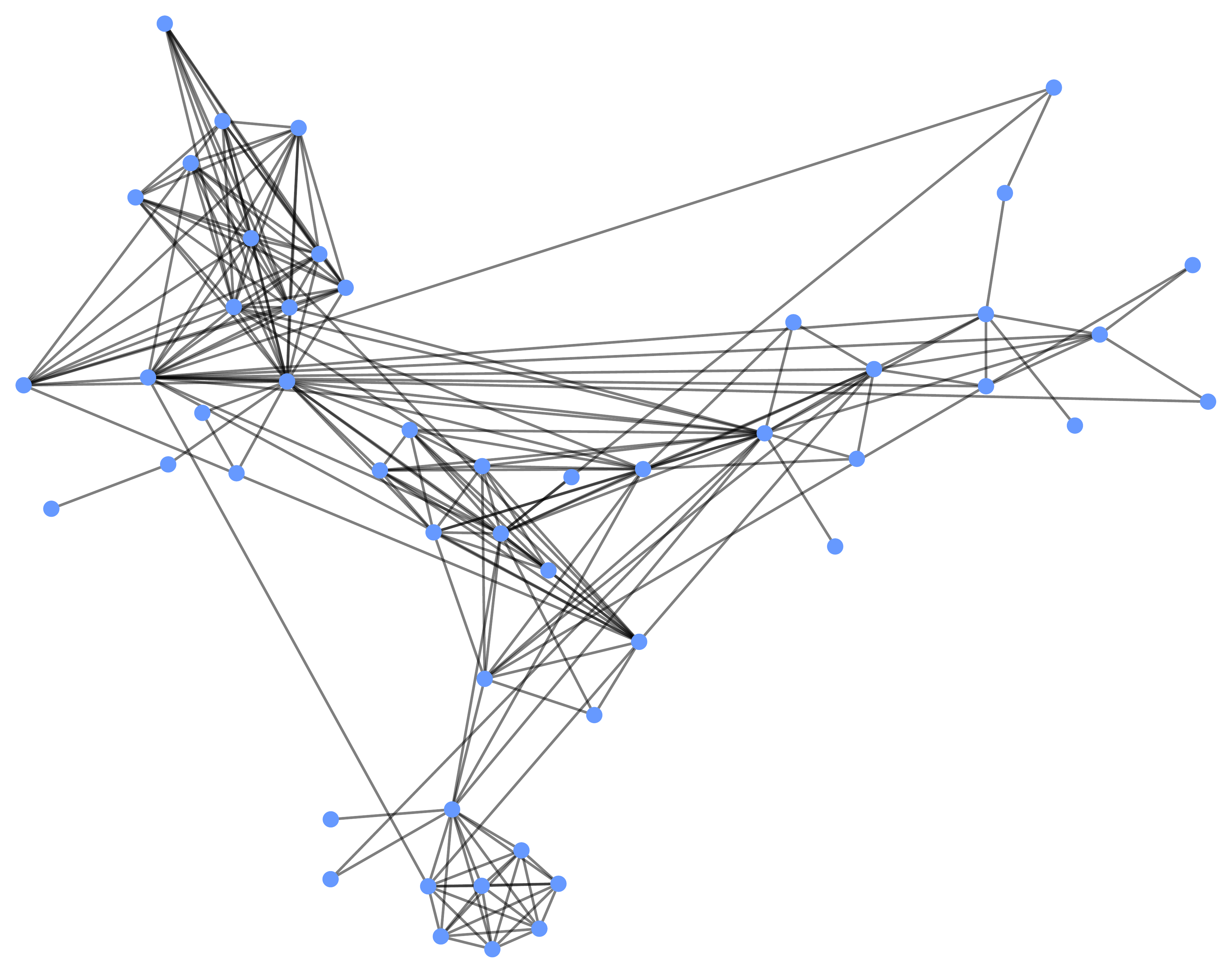}}}$
 &
 $\vcenter{\hbox{\includegraphics[width=0.2\textwidth]{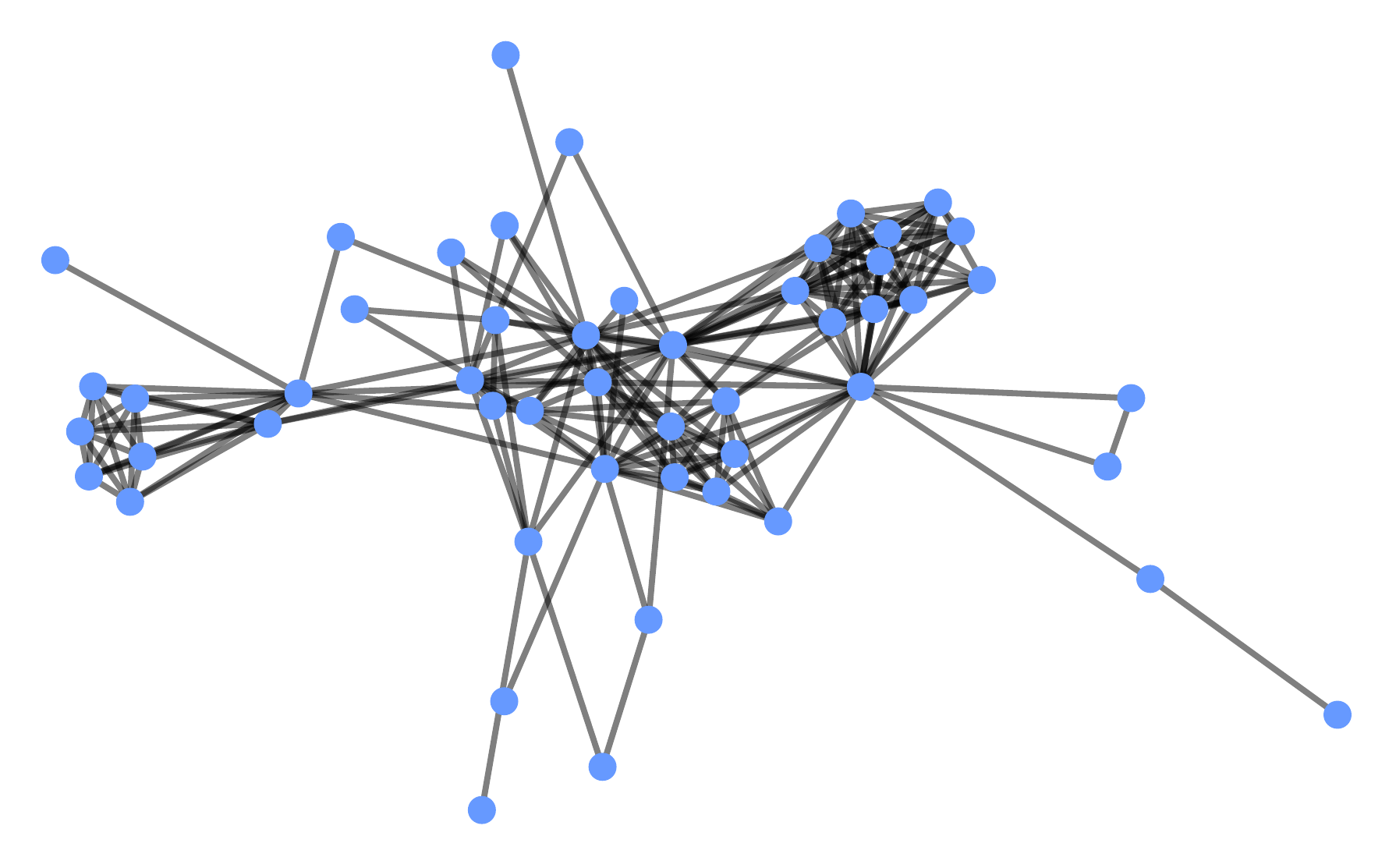}}}$
 \\
 \small(e) Linlog layout & \small(f) Backbone layout & \small(g) sfdp layout\\
 \end{tabular}
 \vspace{-0.5em}
 \caption{A 51-node extract from the co-occurrence network of the characters of \emph{Les Mis\'erables}, the novel of Victor Hugo.}

 \label{fig:miserables}
\end{figure*}

\begin{figure*}[h]
 \centering
 \begingroup
 \setlength{\tabcolsep}{2pt} 
 \begin{tabular}{cccc}
 \includegraphics[height=0.2\textwidth,alt={A 100-node network with four clusters laid out by the Linlog layout. The outlines drawn by the study participants show that most of them saw five clusters, instead of four. It is yet safe to think that they generally understood the cluster identification task.}]{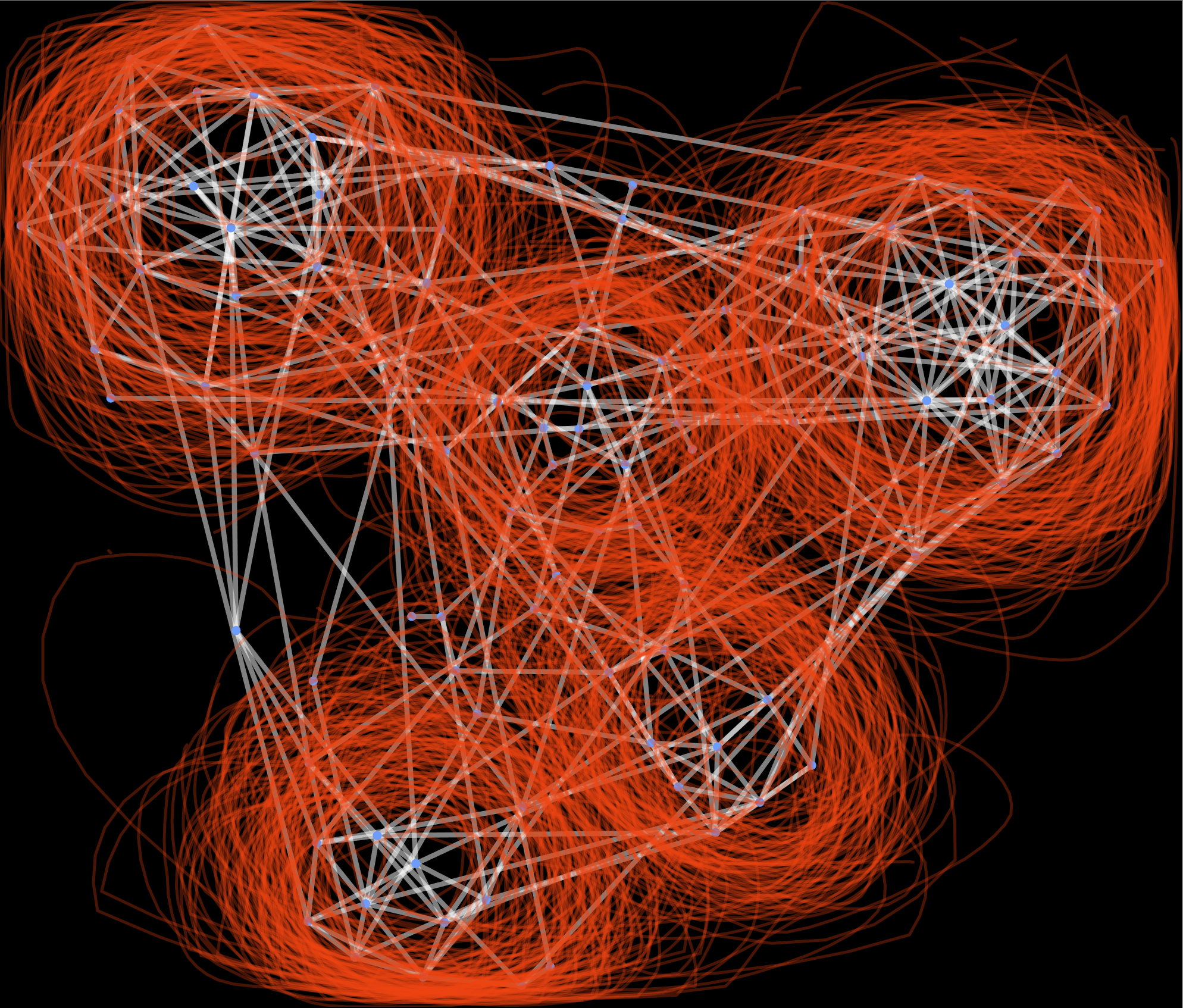}
 & 
 \includegraphics[height=0.2\textwidth,alt={A 100-node network with four clusters laid out by the arc layout. The outlines drawn by the study participants show that most of them saw four clusters, as they should. It is safe to think that they generally understood the cluster identification task.}]{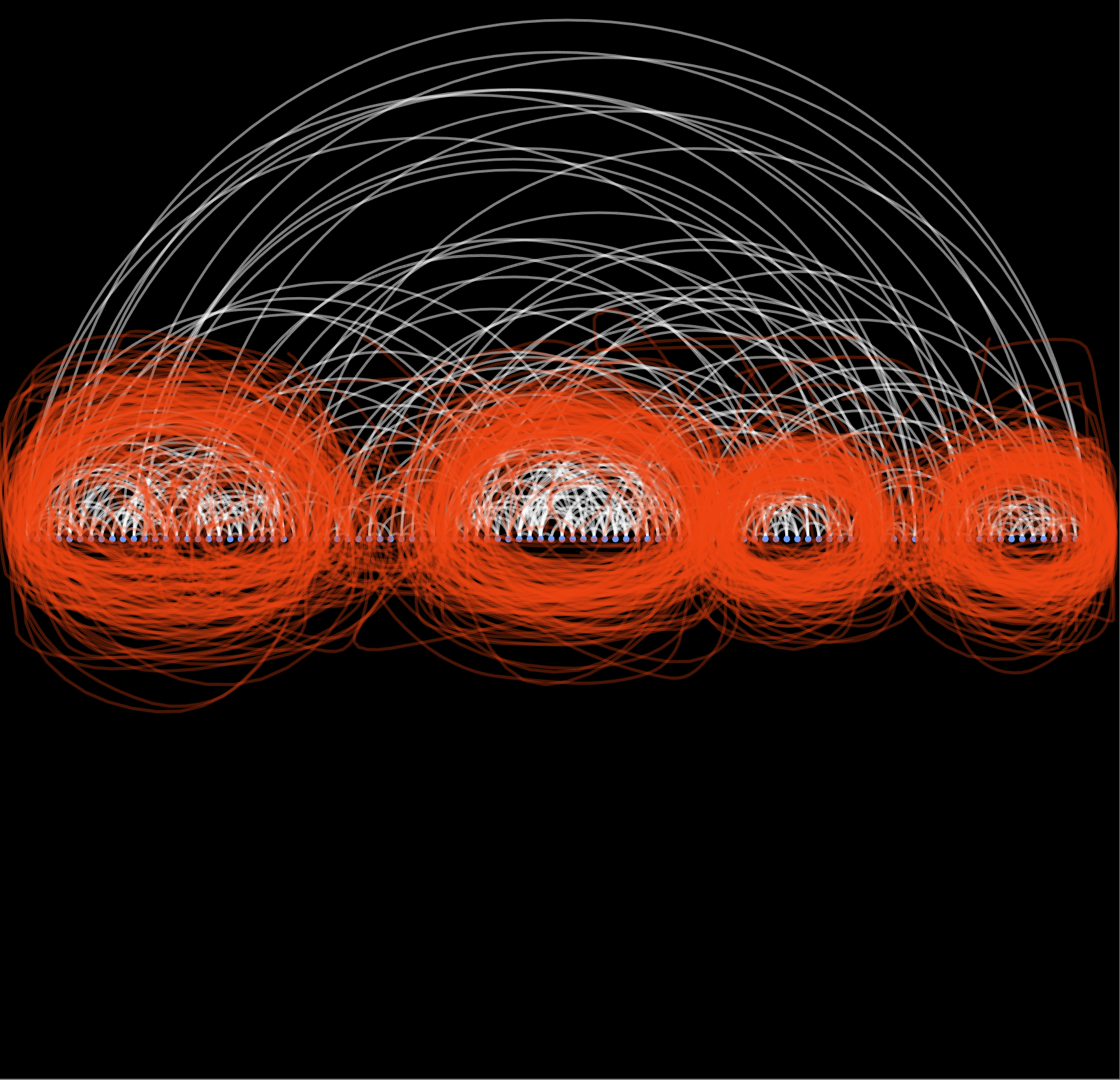}
 &
 \includegraphics[height=0.2\textwidth,alt={A 100-node network with four clusters laid out by the the symmetric arc layout. The outlines drawn by the study participants show that most of them saw four clusters, as they should. It is safe to think that they generally understood the cluster identification task.}]{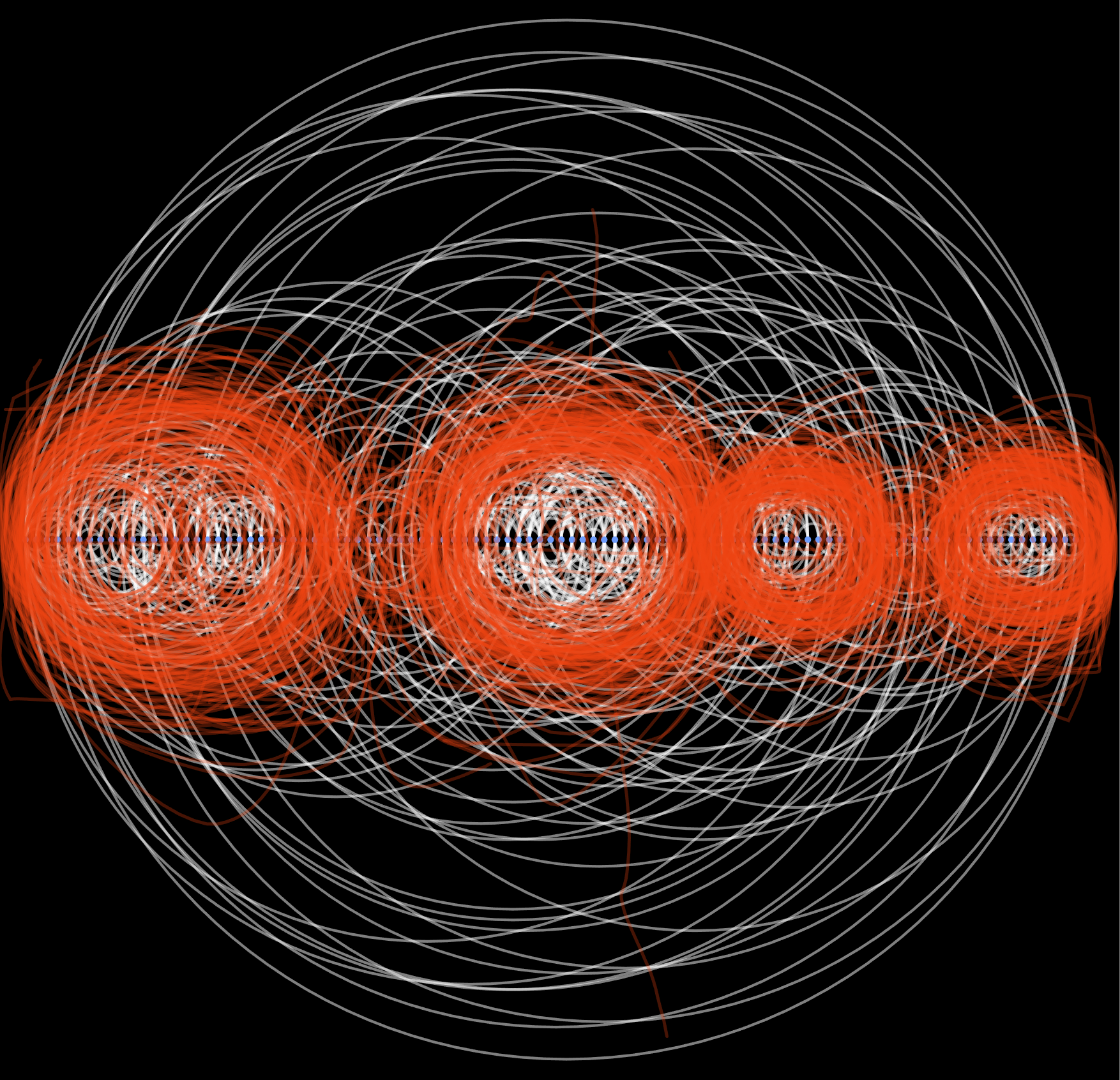}
 &
 \includegraphics[height=0.2\textwidth,alt={A 100-node network with four clusters laid out by the radial layout. The outlines drawn by the study participants show that most of them saw six clusters, instead of four. It is yet safe to think that they generally understood the cluster identification task.}]{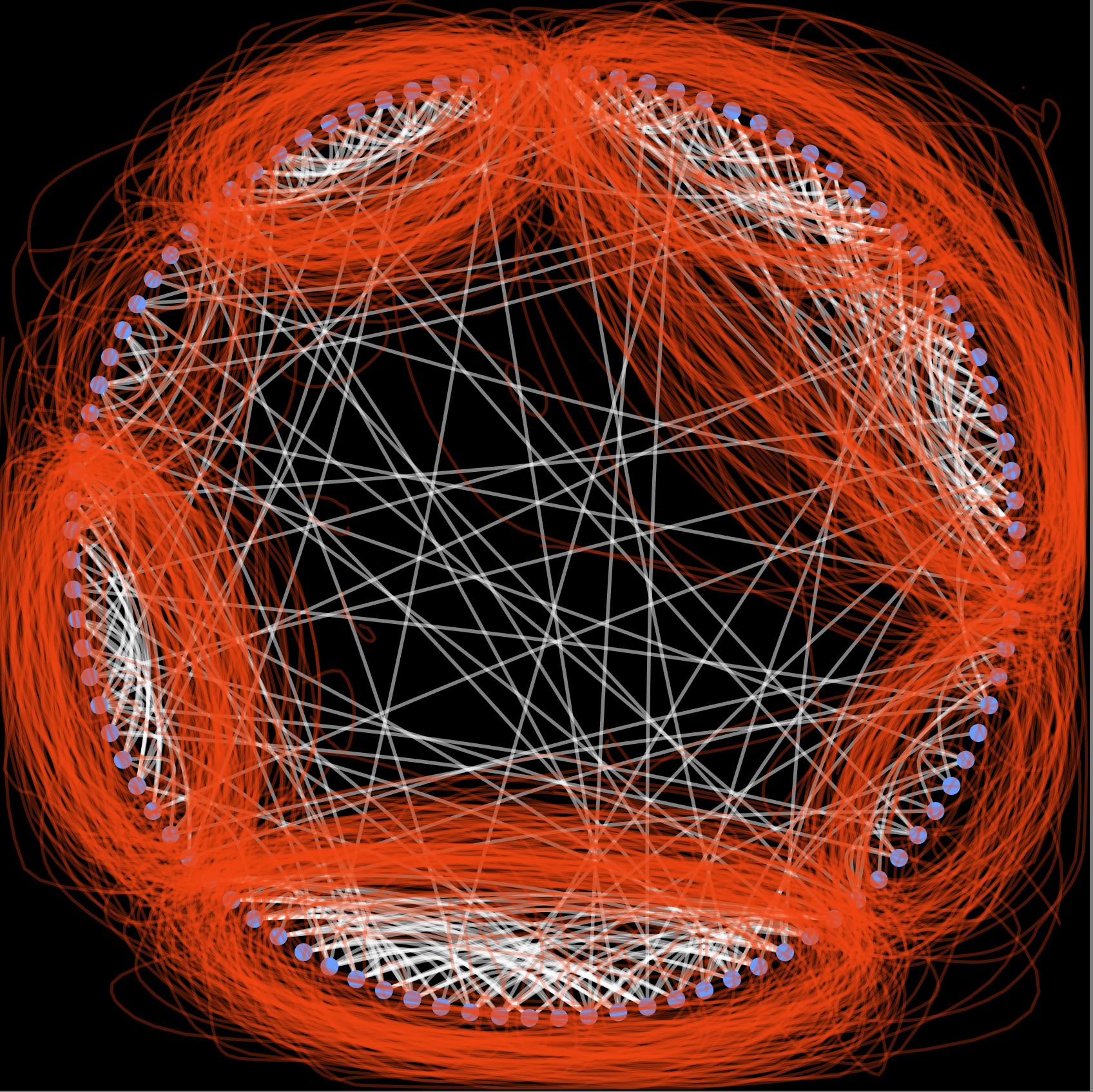}\\
 \small(a) Linlog layout & \small(b) Arc layout & \small(c) Symmetric arc layout & \small(d) Radial layout\\
 \end{tabular}
 \endgroup
 \vspace{-0.5em}
 \caption{The 100-node network used in the drawing task. The cluster outlines show that the users understood the cluster identification task.}

 \label{fig:drawing}
\end{figure*}

After signing the informed consent form, each participant was trained on scribbling on a drawing canvas using their mouse. This was needed later in the procedure to validate their understanding of the cluster identification task. 
\autoref{fig:study_procedure} gives an overview of the study procedure. All participants were exposed to the four visualization types \nestedIcon{img/icons/NL.pdf}\nestedIcon{img/icons/Arc.pdf}\nestedIcon{img/icons/ArcSym.pdf}\nestedIcon{img/icons/radial.pdf} in a random sequence. For each visualization type, participants went through the following steps: 
\begin{enumerate}[leftmargin=*]
 \item \emph{Training.} They watched a one-minute video explaining how to read the visualization, based on a social network scenario, presenting nodes as people, edges as friendship relationships, and clusters as friendship groups. The video showed a 51-node extract of the character co-occurrence network in \emph{Les Mis\'erables} novel~\cite{les_miserable_mbostock} with distinct clusters (see \autoref{fig:miserables}). The network construction was animated to ease comprehension. Users could replay the video. 
 Then, they took a training based on two sample networks: the extract of \emph{Les Mis\'erables} again and the AUCS network~\cite{rossi_towards_2015}. They chose the cluster count among multiple choices between 1 and 8. The correct answer was displayed as an image with an overlay outline of the individual clusters to teach them the visual signature of clusters in the visual metaphor at hand.
 \item \emph{Counting Task.} 
 For each visualization type, participants saw 60 stimuli in random order (4 cluster types $\times$ 5 cluster count values $\times$ 3 variants). The variants correspond either to the three node orderings (GEN, OLO, CR), if applicable, or the three force-directed layouts (Linlog, Backbone, sfdp).

 For each stimulus, users had to count the friendship groups and to choose the right answer between $[1..8]$ (for ground truth $k \in [3..7]$). This task is used to test all three hypotheses.

 \item \emph{Drawing Task.} Finally, users drew an outline around clusters in the 100-node stimuli having four compact and separable clusters (those in \autoref{fig:drawing}). As we did not have direct access to online participants, we needed concrete evidence that they individually understood the task. This might be used typically to exclude anomalous participants, which was not the case here.
\end{enumerate}
After an optional 2-minute break, participants moved on to the next visualization type and repeated the same steps.

Hence, all participants saw 240 stimuli in total ($4$ visualization types $ \times$ 60 stimuli). 
We used Mephisto~\cite{mephisto} to create and deploy the study, and collect data.
For each stimulus, we recorded the cluster count and the response times. 

\begin{figure}[b]
 \centering
 \begin{subfigure}[b]{\columnwidth}
 	\centering
 \includegraphics[width=\textwidth]{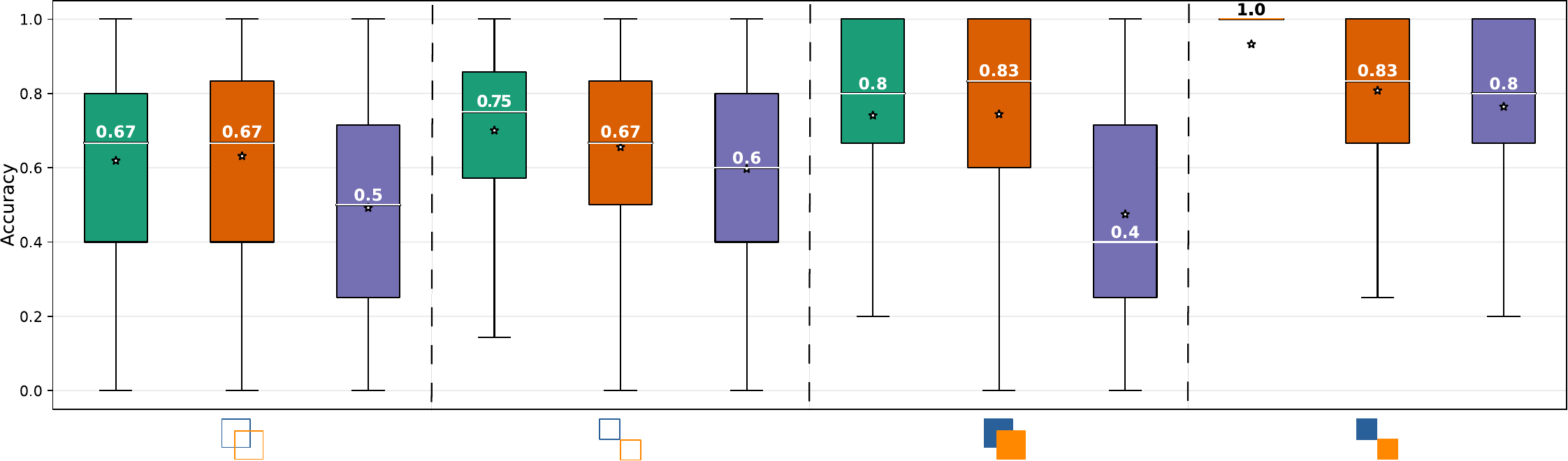}
 \vspace{-1em} 

 	\label{fig:all-nodelinks-acc}
 \end{subfigure}%
 \\
 \begin{subfigure}[b]{\columnwidth}
 	\centering
 \includegraphics[width=\textwidth]{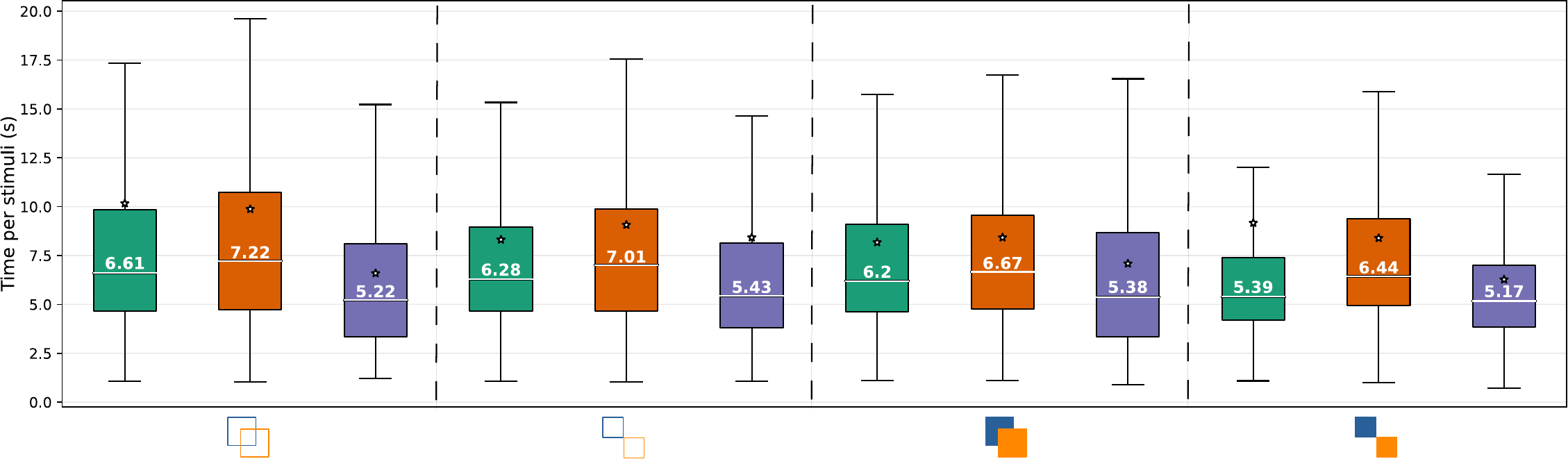}
 \vspace{-1em} 

 	\label{fig:all-nodelinks-time}
 \end{subfigure}%
 \\%
 \subfigsCaption{Cluster count accuracy (top) and completion time (bottom) for sizes=all, clusters=all, layouts=\{Linlog\square{linlog}, Backbone\square{backbone}, sfdp\square{sfdp}\}.}
 \label{fig:all-nodelinks}
\end{figure}

\section{Results}

In \autoref{fig:drawing}, we overlay all the cluster outlines drawn by all study participants in the cluster drawing task, for network size 100. It shows that for all four layouts, participants had a good understanding of what a cluster looked like. For this task, we deliberately chose easy stimuli from the \nestedIcon{img/icons/compact-separable.pdf} category. Choosing more ambiguous stimuli would have complicated our assessment of whether participants understood the task. 
This doubt being cleared, we can proceed with the analysis of results.

The Anderson-Darling test~\cite{anderson-darling} showed that accuracy and task completion time are not normally distributed.

Hence, we performed a one-tailed Kruskal-Wallis test and Bonferroni correction to determine whether the observed differences were statistically significant, with a corrected threshold of $p=0.05$. We discarded one single user who systematically answered 1 or 8 clusters (the two extreme options).

\subsection{Preliminary Remarks}

Regarding force-directed layouts (see~\autoref{fig:all-nodelinks}), users counted clusters faster and less accurately with sfdp than with Linlog and Backbone for all combinations of cluster compactness and separability. Linlog leads to consistently faster cluster count judgements than Backbone, while being statistically significantly more accurate for compact separable clusters and on par for other cluster types. Hence, all plots in this section use Linlog as a baseline for comparison with orderable layouts. 

\begin{figure*}[t]
\centering
 \includegraphics[height=0.21\textwidth]{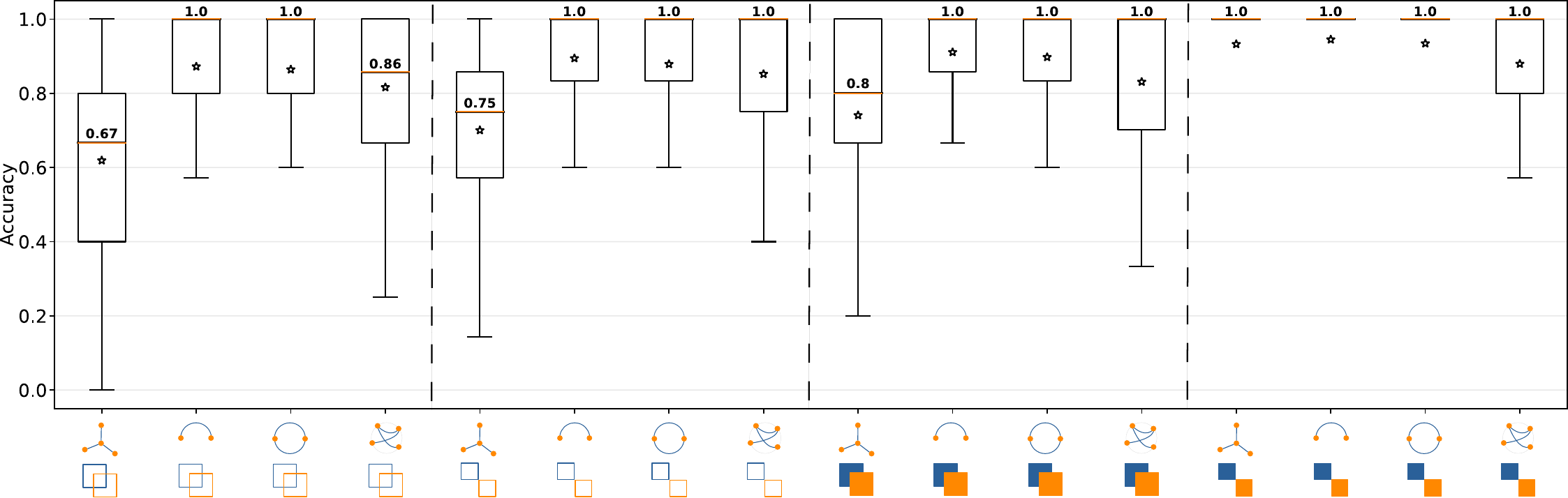}~

 \includegraphics[height=0.21\textwidth]{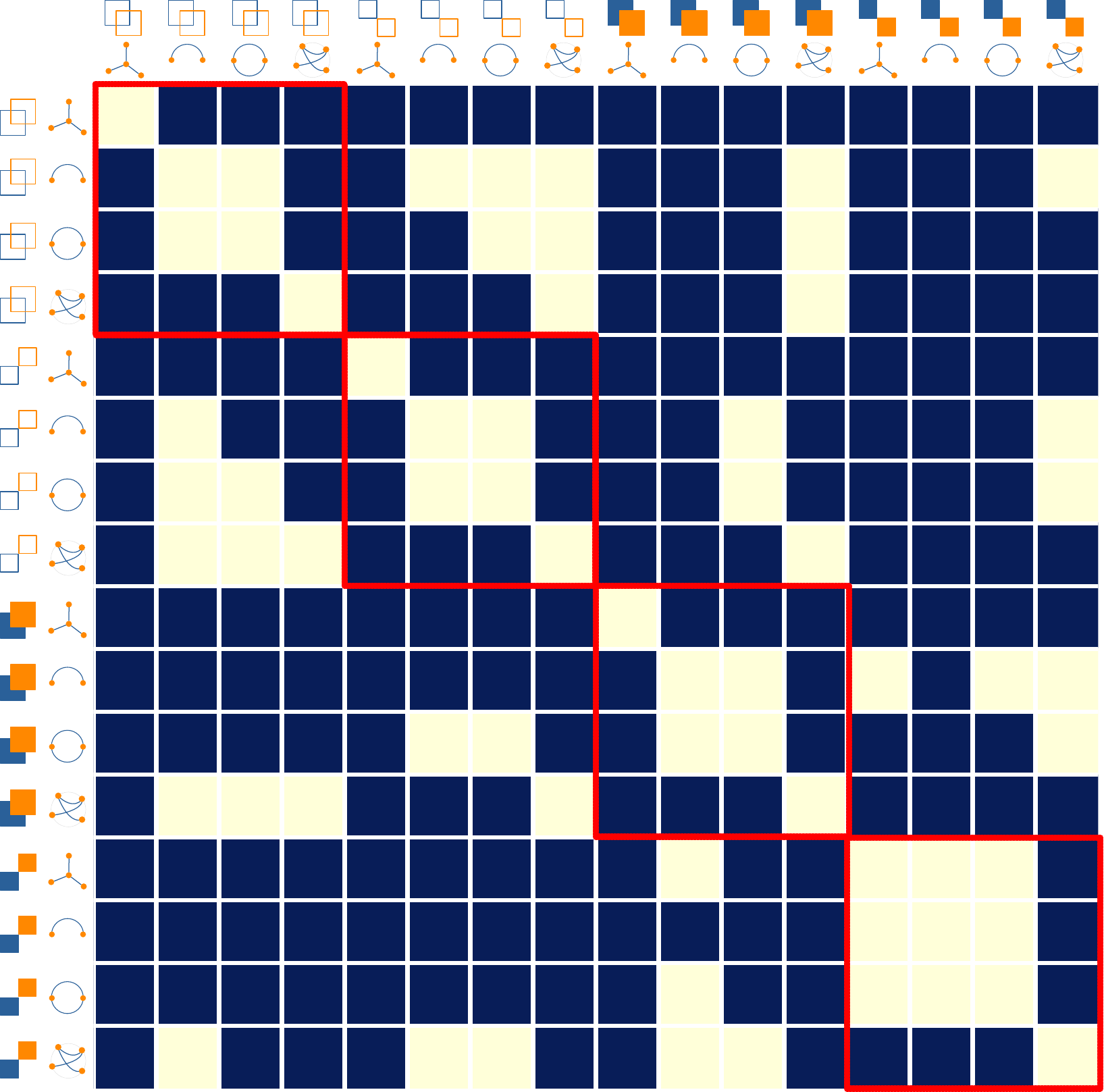}
\vspace{-0.5em} 
 \caption{Cluster count accuracy for sizes = all, clusters = all, \textbf{order = GEN}, baseline = Linlog. The dark matrix cells correspond to $p\leq0.05$.}

 \label{fig:overall_gen_linlog_acc}
\vspace{1em}
\includegraphics[height=0.21\textwidth]{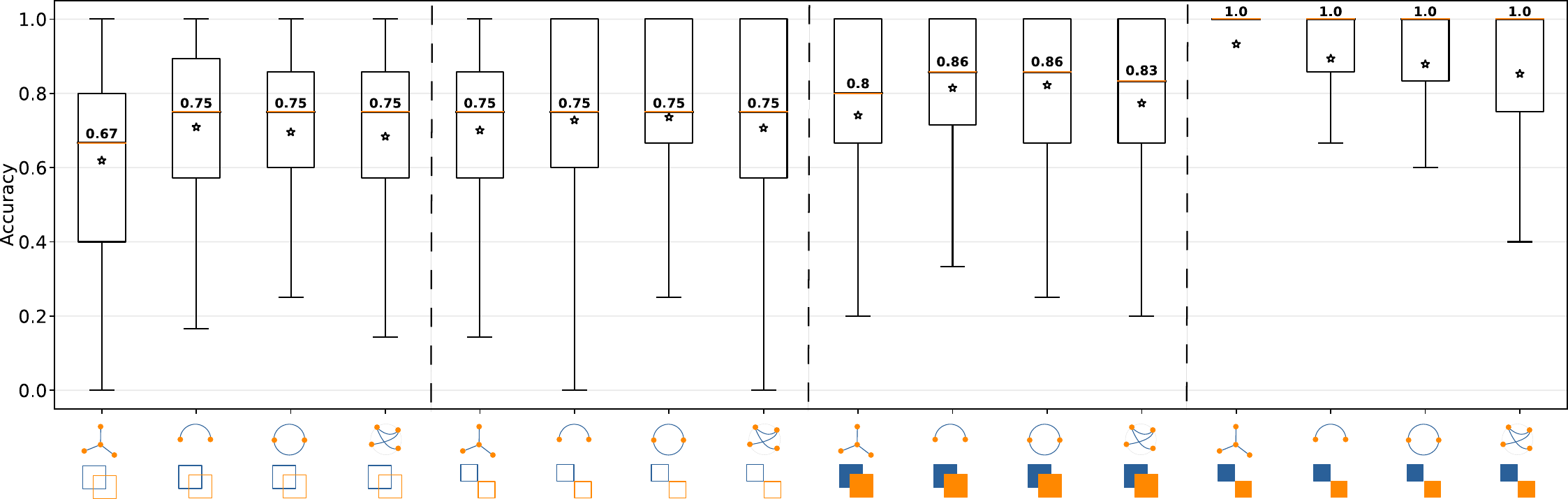}~

 \includegraphics[height=0.21\textwidth]{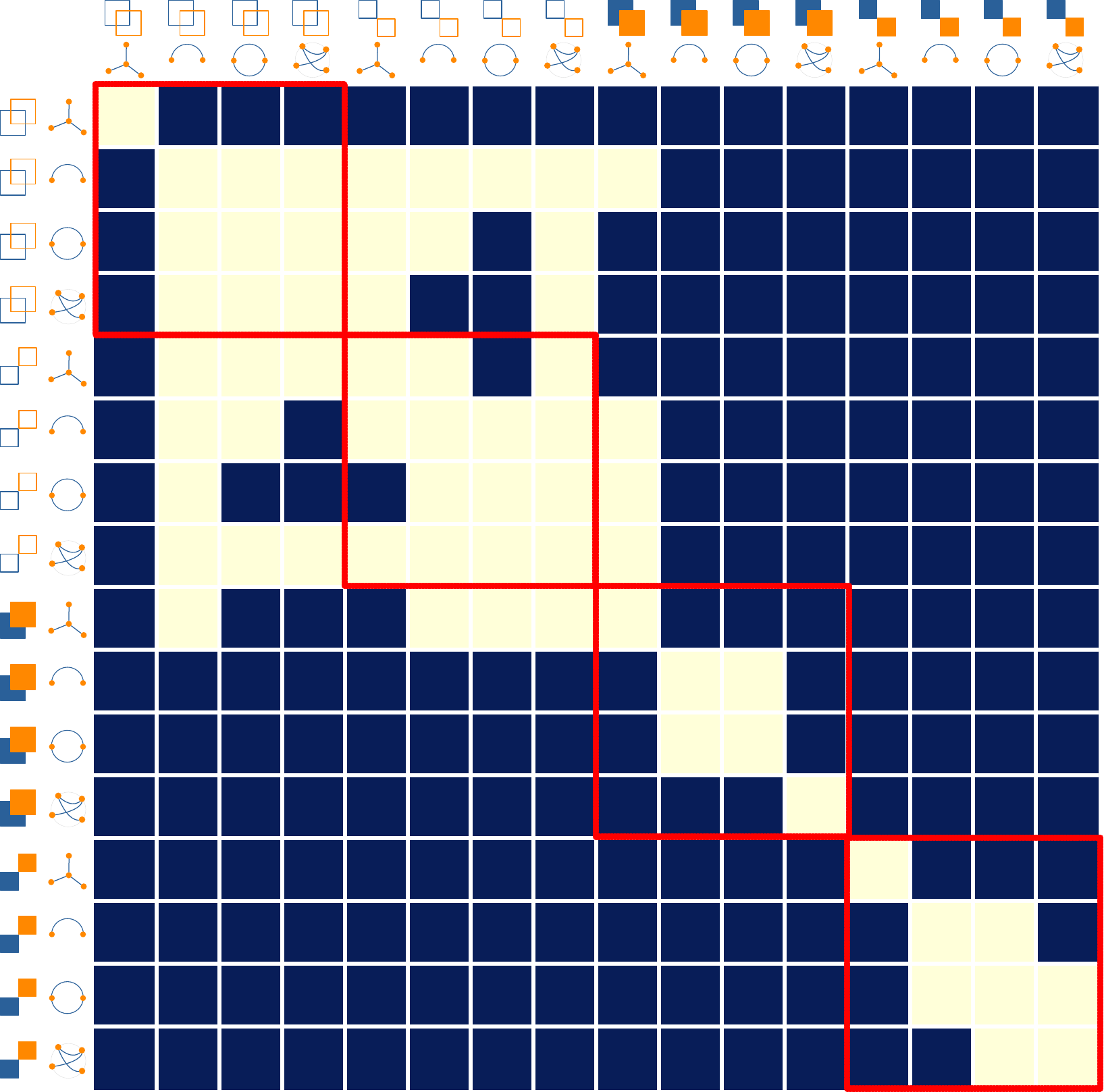}
\vspace{-0.5em} 
 \caption{Cluster count accuracy for sizes = all, clusters = all, \textbf{order = CR}, baseline = Linlog. The dark matrix cells correspond to $p\leq0.05$}

 \label{fig:overall_cr_linlog_acc}
\end{figure*}

In the rest of this section, all figures provided to verify the various hypotheses always include a pair of plots, see \autoref{fig:overall_gen_linlog_acc} for example. On the left, a boxplot comparing the distribution of accuracy in all four $p_{in}\times p_{out}$ quadrants \nestedIcon{img/icons/loose-inseparable.pdf} \nestedIcon{img/icons/loose-separable.pdf} \nestedIcon{img/icons/compact-inseparable.pdf} \nestedIcon{img/icons/compact-separable.pdf}, divided by visualization type \nestedIcon{img/icons/NL.pdf} \nestedIcon{img/icons/Arc.pdf} \nestedIcon{img/icons/ArcSym.pdf} \nestedIcon{img/icons/radial.pdf}. On the right, a matrix visualization shows the p-values of all pairwise comparisons with Bonferroni correction applied across all 16 combinations of layout type and cluster type. Dark cells correspond to statistically significant differences ($p\leq0.05$). We focus on the four $4\times4$ diagonal blocks, outlined in red in the matrix, corresponding to the quadrants. In the interest of space, the plots corresponding to task completion time are provided in supplemental material, as well as plots comparing Backbone or sfdp baselines to orderable layouts.

\begin{figure*}[t]
\centering
 \includegraphics[height=0.21\textwidth]{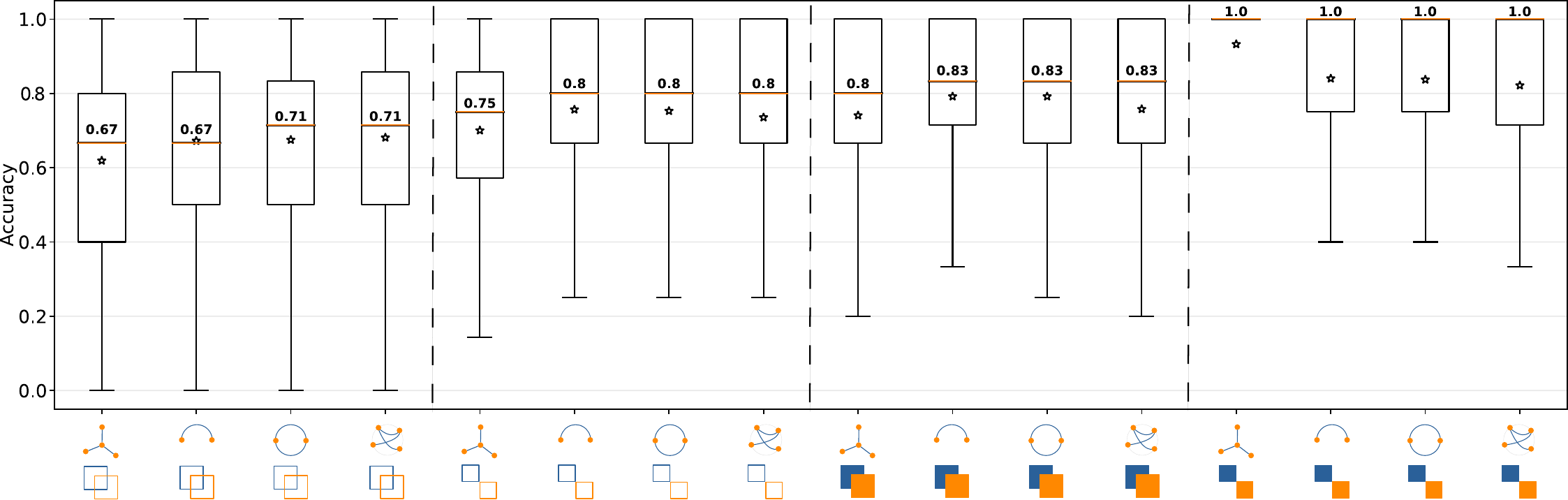}~

 \includegraphics[height=0.21\textwidth]{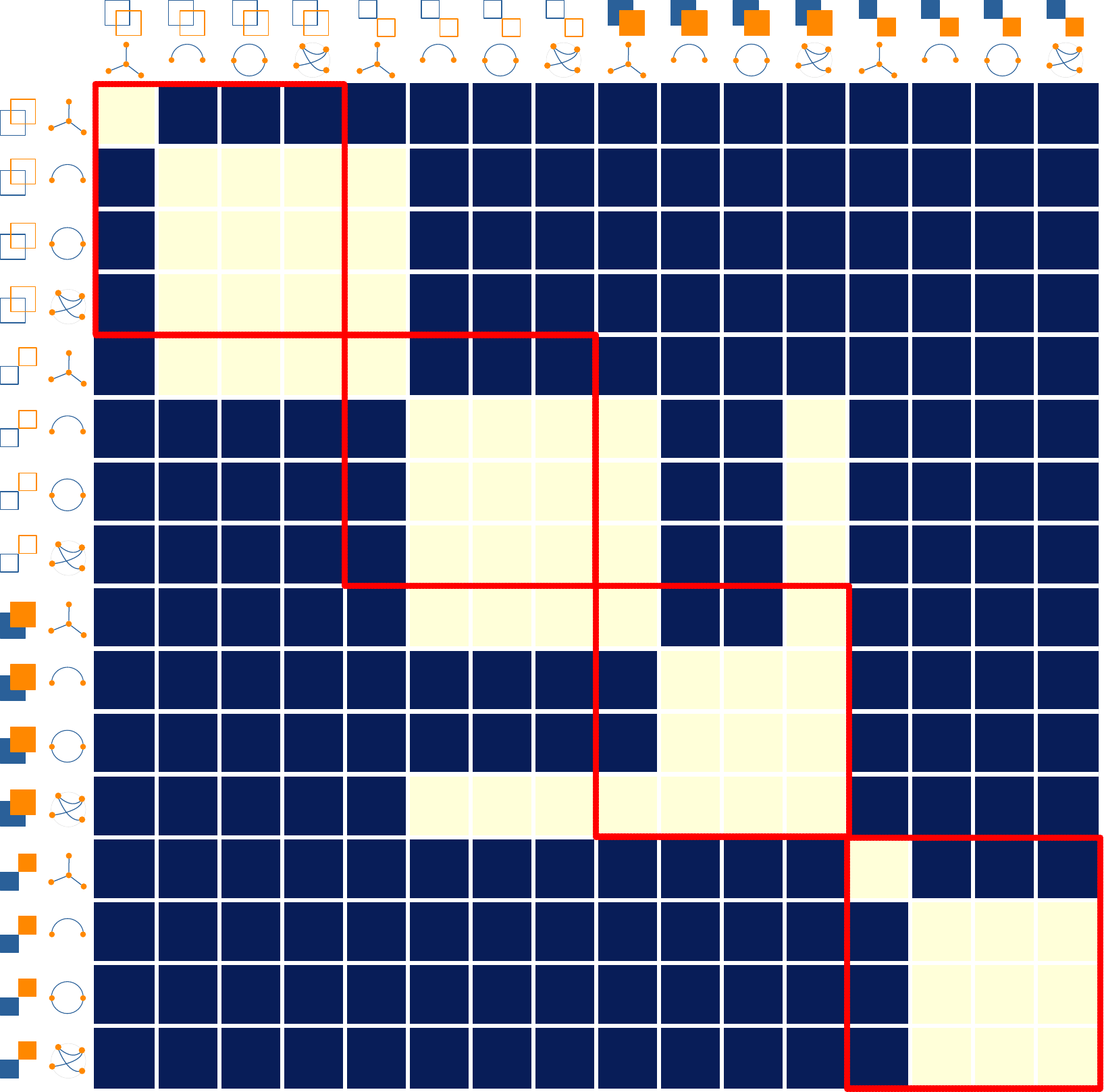}
 \vspace{-0.5em}
 \caption{Cluster count accuracy for sizes = all, clusters = all, \textbf{order = OLO}, baseline = Linlog. The dark matrix cells correspond to $p\leq0.05$}

 \label{fig:overall_olo_linlog_acc}
\end{figure*}

\subsection{Hypothesis H1 (\cmark): Orderable Node-Link Layouts Improve Cluster Saliency Over Linlog Layouts}

\autoref{fig:overall_gen_linlog_acc} shows that we can confirm H1 for loose and/or inseparable clusters \nestedIcon{img/icons/loose-separable.pdf}\nestedIcon{img/icons/loose-inseparable.pdf}\nestedIcon{img/icons/compact-inseparable.pdf} for GEN-ordered orderable layouts, i.e., when the nodes are placed in the same order they were generated. In these three cluster settings, participants using the three orderable layouts \nestedIcon{img/icons/Arc.pdf} \nestedIcon{img/icons/ArcSym.pdf}\nestedIcon{img/icons/radial.pdf} achieve 100\% median accuracy on cluster count judgements, whereas Linlog\nestedIcon{img/icons/NL.pdf} lags significantly behind at 67\% to 80\%. Besides the 33\% median accuracy gains, the orderable layouts achieve a median time up to 2 seconds (25\%) shorter than Linlog (see supplemental material). 

For compact separable clusters\nestedIcon{img/icons/compact-separable.pdf}, Linlog \nestedIcon{img/icons/NL.pdf} and all orderable layouts \nestedIcon{img/icons/Arc.pdf} \nestedIcon{img/icons/ArcSym.pdf} are equally excellent (median accuracy of 100\% with little variance). Except radial \nestedIcon{img/icons/radial.pdf} that has a median accuracy of 100\% but a slightly higher variance. In passing, Backbone\nestedIcon{img/icons/NL.pdf} and sfdp\nestedIcon{img/icons/NL.pdf} lag clearly behind GEN-ordered orderable layouts \nestedIcon{img/icons/Arc.pdf} \nestedIcon{img/icons/ArcSym.pdf}\nestedIcon{img/icons/radial.pdf} even in the case of compact separable clusters\nestedIcon{img/icons/compact-separable.pdf} (see supplemental material).

The situation is more nuanced regarding user performance with the CR and OLO ordering methods. With CR (see \autoref{fig:overall_cr_linlog_acc}), 

orderable layouts\nestedIcon{img/icons/Arc.pdf} \nestedIcon{img/icons/ArcSym.pdf}\nestedIcon{img/icons/radial.pdf} lead to more accurate cluster count judgement for inseparable clusters \nestedIcon{img/icons/loose-inseparable.pdf} \nestedIcon{img/icons/compact-inseparable.pdf}, and only partially for loose separable clusters \nestedIcon{img/icons/loose-separable.pdf}. In the case of compact separable clusters \nestedIcon{img/icons/compact-separable.pdf}, Linlog wins, although all layouts achieve 100\% median accuracy. The CR-ordered layouts\nestedIcon{img/icons/Arc.pdf} \nestedIcon{img/icons/ArcSym.pdf}\nestedIcon{img/icons/radial.pdf} show more variance than Linlog. Participants are consistently statistically significantly slower with CR-ordered radial layout \nestedIcon{img/icons/radial.pdf} than the other layouts, especially with respect to compact clusters\nestedIcon{img/icons/compact-inseparable.pdf}\nestedIcon{img/icons/compact-separable.pdf} (see supplemental Figures \ref{fig:overall_gen_linlog_time}--\ref{fig:overall_gen_cr_olo_compact_time}). With Backbone and sfdp, users are less accurate than with CR-ordered layouts in all four quadrants (see supplemental 
\autoref{fig:overall_backbone_cr_acc} and \autoref{fig:overall_sfdp_cr_acc}).

In \autoref{fig:overall_olo_linlog_acc}, 

OLO-ordered layouts\nestedIcon{img/icons/Arc.pdf} \nestedIcon{img/icons/ArcSym.pdf}\nestedIcon{img/icons/radial.pdf} lead to more accurate cluster counts than Linlog\nestedIcon{img/icons/NL.pdf}, in the case of loose clusters\nestedIcon{img/icons/loose-inseparable.pdf}\nestedIcon{img/icons/loose-separable.pdf}. They are on par regarding compact inseparable clusters\nestedIcon{img/icons/compact-inseparable.pdf}, and Linlog\nestedIcon{img/icons/NL.pdf} wins comfortably in the case of compact separable clusters\nestedIcon{img/icons/compact-separable.pdf}.

\subsection{Hypothesis H2 (\xmark): OLO Beats CR at Cluster Saliency}

\begin{figure*}[htbp]
\centering
 \includegraphics[height=0.21\textwidth]{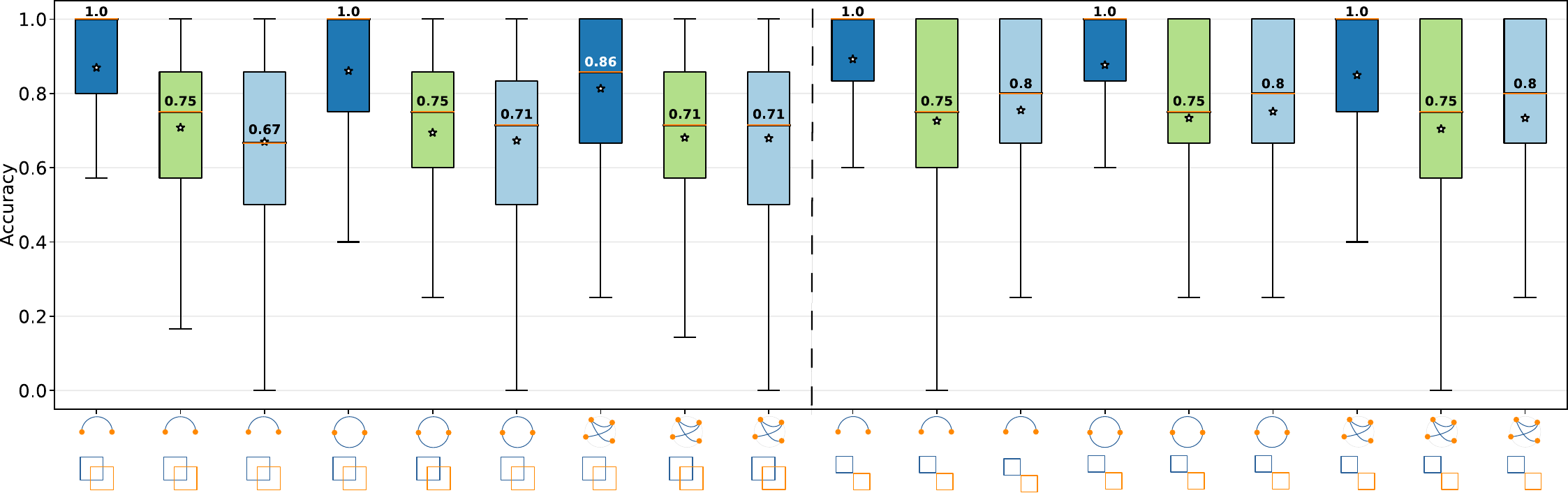}~

 \includegraphics[height=0.21\textwidth]{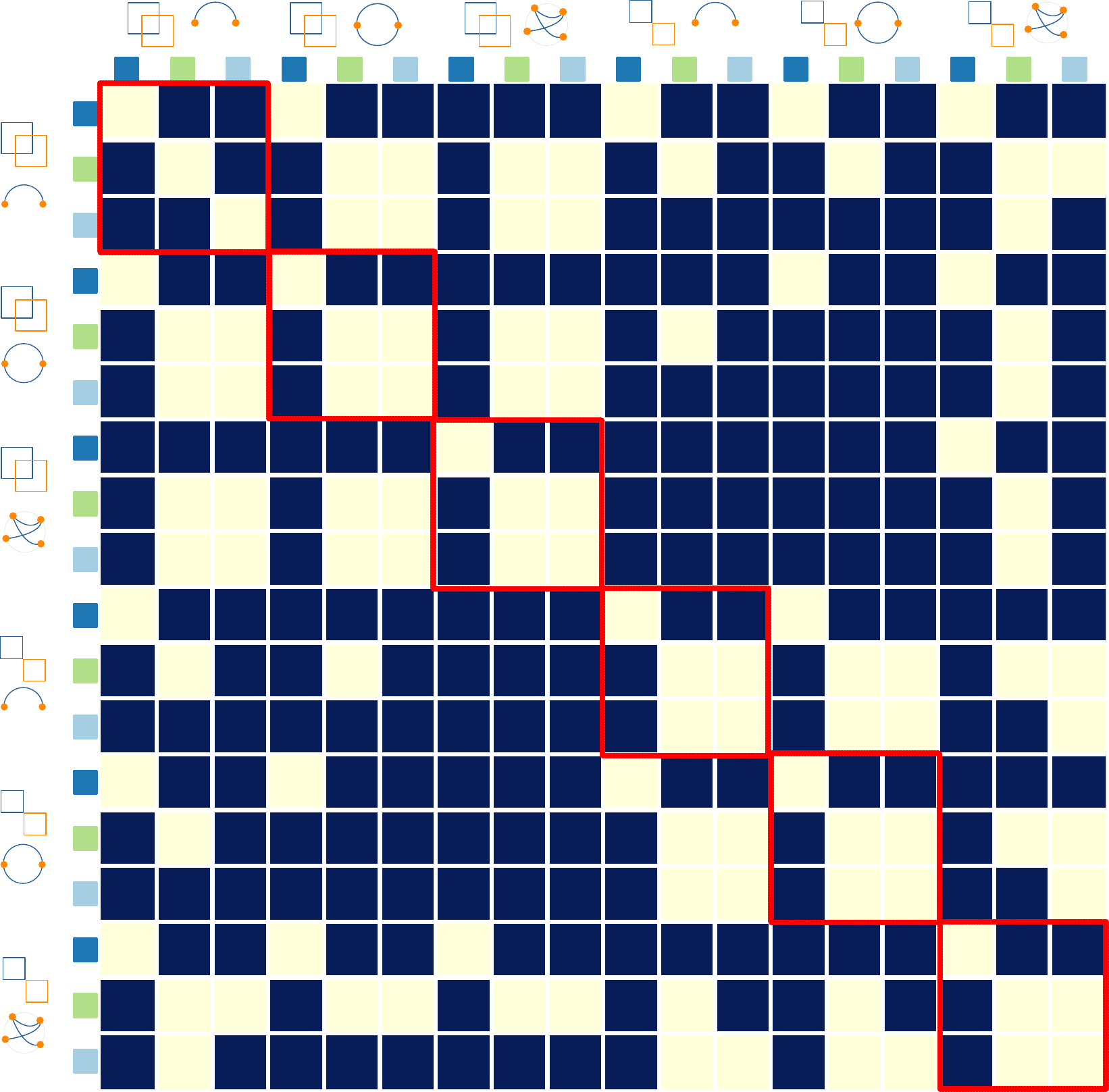}
 \vspace{-0.5em}
 \caption{Cluster count accuracy for sizes = all, \textbf{clusters = loose}, orders = \{ GEN~\square{gen}, CR~\square{cr}, OLO~\square{olo}\}. The dark matrix cells correspond to $p\leq0.05$}

 \label{fig:overall_gen_cr_olo_loose}
\end{figure*}

\begin{figure*}[htbp]
 \centering
 \includegraphics[height=0.21\textwidth]{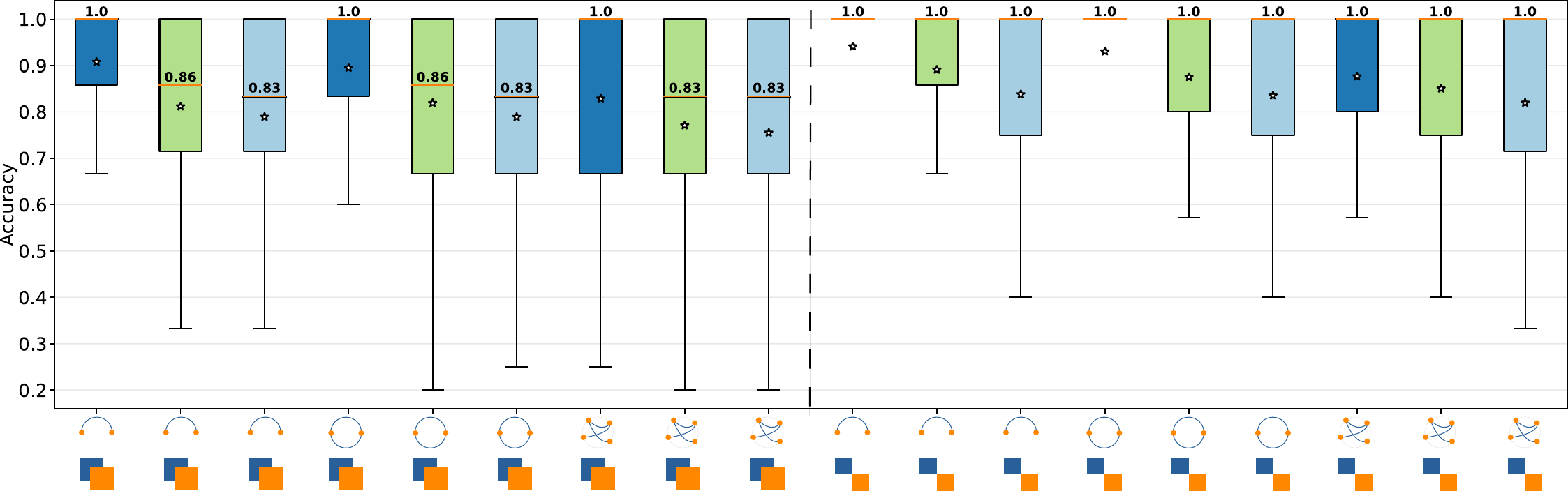}~

 \includegraphics[height=0.21\textwidth]{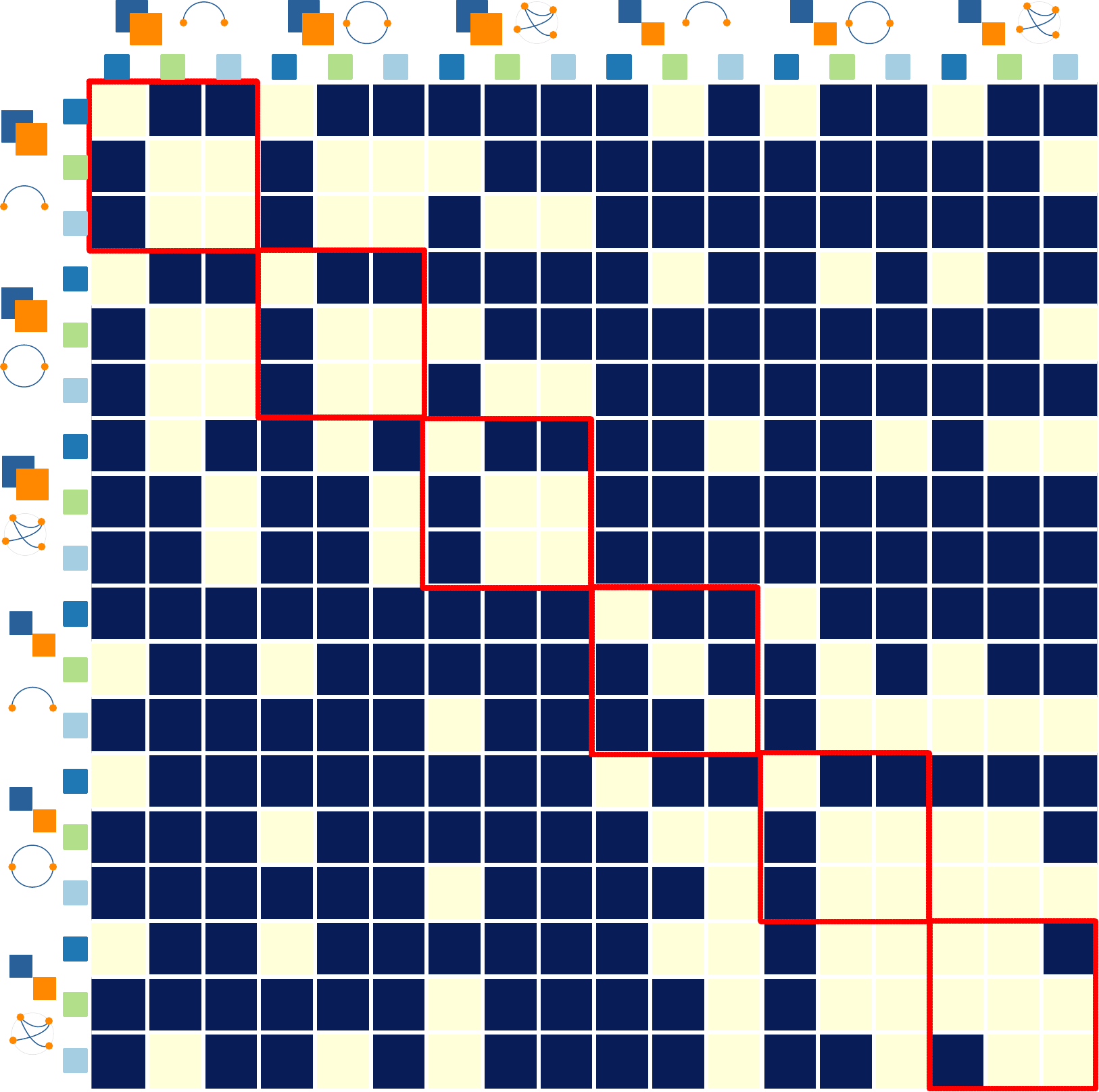}
 \vspace{-0.5em}
 \caption{Cluster count accuracy for sizes = all, \textbf{clusters = compact}, orders = \{ GEN~\square{gen}, CR~\square{cr}, OLO~\square{olo}\}. The dark matrix cells correspond to $p\leq0.05$}

 \label{fig:overall_gen_cr_olo_compact}
\end{figure*}

H2 does not hold. \autoref{fig:overall_gen_cr_olo_loose} and \autoref{fig:overall_gen_cr_olo_compact} show that CR is equivalent to OLO in most cluster settings, except for arc layouts \nestedIcon{img/icons/Arc.pdf} in the case of loose inseparable clusters \nestedIcon{img/icons/loose-inseparable.pdf} and compact separable clusters \nestedIcon{img/icons/compact-separable.pdf} where CR is better. Both CR and OLO are behind GEN, the originally generated node order, by much, across all four quadrants.

\subsection{Hypothesis H3 (\xmark): \nestedIcon{img/icons/ArcSym.pdf} Beats \nestedIcon{img/icons/Arc.pdf} at Cluster Saliency}

H3 does not hold. There is no statistically significant difference between arc \nestedIcon{img/icons/Arc.pdf} and symmetric arc \nestedIcon{img/icons/ArcSym.pdf} layouts, for any cluster type, neither in terms of accuracy (see \autoref{fig:overall_gen_linlog_acc} and \autoref{fig:overall_cr_linlog_acc}), nor in terms of task completion time (see supplemental 
\autoref{fig:overall_gen_linlog_time} and \autoref{fig:overall_cr_linlog_time}). 

\subsection{Additional Findings}
\label{sec:insights}

\paragraph{Network size affects force-directed layouts \nestedIcon{img/icons/NL.pdf}} 
Besides our hypotheses, we found that the accuracy of all three force-directed layouts \nestedIcon{img/icons/NL.pdf} decreases as network size increases (see \autoref{tab:FD_size} in supplemental material).
The impact of network size on cluster count in orderable layouts is more complex \nestedIcon{img/icons/Arc.pdf}\nestedIcon{img/icons/ArcSym.pdf}\nestedIcon{img/icons/radial.pdf} and requires further research. 

\paragraph{Global patterns vs. local patterns.} In \autoref{fig:drawing}, the superposition of cluster outlines drawn by the study participants reveals where they have more often identified clusters. 

While there were four clusters in the ground truth for this particular graph, the participants counted five clusters based on Linlog \nestedIcon{img/icons/NL.pdf}, four based on the arc \nestedIcon{img/icons/Arc.pdf} and symmetric arc layouts \nestedIcon{img/icons/ArcSym.pdf} and six with the radial layout \nestedIcon{img/icons/radial.pdf}. This suggests that arc layouts \nestedIcon{img/icons/Arc.pdf} \nestedIcon{img/icons/ArcSym.pdf} might emphasize more the global cluster structure, while Linlog \nestedIcon{img/icons/NL.pdf} and radial layouts \nestedIcon{img/icons/radial.pdf} might promote local structures. This mere assumption requires experimental validation. 

\section{Discussion}
\label{sec:discussion}

Based on our results, we discuss the performance of orderable layouts at cluster detection, the limitations of this study and formulate guidelines for the use of orderable layouts.

\subsection{Cluster Detection in Orderable Layouts}
In this study, we compared three orderable node-link layouts \nestedIcon{img/icons/Arc.pdf}\nestedIcon{img/icons/ArcSym.pdf}\nestedIcon{img/icons/radial.pdf} to three force-directed node-link layouts\nestedIcon{img/icons/NL.pdf}, Linlog, Backbone and sfdp. Our empirical results confirm previous findings that Linlog and Backbone outperform sfdp for the cluster perception task, and that Linlog is better or equivalent to Backbone~\cite{Meidiana2019AQM}.

Consistently with prior work on cluster perception in matrix visualizations~\cite{Riche2006MatrixExplorerAD, Riche2007MatLinkEM, Nobre2020EvaluatingMN, Okoe2019NodeLinkOA}, our results also show that the performance of orderable layouts at the cluster detection task depends on the seriation method. We extend this knowledge to understudied orderable node-link layouts and provide an empirical validation across the cluster compactness and separability quadrants. In our study, the original node order (GEN) resulting from the graph generator leads to a much more accurate and faster cluster count judgement than the node orders given by CR and OLO (with Jaccard distance and average linkage). In a sense, GEN is a very good reflection of the known ground truth. Both CR and OLO achieve a high median accuracy from 75\% to 85\% for the hardest, loose and/or inseparable cluster types \nestedIcon{img/icons/loose-inseparable.pdf}\nestedIcon{img/icons/loose-separable.pdf}\nestedIcon{img/icons/compact-inseparable.pdf}. In the easy case of compact separable clusters\nestedIcon{img/icons/compact-separable.pdf}, both CR and OLO achieve a median accuracy of 100\%, with a statistically significant edge for CR which has a smaller variance than OLO. In the absence of a known ground truth, CR is hence an excellent seriation method for visual cluster detection.

Our results also show that, subject to a suitable node ordering, orderable node-link diagrams \nestedIcon{img/icons/Arc.pdf}\nestedIcon{img/icons/ArcSym.pdf} \nestedIcon{img/icons/radial.pdf} outperform Linlog \nestedIcon{img/icons/NL.pdf}, the best force-directed node-link layout, for the cluster identification task, in the case of inseparable clusters, both compact \nestedIcon{img/icons/compact-inseparable.pdf} and loose \nestedIcon{img/icons/loose-inseparable.pdf}, and also in the case of separable loose clusters \nestedIcon{img/icons/loose-separable.pdf}. This is the case of GEN-ordered, CR-ordered and OLO-ordered orderable layouts (see \autoref{fig:overall_gen_linlog_acc}, \autoref{fig:overall_cr_linlog_acc} and \autoref{fig:overall_olo_linlog_acc}). Users achieve very high accuracies with the GEN ordering of nodes, with median accuracy scores of 100\% in all quadrants, up to 33\% better than Linlog. 
These new findings suggest that orderable node-link layouts \nestedIcon{img/icons/Arc.pdf}\nestedIcon{img/icons/ArcSym.pdf}\nestedIcon{img/icons/radial.pdf} combined with a suitable node ordering method can complement the analyst's toolbox for cluster identification beyond the pristine case of compact separable clusters \nestedIcon{img/icons/compact-separable.pdf}. 
For this latter case, we find that CR-ordered orderable node-link layouts and the Linlog layout achieve 100\% median accuracy for the cluster identification task, while the orderable layouts are better for the more challenging cluster types (see \autoref{fig:overall_cr_linlog_acc}).

Based on the computational benchmark of Behrisch et~al.~\cite{behrisch2016ordering} for adjacency matrices, cluster perception is well served if the ordering method is based on a clustering approach. 
This is confirmed empirically by our observation that the orderable layouts \nestedIcon{img/icons/Arc.pdf} \nestedIcon{img/icons/ArcSym.pdf} \nestedIcon{img/icons/radial.pdf} ordered by OLO achieve between 70\% and 85\% median accuracy (\autoref{fig:overall_gen_cr_olo_compact}).

As long as the analyst needs to identify clusters in an overview visualization, any clustering-based order applied to nodes will induce the appearance of clusters in orderable layouts \nestedIcon{img/icons/Arc.pdf} \nestedIcon{img/icons/ArcSym.pdf} \nestedIcon{img/icons/radial.pdf}. 
Since, among other aspects, clusters may vary in shape, size, number and density, different clustering methods will lead to more or less useful node orders/overview visualizations. 
Similarly, different force-directed approaches lead to more or less useful layouts.
In our study, CR, a seriation method not based on clustering, did better than OLO, which is based on agglomerative hierarchical clustering, with respect to arc diagrams \nestedIcon{img/icons/Arc.pdf} with compact separable clusters \nestedIcon{img/icons/compact-separable.pdf} (\autoref{fig:overall_gen_cr_olo_compact}). A possible explanation is that the Jaccard distance employed by OLO in this experiment did not cope well with the many between-cluster links, while the objective to reduce edge-crossings of CR is less sensitive to this type of noise.

Besides the global patterns revealed by clustering-based seriation methods, some of these methods can also optimize for more local patterns, e.g., OLO optimizes node order locally. Local optimization might address gaps identified by Noack, such as promoting the perception of missing details concerning the internal structure of clusters and relationships between clusters. Also, the different link shapes (arcs, symmetric arcs, straight lines) and node positions inherent to each layout seem to have different effects on the perception of (sub-)clusters, as shown by the annotated cluster areas in \autoref{fig:drawing}. 

A formal validation is still needed for such sub-cluster level tasks, though.

Earlier user studies by Okoe et al.~\cite{Okoe2019NodeLinkOA} and Nobre et al.~\cite{Nobre2020EvaluatingMN} found that ordered adjacency matrices outperform force-directed layouts, for cluster identification. 
We introduce a new possibility, consisting in using orderable node-link layouts \nestedIcon{img/icons/Arc.pdf}\nestedIcon{img/icons/ArcSym.pdf}\nestedIcon{img/icons/radial.pdf} for this task. 
To keep the size of our experiment manageable, and the duration of participant sessions reasonable, we left adjacency matrices out of the scope of this study. 
Now that it has become clear that orderable layouts \nestedIcon{img/icons/Arc.pdf}\nestedIcon{img/icons/ArcSym.pdf}\nestedIcon{img/icons/radial.pdf} can outperform by much some of the best force-directed node-link layouts\nestedIcon{img/icons/NL.pdf}, subject to a suitable node order, a future user study could compare them to similarly ordered matrices in terms of cluster perception and explore a larger number of ordering methods.
Yet, adjacency matrices rely on a visual metaphor deemed unfamiliar by most users who often prefer node-link diagrams \nestedIcon{img/icons/NL.pdf}~\cite{Ghoniem2005OnTR}.
The path finding task is also challenging with matrix visualizations, due to the duplication of nodes on the sides of the matrix.
 Whether this task is better supported by orderable node-link layouts \nestedIcon{img/icons/Arc.pdf} \nestedIcon{img/icons/ArcSym.pdf} \nestedIcon{img/icons/radial.pdf} thanks to the node-link metaphor is an open question. In general, more work is needed on the evaluation of orderable node-link layouts \nestedIcon{img/icons/Arc.pdf} \nestedIcon{img/icons/ArcSym.pdf} \nestedIcon{img/icons/radial.pdf} to determine which other graph analysis tasks they can effectively support.

For the cluster detection task, Abdelaal et al.~\cite{Abdelaal2022ComparativeEO} found that bipartite diagrams ordered according to agglomerative hierarchical clustering perform worse than force-directed node-link layouts and matrices, especially for sparse networks. 
Unlike previous work by Okoe et al.~\cite{Okoe2019NodeLinkOA} and Nobre et al.~\cite{Nobre2020EvaluatingMN}, Abdelaal et al.~\cite{Abdelaal2022ComparativeEO} did not report any statistically significant difference in terms of cluster perception between matrices and node-link layouts generated by neato~\cite{graphviz} (layout based on multi-dimensional scaling). 
Such a disagreement between different studies may usually be explained by differences in the experimental protocol. 
This points to the importance of replication studies in the visualization field, and the necessity to share openly as much information as possible, including training materials, data sets etc.
Towards this, we release in supplementary material all the stimuli used in this user study, along with the corresponding graph data sets, and the training material.

From a visual perception perspective, the saliency of clusters in force-directed node-link layouts, e.g., Linlog \nestedIcon{img/icons/NL.pdf}, exploits two principles: the Gestalt proximity principle entailing that nodes whose position is close in the 2D plane tend to be seen as a cluster, and the difference in luminosity induced by the locally high edge clutter within clusters in contrast to the rather sparse link connectivity elsewhere, especially in many real network data sets. In this experiment, we did our best to enhance the readability of the stimuli resulting from Linlog \nestedIcon{img/icons/NL.pdf}, for example, by increasing node size and link thickness in the larger 300-node networks to compensate for the zoom out effect resulting from fitting a larger drawing in a fixed-size display and, hence, to ensure a good readability. In contrast, our study used a basic implementation of orderable node-link layouts \nestedIcon{img/icons/Arc.pdf}\nestedIcon{img/icons/ArcSym.pdf}\nestedIcon{img/icons/radial.pdf}, where clusters are visually characterized by edge concentrations only. Orderable layouts \nestedIcon{img/icons/Arc.pdf}\nestedIcon{img/icons/ArcSym.pdf}\nestedIcon{img/icons/radial.pdf} could be further enhanced by exploiting the Gestalt proximity principle too, i.e., by introducing gaps between clusters along the underlying geometric locus, when the clusters are known. We expect such gaps to improve the visual saliency of clusters in orderable layouts \nestedIcon{img/icons/Arc.pdf}\nestedIcon{img/icons/ArcSym.pdf}\nestedIcon{img/icons/radial.pdf}, which is supported by a prior qualitative study on graph layouts based on space-filling curves~\cite{spacefilling}.
One might also study the effect of edge bundling on the cluster identification task. Edge bundling has initially been applied on radial diagrams, given an overarching hierarchy~\cite{bundlingHolten}. It has later been extended to preserve the perception of paths~\cite{pathBundling}. Since edge bundling affects drastically the way edges are routed, and the overall edge clutter, it might as well impact cluster perception. While prior work has explored edge bundling for radial graphs and general 2D graph layouts, work on edge bundling for arc diagrams is scarce. The results of this study constitute a stepping stone in this direction, and a baseline against which future work could compare cluster perception for both edge-bundled orderable and force-directed node-link layouts.

Finally, an inconvenient of orderable layouts \nestedIcon{img/icons/Arc.pdf}\nestedIcon{img/icons/ArcSym.pdf}\nestedIcon{img/icons/radial.pdf} is their relatively inefficient use of display space. Indeed, minimizing the area of the drawing is a known graph drawing aesthetic rule~\cite{graph-drawing-book}. In this respect, vanilla arc diagrams \nestedIcon{img/icons/Arc.pdf} obviously use half the area of their symmetric variant \nestedIcon{img/icons/ArcSym.pdf}, but much more than the equivalent radial layout \nestedIcon{img/icons/radial.pdf} (it is easy to prove that arc diagrams \nestedIcon{img/icons/Arc.pdf} use $\frac{\pi^2}{2}\approx5$ times the area of equivalent radial diagrams \nestedIcon{img/icons/radial.pdf}). This might matter when integrating orderable layouts in a broader context, for example in a multiple coordinated view setting with other visualizations. On the one hand, arc diagrams \nestedIcon{img/icons/Arc.pdf} can be easily displayed on the side of a matrix view, like in MatLink~\cite{Riche2007MatLinkEM}. But, in any case, their nodes may need to be scaled up to ensure sufficient visibility. In this study, we scaled the different layouts to use all the available display area, and ensured a good visibility of the nodes and links in all layouts (see Section~\ref{sec:visual_stimuli}).

\subsection{Limitations}

\paragraph{Data generation} Much research has gone into generating synthetic graphs meeting certain requirements, e.g., structural patterns characterized by common graph metrics such as degree distribution, or clustering coefficient~\cite{generators}. 
The use of synthetic graphs is of practical value for large-scale benchmarking activities, for which it is difficult to gather enough real graph data sets with comparable features, and also to be able to control the key properties of graphs for a specific investigation. 
Also, similar to Anscombe's quartet, previous work by Chen et~al. has shown that very different graphs might share the same summary statistics~\cite{graph-stats}. 
So, even when researchers use popular graph generators, it is difficult in practice to control for everything, and there is a large overhead associated with the preparation of graph data sets. 

In this study, we used the Gaussian random partition graph generator~\cite{Brandes2003ExperimentsOG} of NetworkX~\cite{Hagberg2008ExploringNS}, to control the number of nodes, the number of clusters and the internal and external link probabilities of these clusters. 
We wanted to have small and large clusters in the same network, to better mimic real networks. 
Still, one limitation of our approach is that, in any given graph, we have one cluster type \nestedIcon{img/icons/loose-inseparable.pdf}\nestedIcon{img/icons/loose-separable.pdf}\nestedIcon{img/icons/compact-inseparable.pdf}\nestedIcon{img/icons/compact-separable.pdf}. In real graphs, a mix of compact and loose clusters exist, with more or less coupling between them.
 An alternate data generation approach could consist in creating the clusters separately with the desired mix of inner link densities and cluster sizes, before injecting external links, as needed.
Yet, our experiment provides a good complement to existing work on cluster perception, and a baseline for future inquiries. 

\paragraph{Cluster validity} 
When generating graphs with loose and/or inseparable clusters, given the stochastic nature of the generative process some clusters may be so loose and inseparable not to qualify as clusters. One would ideally have a measure of the clustering tendency of the generated data, and only keep graphs with a sufficient clustering tendency. This is however a complex task. Global measures associated to the presence of clusters, such as clustering coefficient or maximum modularity, are convenient, but they cannot tell whether individual clusters are too loose or inseparable. Also, maximum modularity can be very high even in random networks~\cite{guimera_modularity_2004}, a problem shared with other descriptive methods for cluster detection~\cite{peixoto_descriptive_2023}. 

A principled way to assess clustering tendency would be to compute statistical significance~\cite{lancichinetti_finding_2011}, which is also a complex task with strong assumptions. First, some work proposing significance testing for clustering focus on the whole set of clusters, e.g.,~\cite{zhang_scalable_2014}, where individual non-significant clusters might occur within significant clusterings~\cite{peixoto_descriptive_2023}. Recent work explores how to test individual clusters~\cite{palowitch_computing_2019}, but there is no standard way of doing it yet, and these methods rely themselves on assumptions such as the null model against which statistical significance is computed. Since different cluster types may exist, and different clustering methods are based on different definitions of what a cluster is, we conclude that clustering tendency should rather be considered while interpreting the results of the empirical study, rather than as a hard filter to decide which stimuli should be included in the study. In particular, if for some stimuli the participants fail to identify clusters for all the tested layout methods, an explanation may be that those clusters are too loose and inseparable to even qualify as clusters. If the correct number of clusters is consistently identified using one method, it seems reasonable to consider this as empirical evidence that the generated clusters are valid. As this paper aims to compare different layouts, and not to study the level of compactness/separability where specific methods start failing, this evidence suffices to reach our conclusions. 

\enlargethispage{10pt}
Like in \autoref{fig:VAT}, we used matrix visualizations to check the existence of clusters in every generated graph. 
Future work may seek to determine the frontiers of the cluster validity space empirically by generating many graphs with decreasing cluster compactness and/or separability until users cease to catch a signal in the stimuli. 
Such a principled approach would provide useful insights, to guide graph generation for similar studies on cluster perception.
In this study, we have seen that GEN, the node generation order, leads to significantly faster and more accurate cluster count judgements, than OLO and CR. This may occur when the clusters are generated one after the other.
The 20\% accuracy gap between GEN and OLO/CR calls for
more adapted clustering methods for the relatively loose and/or inseparable clusters considered here. 

\paragraph{Layout optimization}
 Since the output of most graph layout algorithms depends on hyperparameter values, 
researchers should strive to evaluate algorithms at their best. 
For example, a computational study might seek to find the optimal hyperparameter values to refine each of the three force-directed layouts for each of the 60 graphs included in our study. This would yet require establishing which aesthetic rules best capture cluster saliency, which is an open question. 
In this study, we have relied on published best practice, e.g., the use of \emph{quadrilateral Simmelian} Backbone, and the fact that default settings are usually chosen to give good results in general. 
Lastly, OLO and CR are both linear seriation methods, which may be suboptimal for radial layouts \nestedIcon{img/icons/radial.pdf}. Future work may consider circular seriation methods too~\cite{circularSeriation}.

\paragraph{Crowdsourcing} One benefit of crowdsourced studies is the access to a large and diverse participant sample. While the researcher has limited control over the setup, e.g., software, hardware and ergonomic aspects, we selected participants with a suitable display size: laptops and desktop monitors, excluding all handheld devices. The high levels of accuracy, and short task completion times, achieved by the study participants make us believe that the experimental setting was suitable. We were also keen to ensure that the participants understood the task at hand, without being able to observe them directly. To mitigate any issues, we had several iterations over the training material to make the video tutorials as clear and as short as possible, prior to the study. Ultimately, the short drawing task at the end of each stimulus batch, i.e., for each visualization type, allowed us, post hoc, to quality control and check for any major misunderstanding of the task. 

\subsection{Guidelines}

Our study compared three force-directed layouts\nestedIcon{img/icons/NL.pdf} (ForceAtlas2 with Linlog energy, Backbone and sfdp) to three orderable layouts\nestedIcon{img/icons/Arc.pdf}\nestedIcon{img/icons/ArcSym.pdf}\nestedIcon{img/icons/radial.pdf} in terms of cluster perception in graphs having four different types of clusters \nestedIcon{img/icons/loose-inseparable.pdf}\nestedIcon{img/icons/loose-separable.pdf} \nestedIcon{img/icons/compact-inseparable.pdf}\nestedIcon{img/icons/compact-separable.pdf} (summarized in \autoref{fig:cluster_settings}). 
The study revealed how, depending on cluster type, certain visualizations outperform others. Hence, we can formulate the following guidelines:\vspace{0.3em}

\noindent\ding{110}\,\textbf{G1} When node clusters are of analytical interest, use an orderable node-link layout to get an overview of the cluster structure. \textbf{Observation:} For graphs having loose and/or inseparable clusters\nestedIcon{img/icons/loose-inseparable.pdf}\nestedIcon{img/icons/loose-separable.pdf}\nestedIcon{img/icons/compact-inseparable.pdf}, orderable layouts \nestedIcon{img/icons/Arc.pdf}\nestedIcon{img/icons/ArcSym.pdf}\nestedIcon{img/icons/radial.pdf}, subject to node order, can outperform the best force-directed layouts\nestedIcon{img/icons/NL.pdf}. Therefore, only using e.g.~a force-based layout may lead to a wrong perception of the cluster structure.

\noindent\ding{110}\,\textbf{G2} Check the agreement between force-directed and orderable node-link layouts. \textbf{Observation:} When the graph has ideal clusters, i.e., compact and separable \nestedIcon{img/icons/compact-separable.pdf}, all the tested layouts promote the emergence of clusters. Therefore, if force-directed and orderable node-link layouts show the same cluster structure, the choice of visualization can then be based on additional tasks to be addressed.

\noindent\ding{110}\,\textbf{G3} When an orderable node-link diagram is chosen as a visualization, use CR or OLO. \textbf{Observation:} Both CR and OLO seriation algorithms achieve good results in terms of cluster identification.

These guidelines are valid within the experimental conditions used in our study. While our results cannot be generalized to all layout algorithms under all choices of hyperparameters, the force-directed layouts we used were informed by the literature and are commonly used. We also note that each of our stimuli only contains a single type of clusters, so that we could use this as an independent variable. While we cannot generalize our results to cases with different types of clusters in the same graph, we believe that observations about specific cluster types will still apply to these cases.

\section{Conclusion}
\label{sec:conclusion}

Clusters are common and important network patterns. We presented the first user study investigating the visual saliency of graph clusters in orderable node-link layouts under varying conditions of cluster compactness and separability. We found empirically that the use of orderable node-link layouts \nestedIcon{img/icons/Arc.pdf}\nestedIcon{img/icons/ArcSym.pdf}\nestedIcon{img/icons/radial.pdf} with an appropriate node ordering algorithm achieves fast and accurate cluster detection, surpassing significantly several force-directed layouts \nestedIcon{img/icons/NL.pdf} (Linlog, Backbone and sfdp), when dealing with loose and/or inseparable clusters \nestedIcon{img/icons/loose-inseparable.pdf}\nestedIcon{img/icons/loose-separable.pdf}\nestedIcon{img/icons/compact-inseparable.pdf}. Only in the ideal case of compact and separable clusters \nestedIcon{img/icons/compact-separable.pdf}, the Linlog node-link layout \nestedIcon{img/icons/NL.pdf} is on par with or better than orderable node-links diagrams \nestedIcon{img/icons/Arc.pdf} \nestedIcon{img/icons/ArcSym.pdf} \nestedIcon{img/icons/radial.pdf}. Moreover, the crossing reduction heuristic (i.e., the barycenter heuristic) outperforms the optimal leaf ordering seriation method in the case of loose and/or inseparable clusters \nestedIcon{img/icons/loose-inseparable.pdf}\nestedIcon{img/icons/loose-separable.pdf}\nestedIcon{img/icons/compact-inseparable.pdf}.

Future perceptual studies might explore other graph motifs, e.g., hubs, and other tasks like path finding to better understand the pros and cons of orderable node-link layouts \nestedIcon{img/icons/Arc.pdf}\nestedIcon{img/icons/ArcSym.pdf}\nestedIcon{img/icons/radial.pdf}. One might also extend the study to include synthetic data sets with overlapping clusters, and real data sets with application-driven analytical goals.
Finally, one might investigate why the orderable layouts are good at showing clusters. The rationale for H1 is that these layouts create locally high link concentrations. 
Insights into the specific features associated to better cluster perception with orderable layouts may lead to new methods, e.g., to quantify layout quality for this task automatically.

%% file: inc-appendices.tex
\newpage
\onecolumn
\section*{Appendices}
\label{sec:appendices}

\subsection*{Task completion time of Linlog versus orderable layouts}

\begin{figure*}[h]
\centering
    \includegraphics[height=0.235\textwidth]{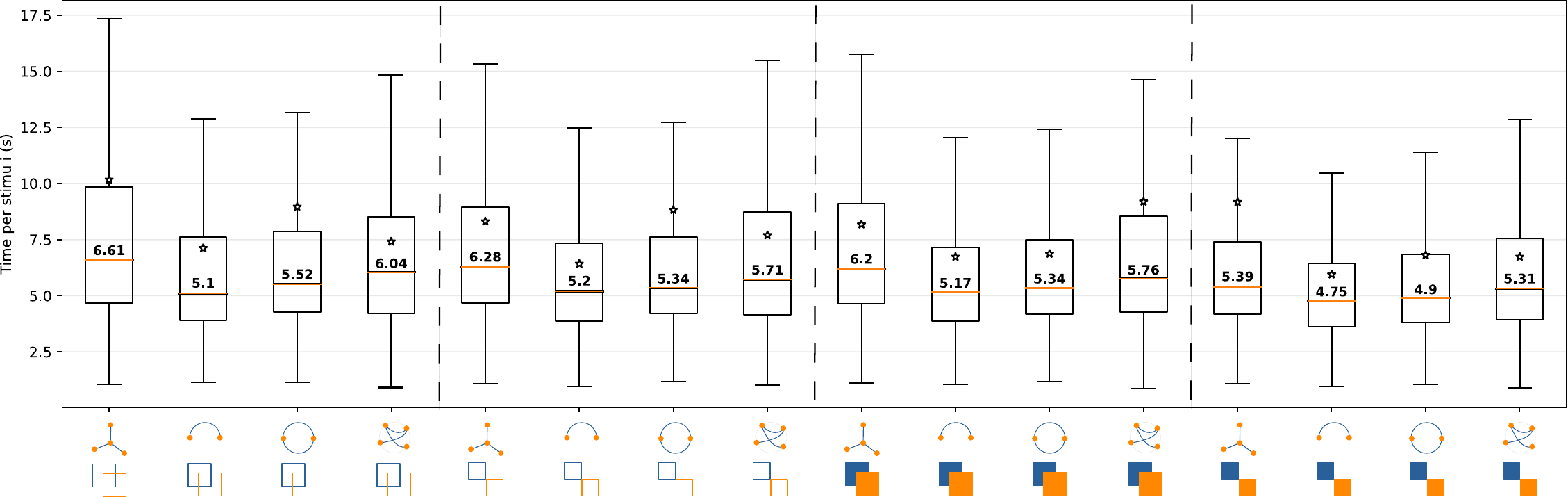}
    \hfill        
    \includegraphics[height=0.235\textwidth]{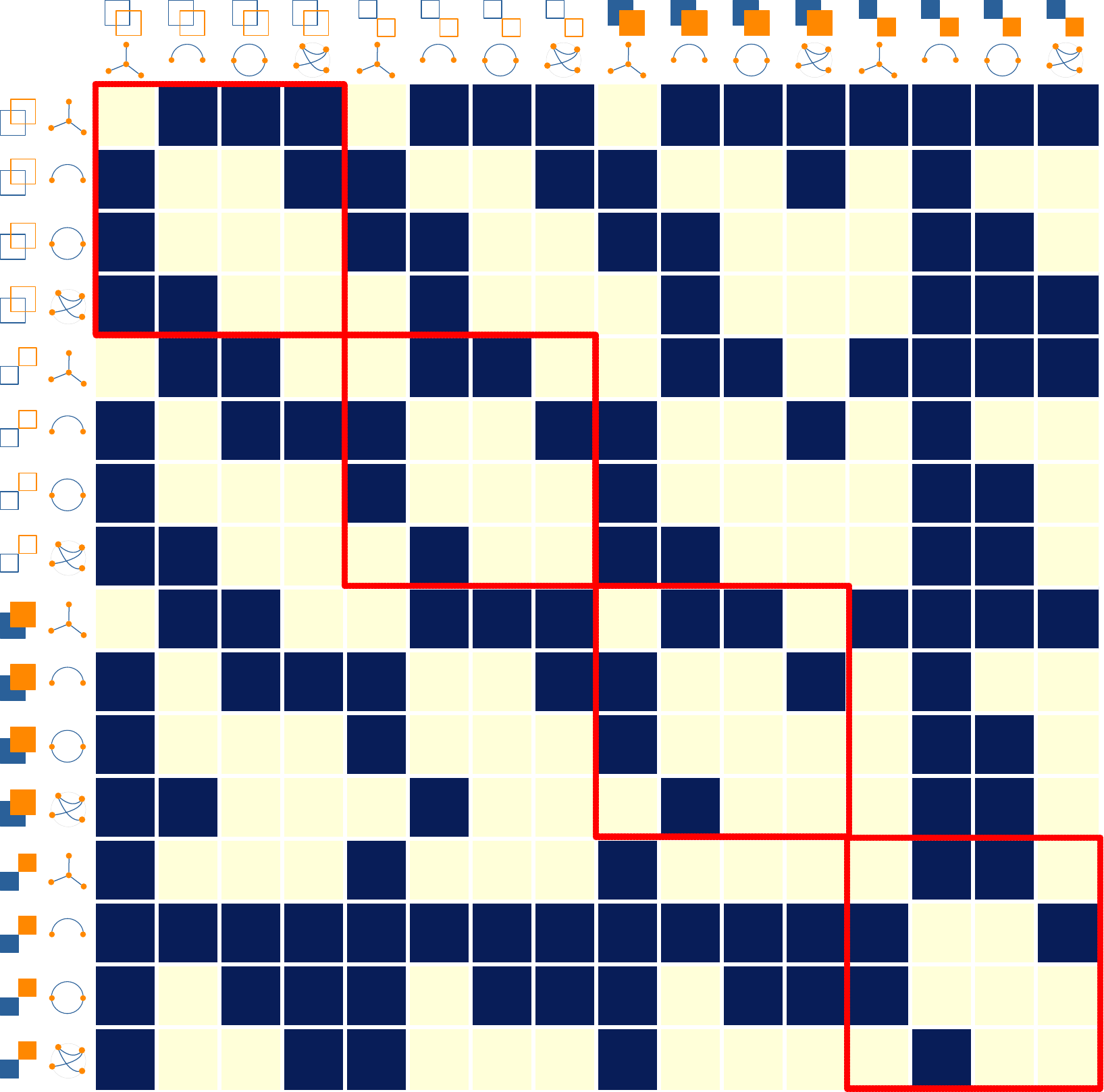}
\vspace{-0.5em}    
    \caption{Overall task completion time for sizes = all, clusters = all, \textbf{order = GEN}, baseline = Linlog. On the right, the dark blue matrix cells correspond to the Bonferroni corrected $p\leq0.05$ for all pairwise comparisons. Linlog is significantly (up to 25\%) slower than GEN-ordered Arc and Arc symmetric in all quadrants, and slower than radial layouts for compact inseparable clusters.}

   \label{fig:overall_gen_linlog_time}
\end{figure*}

\begin{figure*}[h]
\centering
    \includegraphics[height=0.235\textwidth]{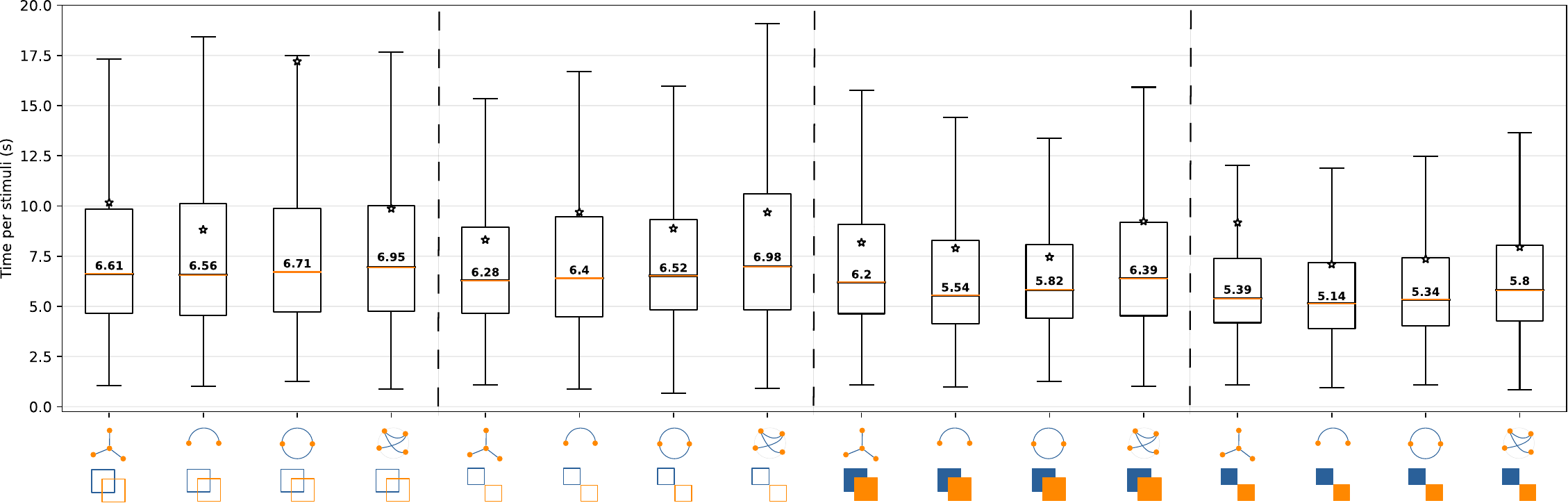}
    \hfill        
    \includegraphics[height=0.235\textwidth]{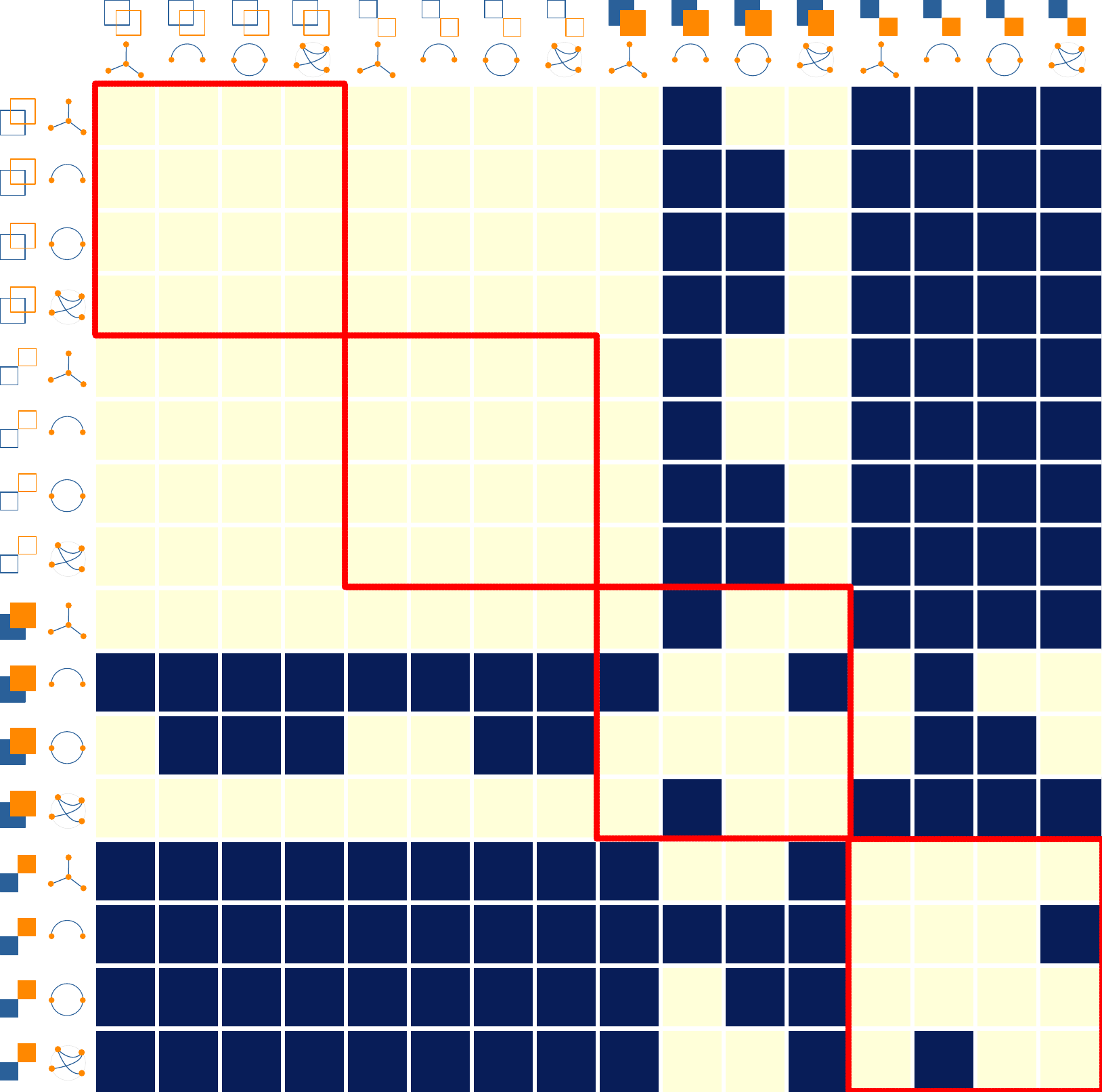}
\vspace{-0.5em}    
    \caption{Overall task completion time for sizes = all, clusters = all, \textbf{order = CR}, baseline = Linlog. On the right, the dark blue matrix cells correspond to the Bonferroni corrected $p\leq0.05$ for all pairwise comparisons. Linlog is generally as fast as the CR-ordered orderable layouts, except for compact inseparable clusters where arc does significantly better.}

   \label{fig:overall_cr_linlog_time}
\end{figure*}

\begin{figure*}[h]
\centering
    \includegraphics[height=0.235\textwidth]{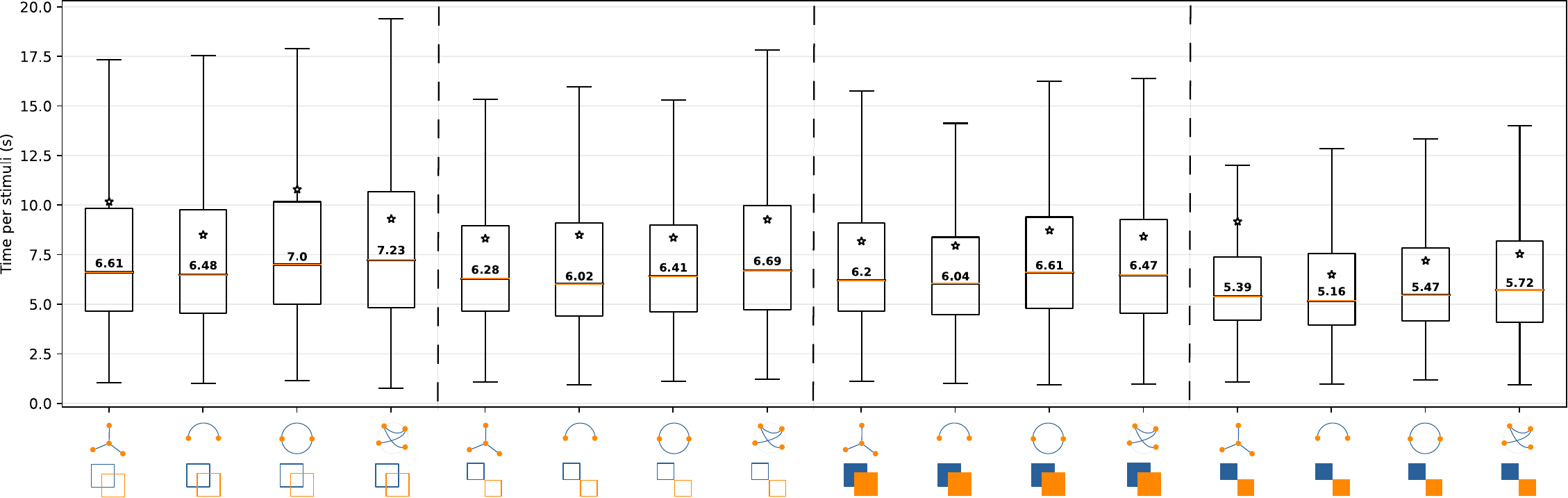}
    \hfill
    \includegraphics[height=0.235\textwidth]{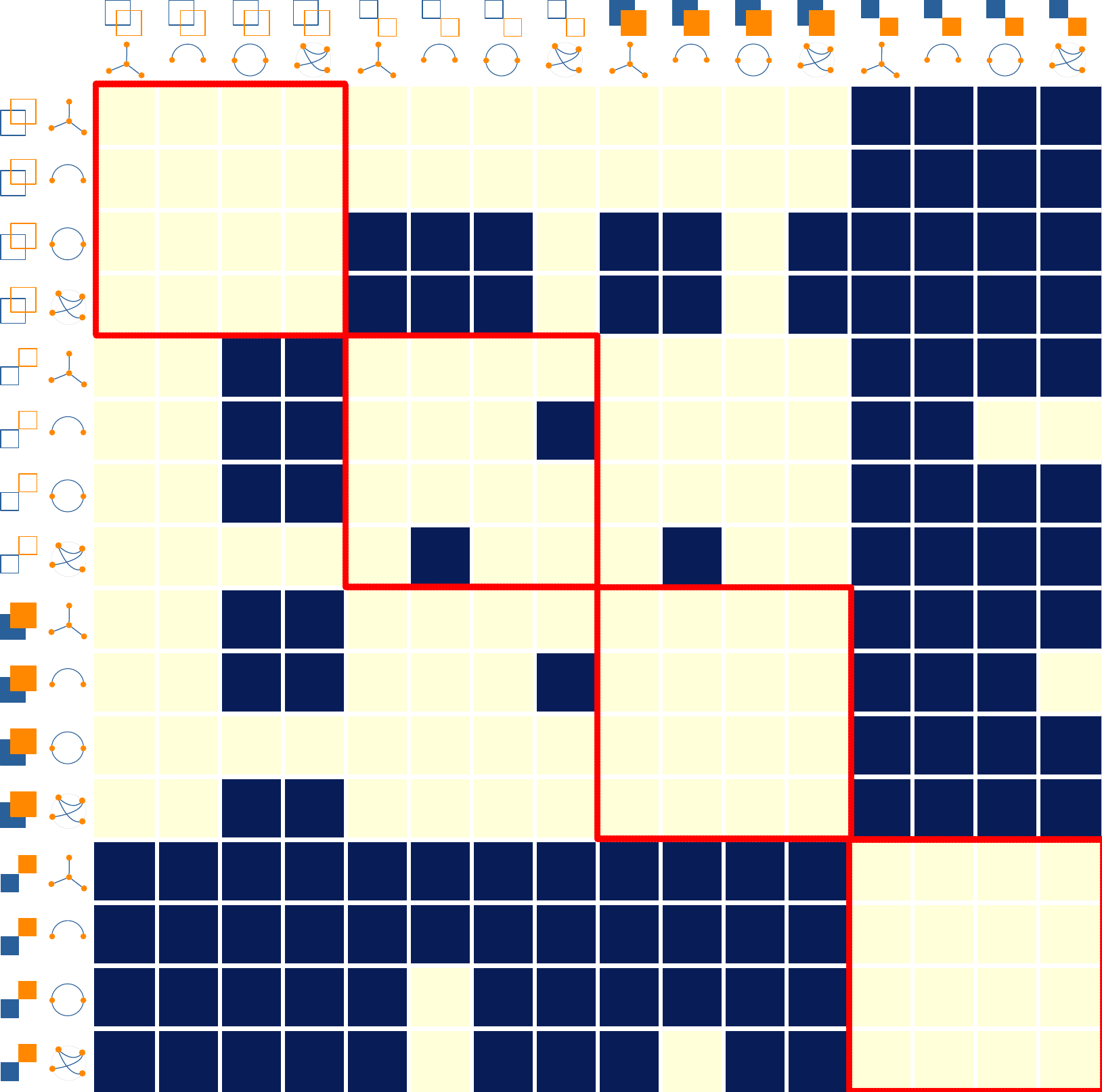}
    \vspace{-0.5em}
    \caption{Overall task completion time of cluster count for sizes = all, clusters = compact, \textbf{order = OLO}, baseline = Linlog. The dark matrix cells correspond to Bonferroni corrected $p\leq0.05$ for all pairwise comparisons. Linlog is not significantly different from OLO-ordered orderable layouts.}

    \label{fig:overall_olo_linlog_time}
\end{figure*}

\newpage

\subsection*{Task completion time of orderable layouts for loose and compact clusters}

\begin{figure*}[h]
\centering
        \includegraphics[height=0.235\textwidth]{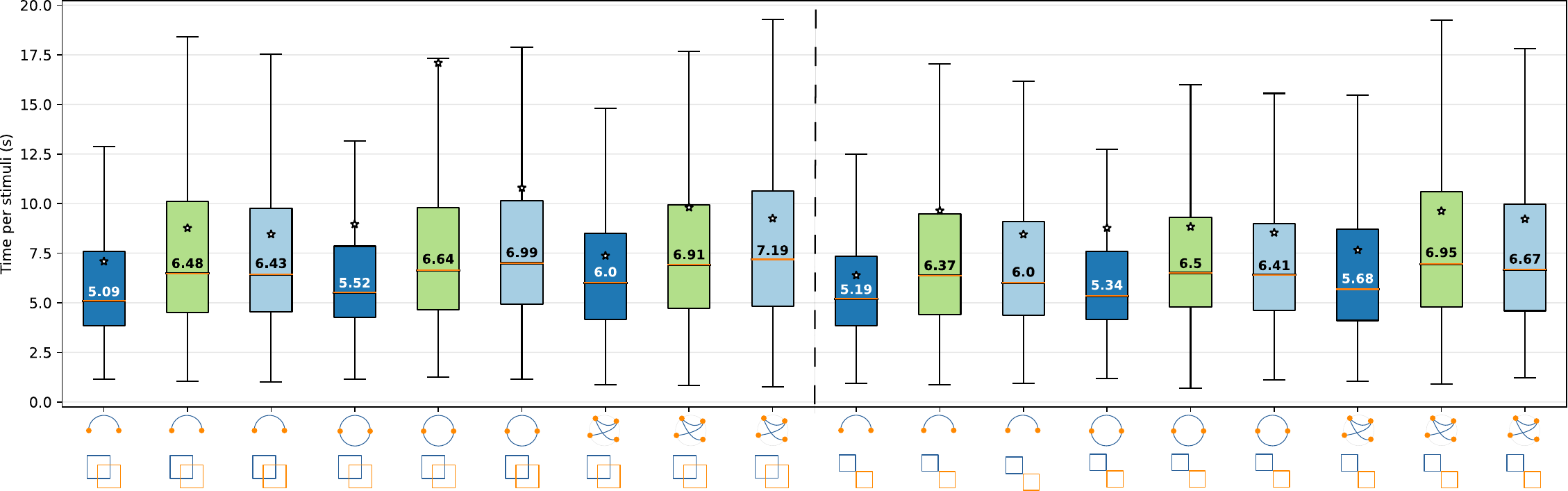}
    \hfill
    \includegraphics[height=0.235\textwidth]{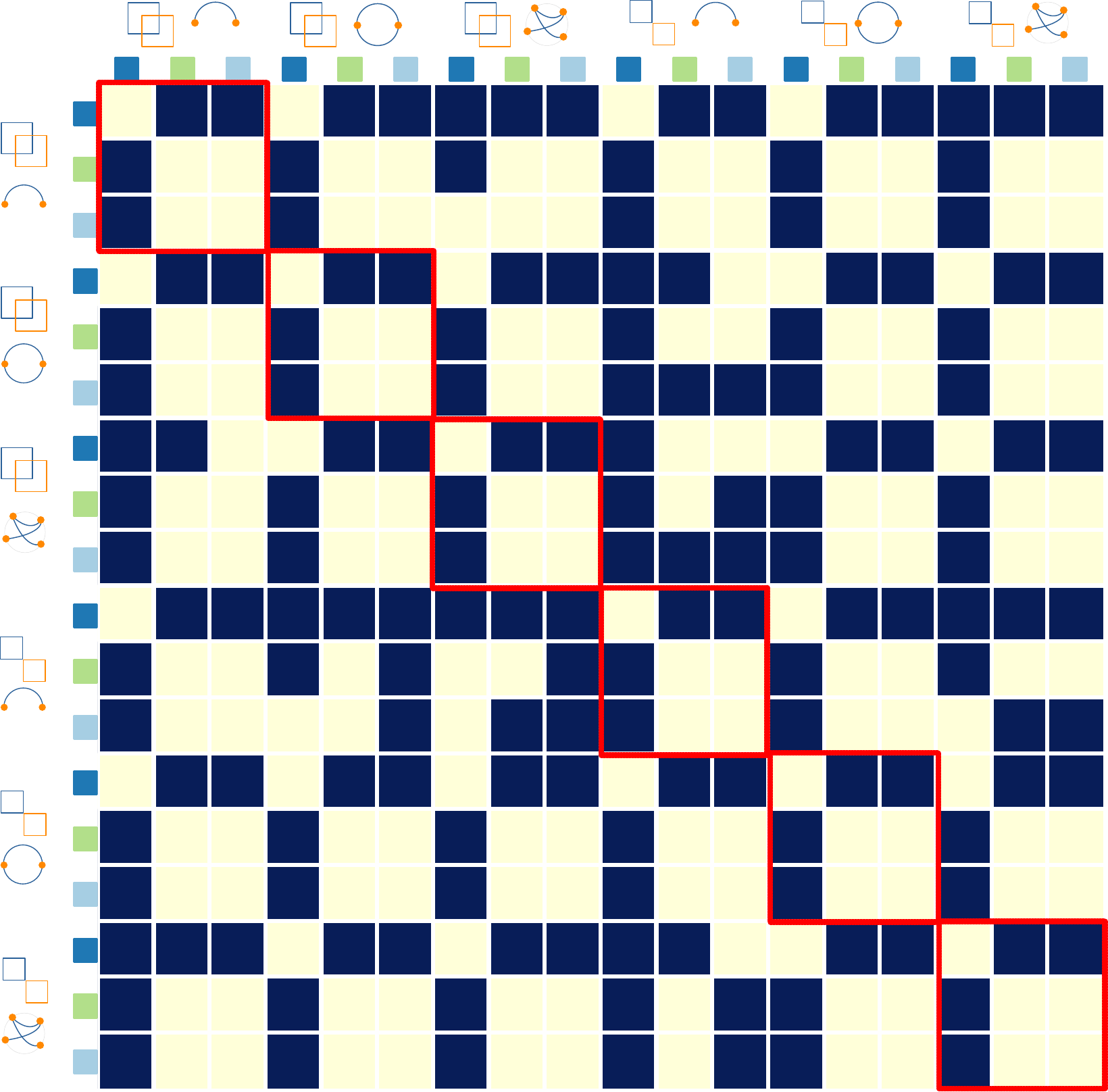}
    \vspace{-0.5em}
    \caption{Task completion time for network sizes = all, \textbf{clusters = loose}, orders = \{ GEN~\square{gen}, CR~\square{cr}, OLO~\square{olo}\}. On the right, the dark blue matrix cells correspond to the Bonferroni corrected $p\leq0.05$ for all pairwise comparisons. GEN-ordered orderable layouts lead to consistently faster cluster count judgements compared to those ordered by OLO and CR with respect to loose clusters.}

    \label{fig:overall_gen_cr_olo_loose_time}
\end{figure*}    

\begin{figure*}[h]
\centering
    \includegraphics[height=0.235\textwidth]{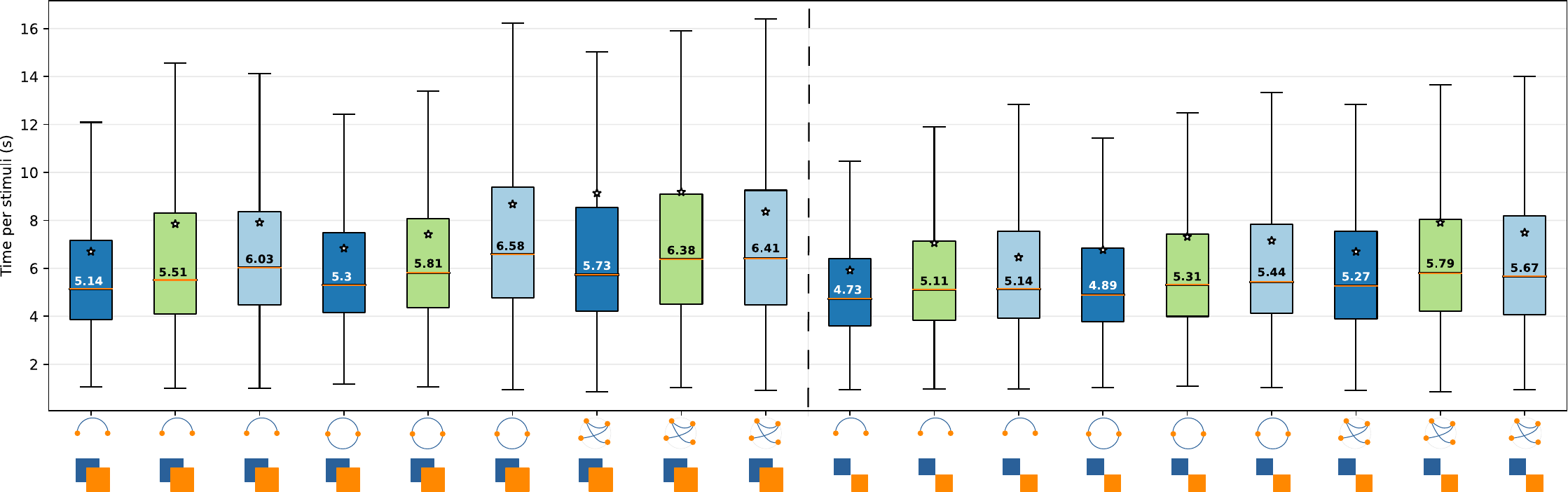}
    \hfill
    \includegraphics[height=0.235\textwidth]{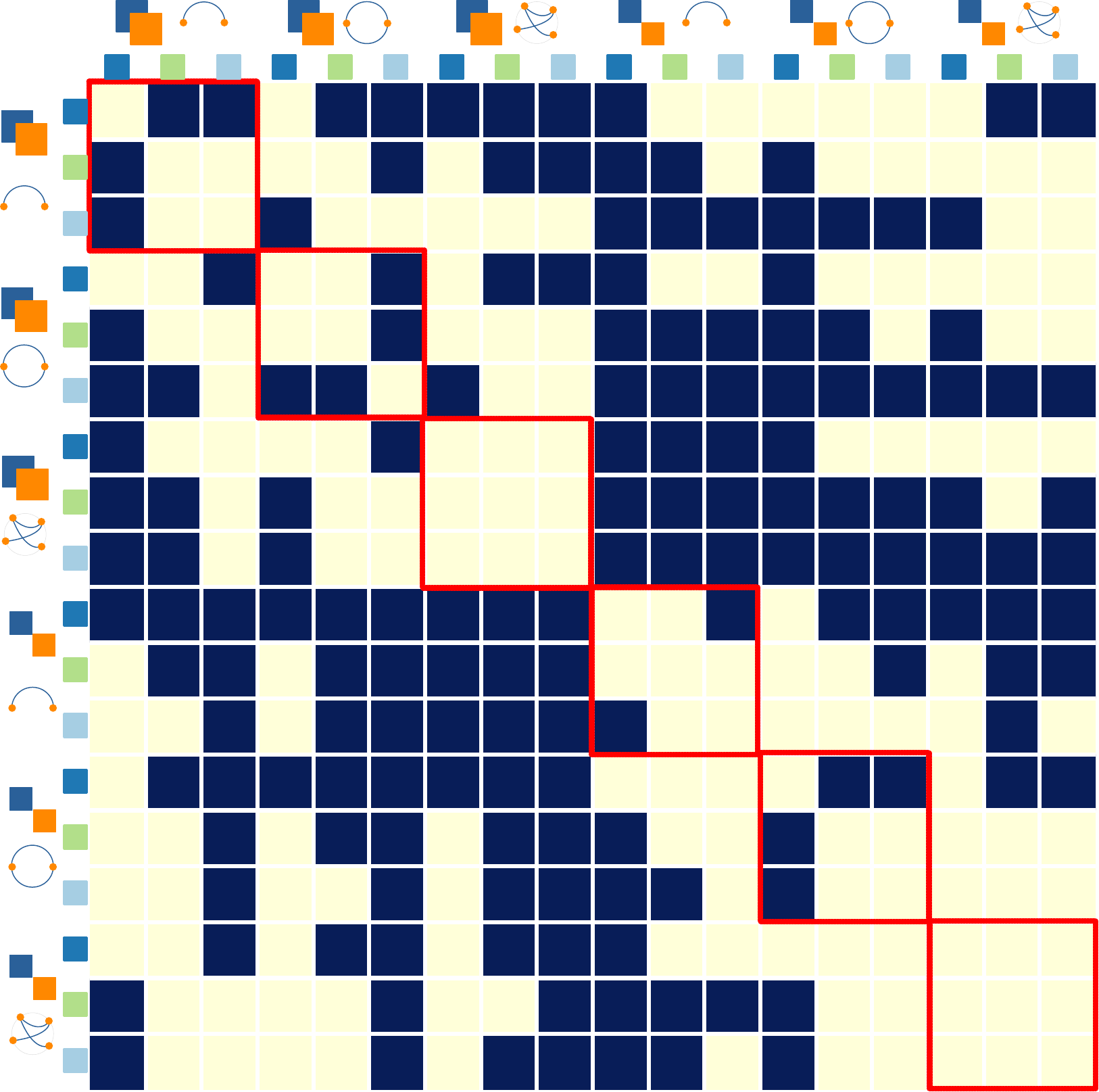}
    \vspace{-0.5em}
    \caption{Task completion time for network sizes = all, \textbf{clusters = compact}, orders = \{ GEN~\square{gen}, CR~\square{cr}, OLO~\square{olo}\}. On the right, the dark blue matrix cells correspond to the Bonferroni corrected $p\leq0.05$ for all pairwise comparisons. OLO and CR are not statistically significantly different in any of the compact cluster settings. GEN is faster than OLO when combined with arc and symmetric arc layouts. GEN beats CR occasionally with arc and symmetric arc in some compact cluster settings. GEN, CR and OLO are on par when it comes to radial layouts.}

    \label{fig:overall_gen_cr_olo_compact_time}
\end{figure*}
\newpage 
\subsection*{Comparison of backbone and orderable layouts}

\begin{figure*}[h]
\centering
    \includegraphics[height=0.235\textwidth]{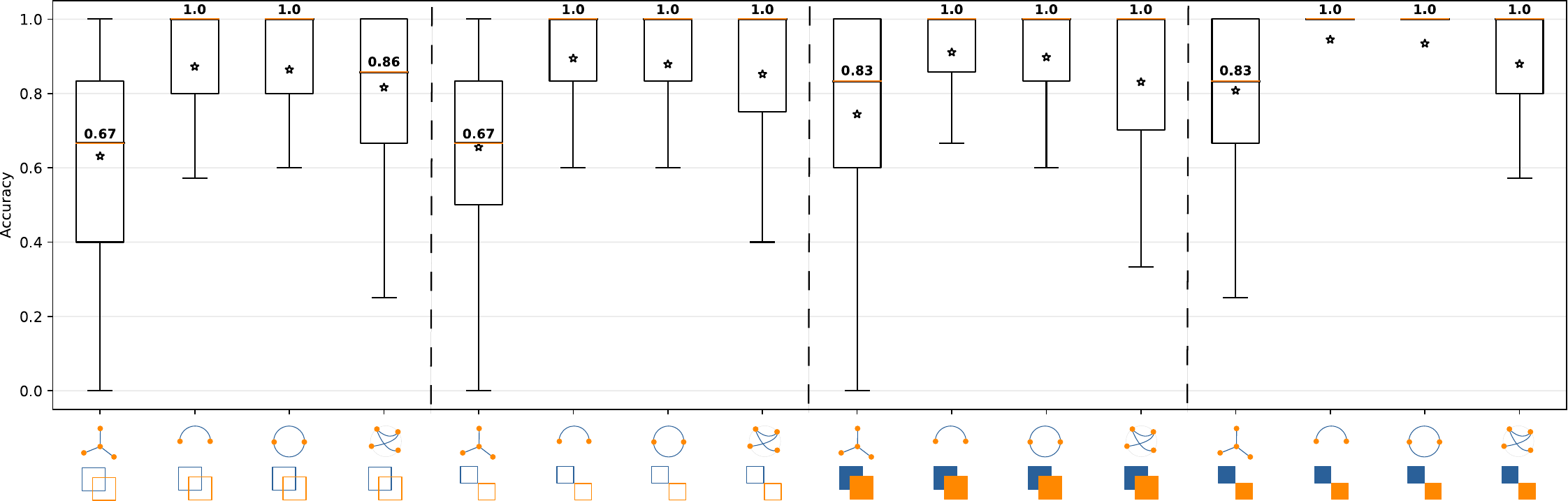}
    \hfill        
    \includegraphics[height=0.235\textwidth]{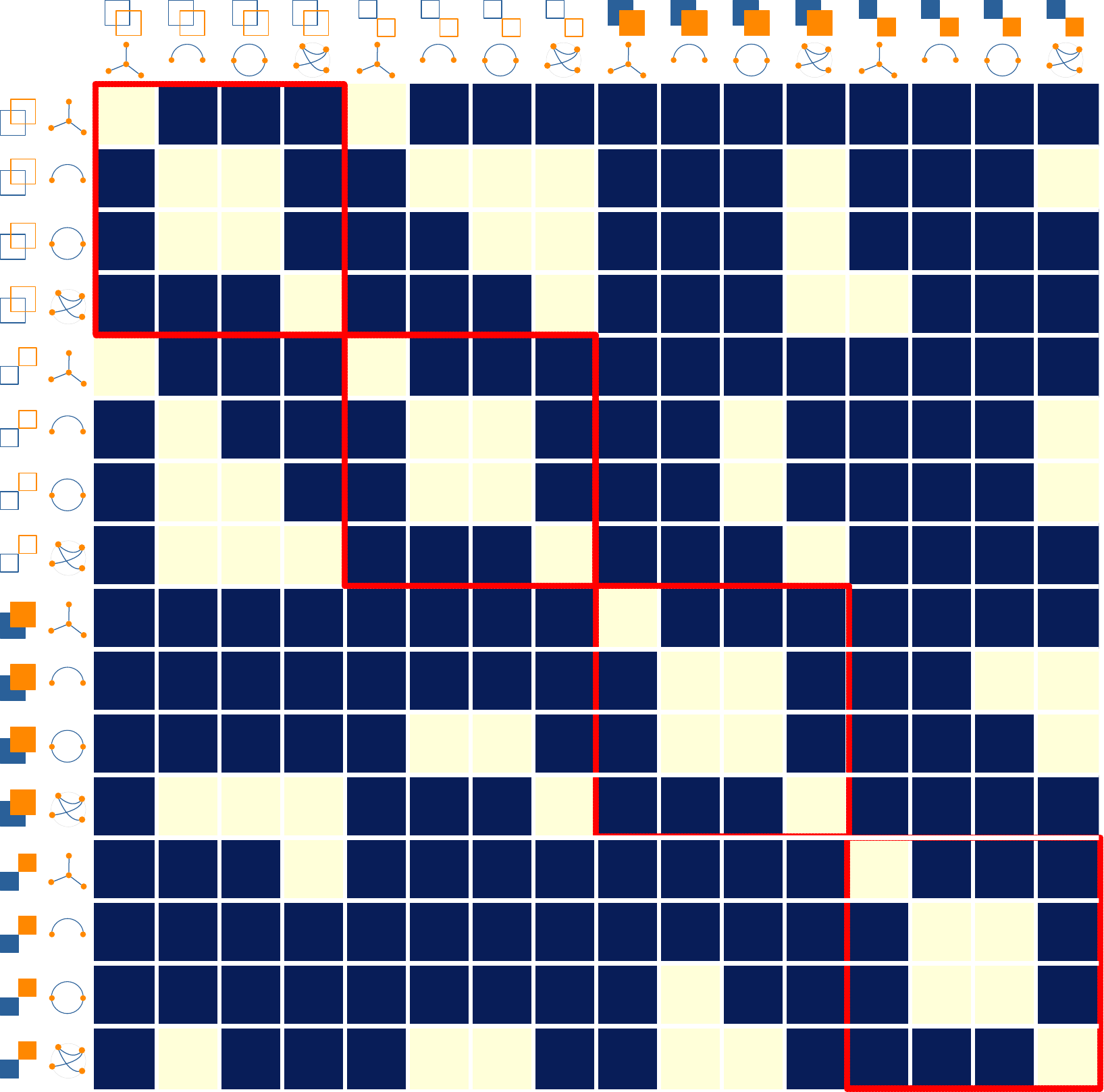}
\vspace{-0.5em}    
    \caption{Overall \textbf{accuracy} of cluster count judgments for sizes = all, clusters = all, \textbf{order = GEN}, baseline = backbone. On the right, the dark blue matrix cells correspond to the Bonferroni corrected $p\leq0.05$ for all pairwise comparisons. GEN-ordered orderable layouts are significantly (up to 33\%) more accurate than backbone for all cluster types. the orderable layouts achieve 100\% accuracy, in almost all cluster settings. Radial layouts have a greater variance than arc and arc symmetric.}

   \label{fig:overall_backbone_gen_acc}
\vspace{1.5em}

    \includegraphics[height=0.235\textwidth]{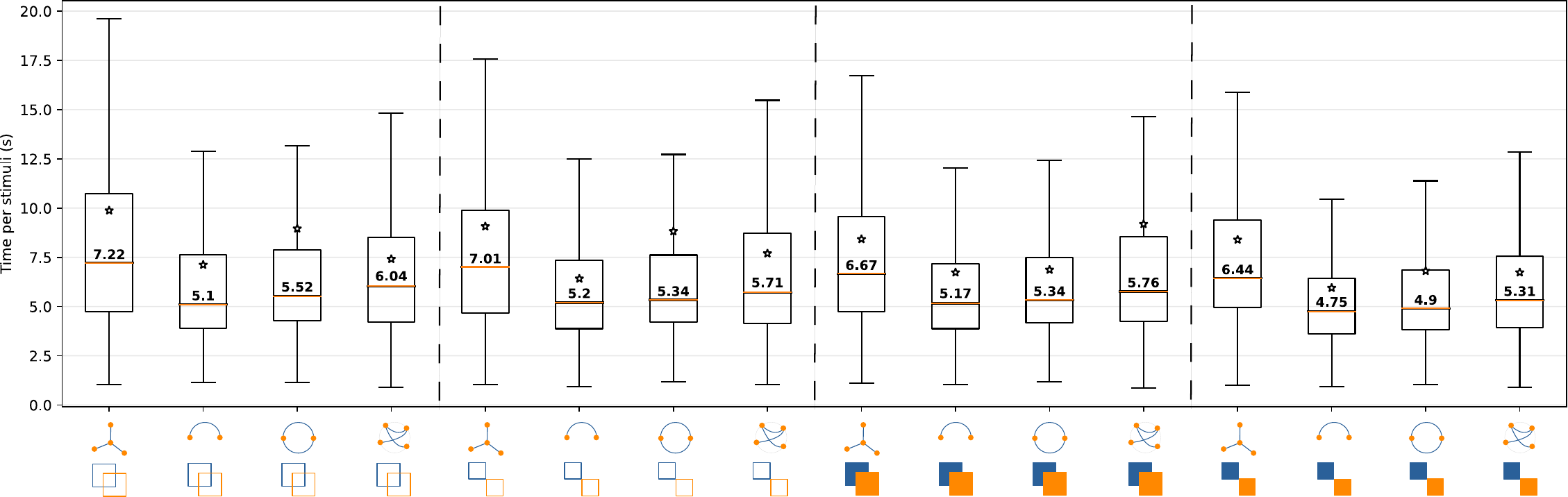}
    \hfill        
    \includegraphics[height=0.235\textwidth]{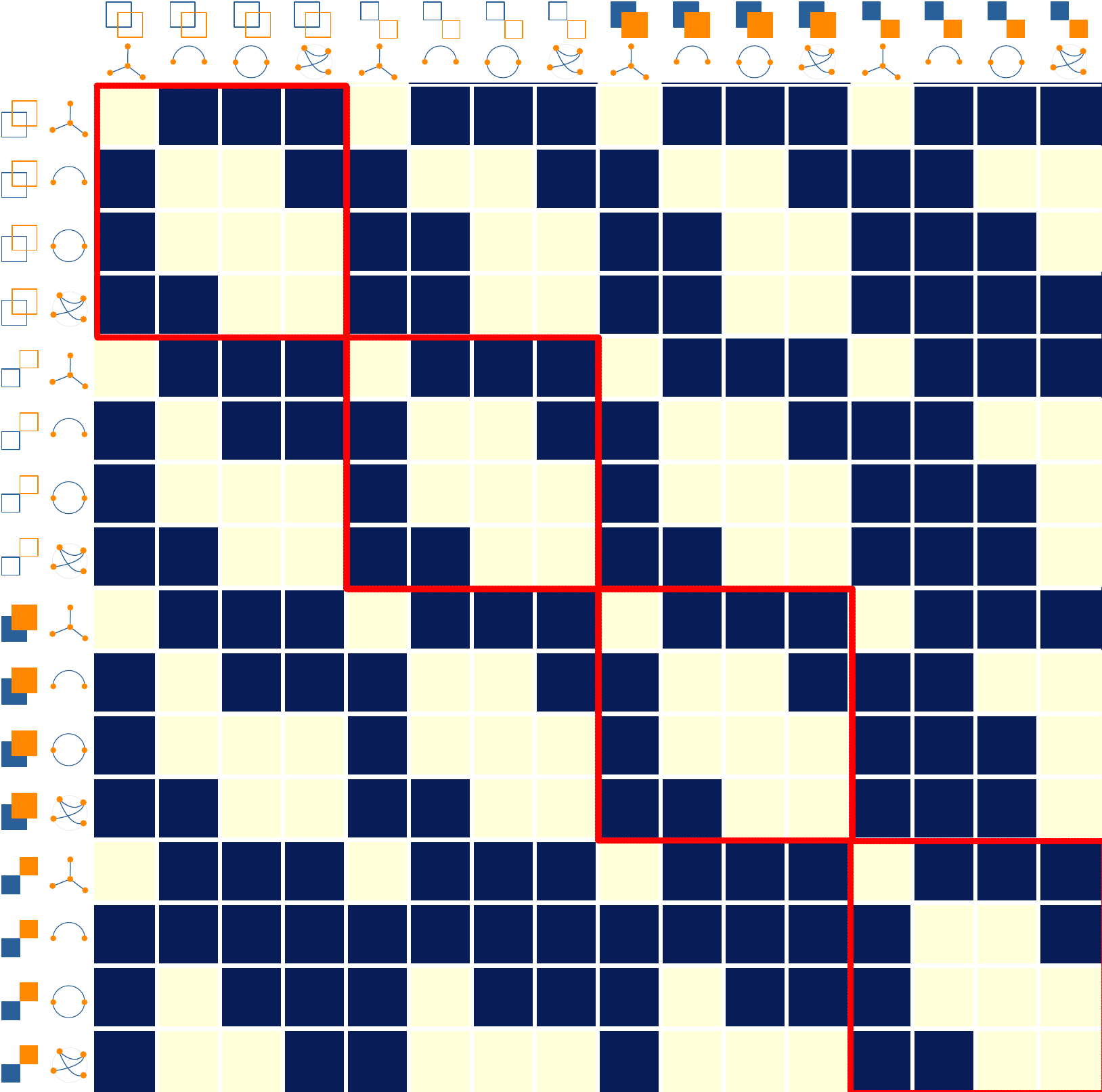}
\vspace{-0.5em}    
    \caption{Task \textbf{completion time} for sizes = all, clusters = all, \textbf{order = GEN}, baseline = backbone. On the right, the dark blue matrix cells correspond to the Bonferroni corrected $p\leq0.05$ for all pairwise comparisons. The lack of accuracy of backbone seen in \autoref{fig:overall_backbone_gen_acc} also translates into slower cluster count judgments in all cluster settings. Likewise, radial is always slower than arc layouts.}
   \label{fig:overall_backbone_gen_time}
\vspace{1.5em}

    \includegraphics[height=0.235\textwidth]{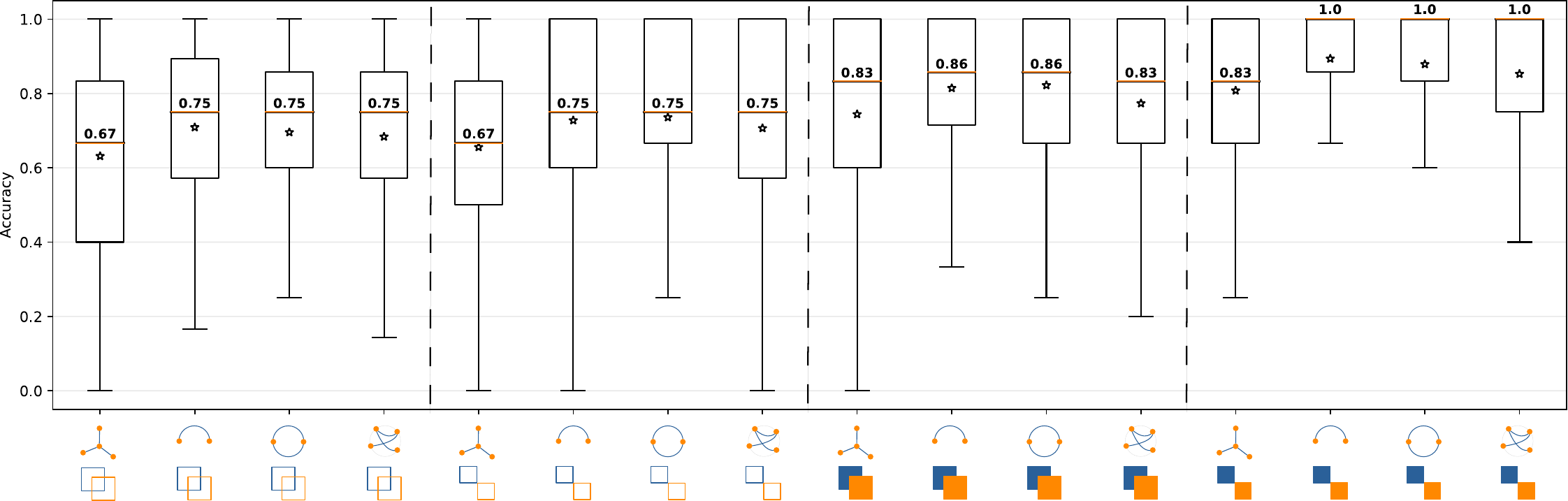}
    \hfill        
    \includegraphics[height=0.235\textwidth]{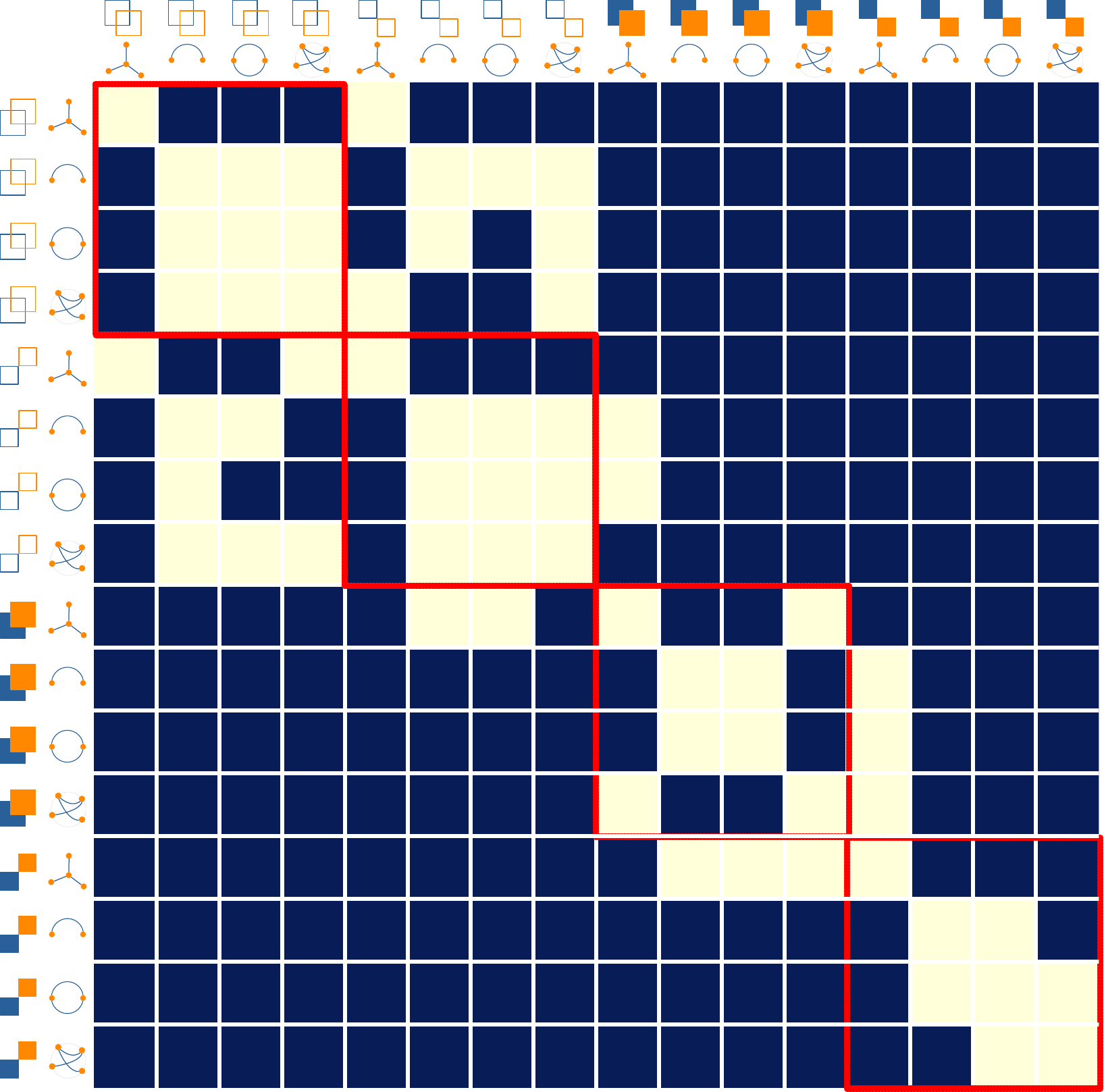}
\vspace{-0.5em}    
    \caption{Overall \textbf{accuracy} of cluster count judgments for sizes = all, clusters = all, \textbf{order = CR}, baseline = backbone. On the right, the dark blue matrix cells correspond to the Bonferroni corrected $p\leq0.05$ for all pairwise comparisons. Backbone is less accurate than the CR-ordered orderable layouts in all cluster settings, except in the case compact inseparable clusters where it cannot be distinguished from radial.}

   \label{fig:overall_backbone_cr_acc}
\vspace{0.5em}

\end{figure*}

\begin{figure*} [h]
\centering
    \includegraphics[height=0.235\textwidth]{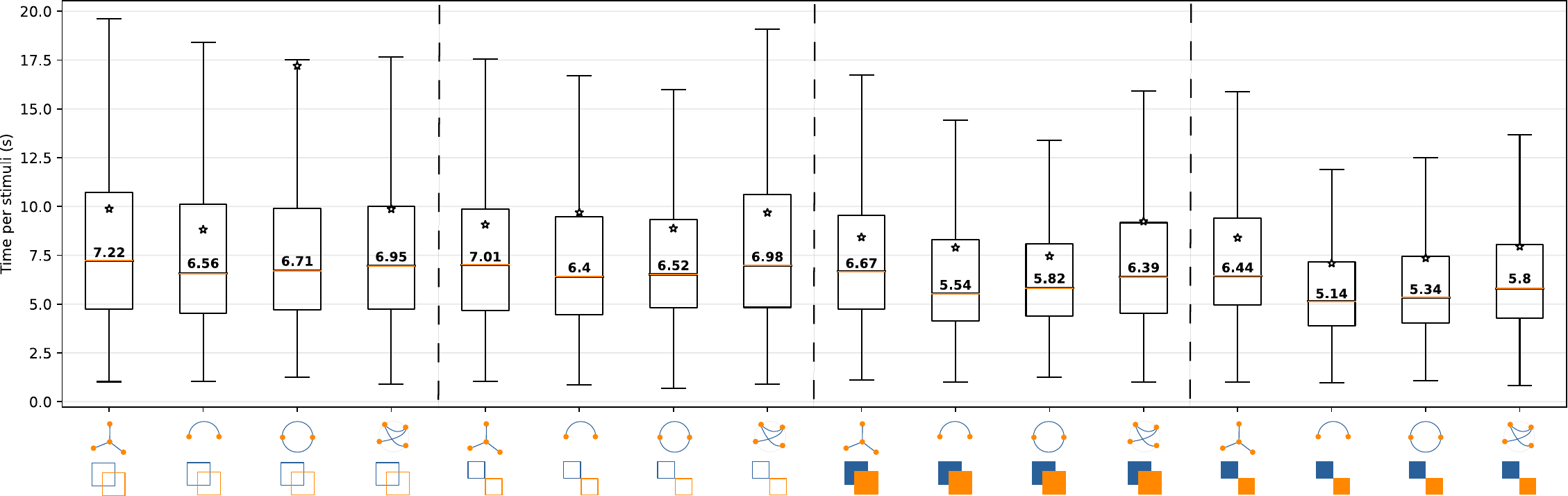}
    \hfill        
    \includegraphics[height=0.235\textwidth]{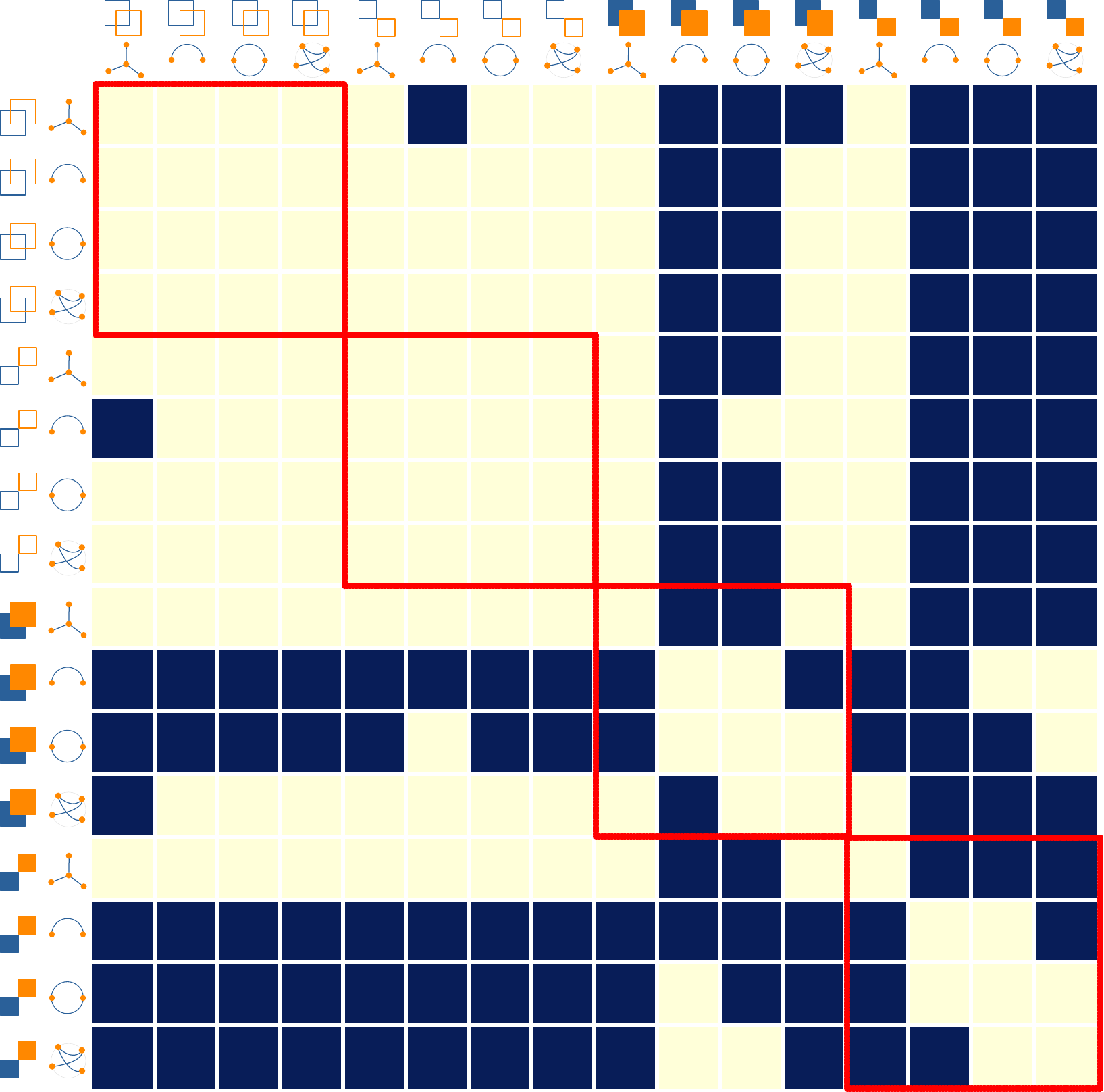}
\vspace{-0.5em}    
    \caption{Overall task \textbf{completion time} for sizes = all, clusters = all, \textbf{order = CR}, baseline = backbone. On the right, the dark blue matrix cells correspond to the Bonferroni corrected $p\leq0.05$ for all pairwise comparisons. Backbone is significantly slower than CR-ordered orderable layouts for compact clusters, and comparable in the case of loose clusters.}
   \label{fig:overall_backbone_cr_time}
\vspace{1.5em}       
    \includegraphics[height=0.235\textwidth]{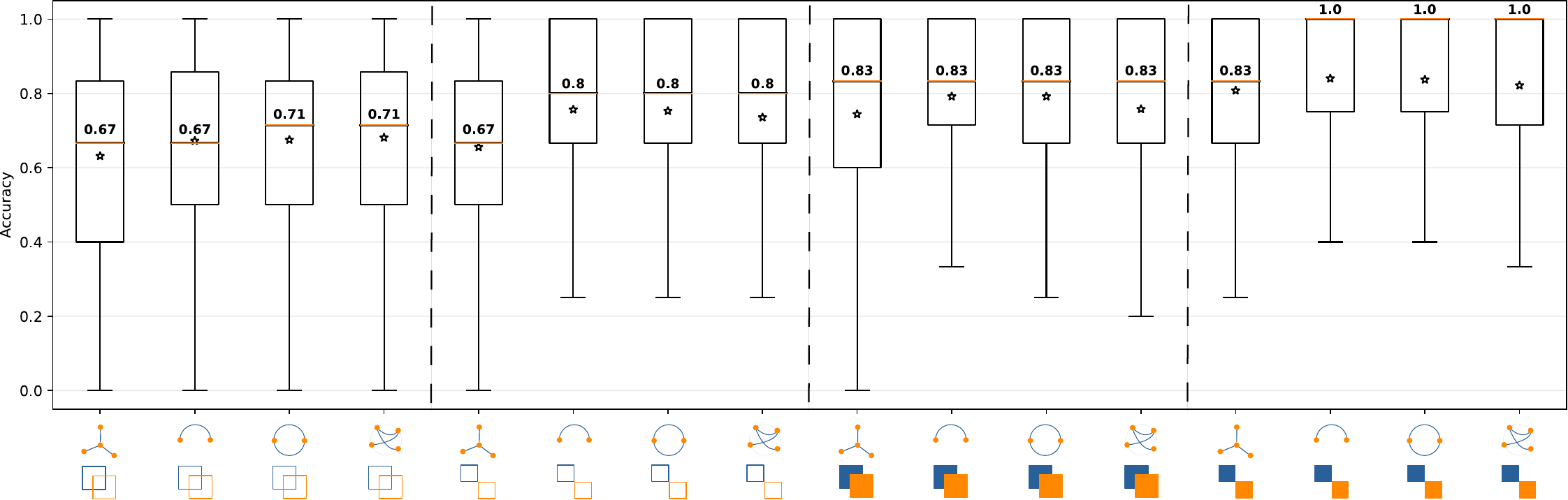}
    \hfill        
    \includegraphics[height=0.235\textwidth]{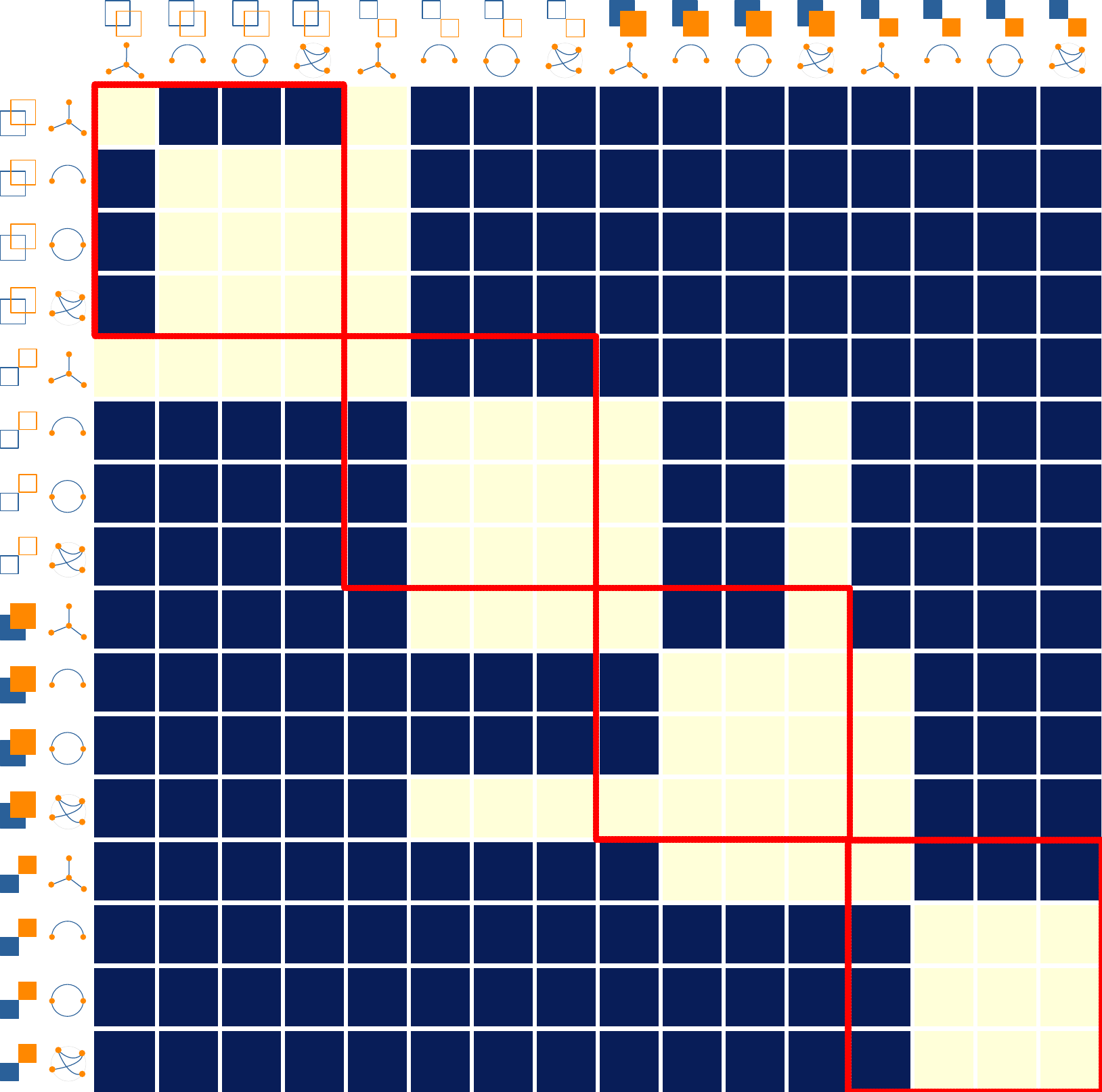}
\vspace{-0.5em}    
    \caption{Overall \textbf{accuracy} of cluster count judgments for sizes = all, clusters = all, \textbf{order = OLO}, baseline = backbone. On the right, the dark blue matrix cells correspond to the Bonferroni corrected $p\leq0.05$ for all pairwise comparisons. Backbone is less accurate than the OLO-ordered orderable layouts in all cluster settings, except in the case compact inseparable clusters where it cannot be distinguished from radial.}
   \label{fig:overall_backbone_olo_acc}
   \vspace{1.5em}

    \includegraphics[height=0.235\textwidth]{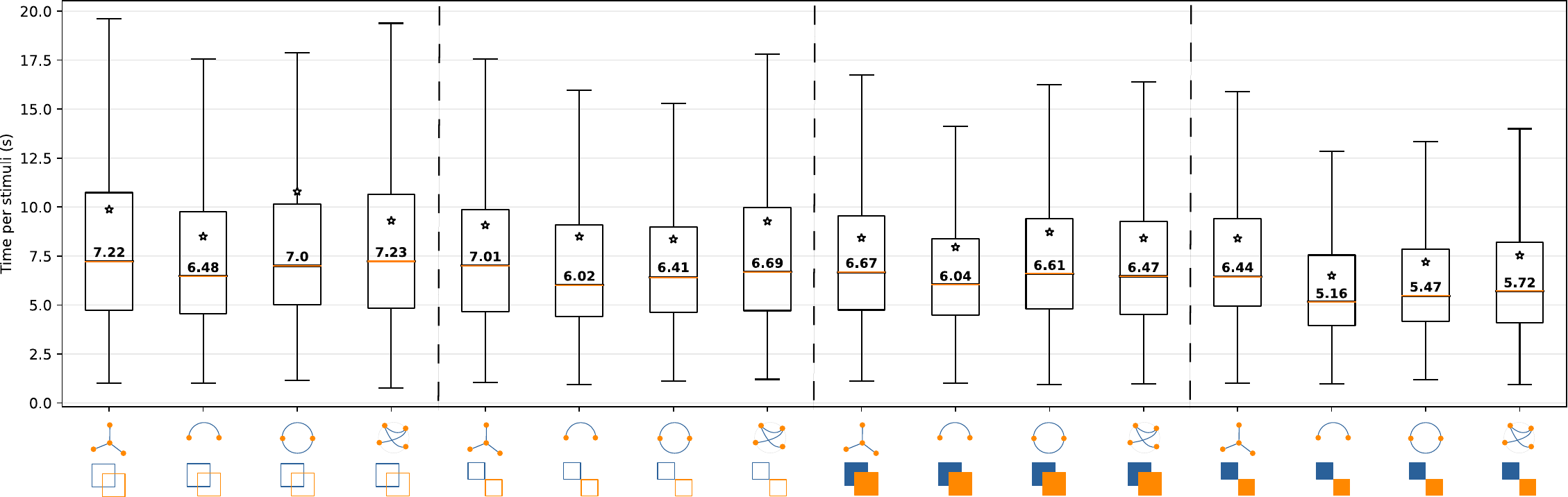}
    \hfill        
    \includegraphics[height=0.235\textwidth]{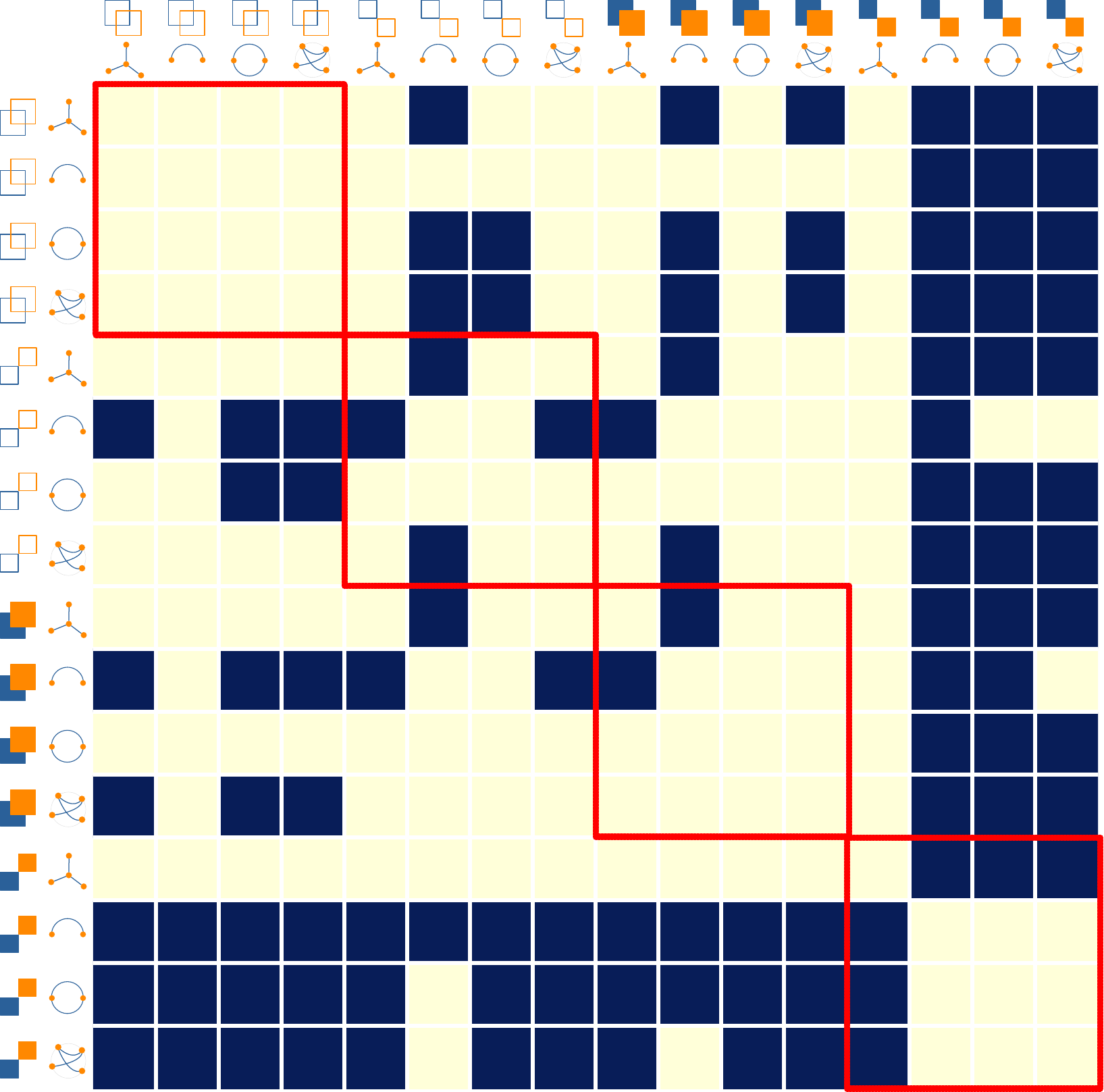}
\vspace{-0.5em}    
    \caption{Overall task \textbf{completion time} for sizes = all, clusters = all, \textbf{order = OLO}, baseline = backbone. On the right, the dark blue matrix cells correspond to the Bonferroni corrected $p\leq0.05$ for all pairwise comparisons. Backbone is slower than OLO-ordered orderable layouts for compact and separable clusters. It is comparable to orderable layouts in the other cluster types, except for loose separable and compact inseparable clusters where arc layouts are significantly faster.}
   \label{fig:overall_backbone_olo_time}
\end{figure*}

\clearpage
\subsection*{Comparison of sfdp and orderable layouts}

\begin{figure*}[h]
\centering
    \includegraphics[height=0.235\textwidth]{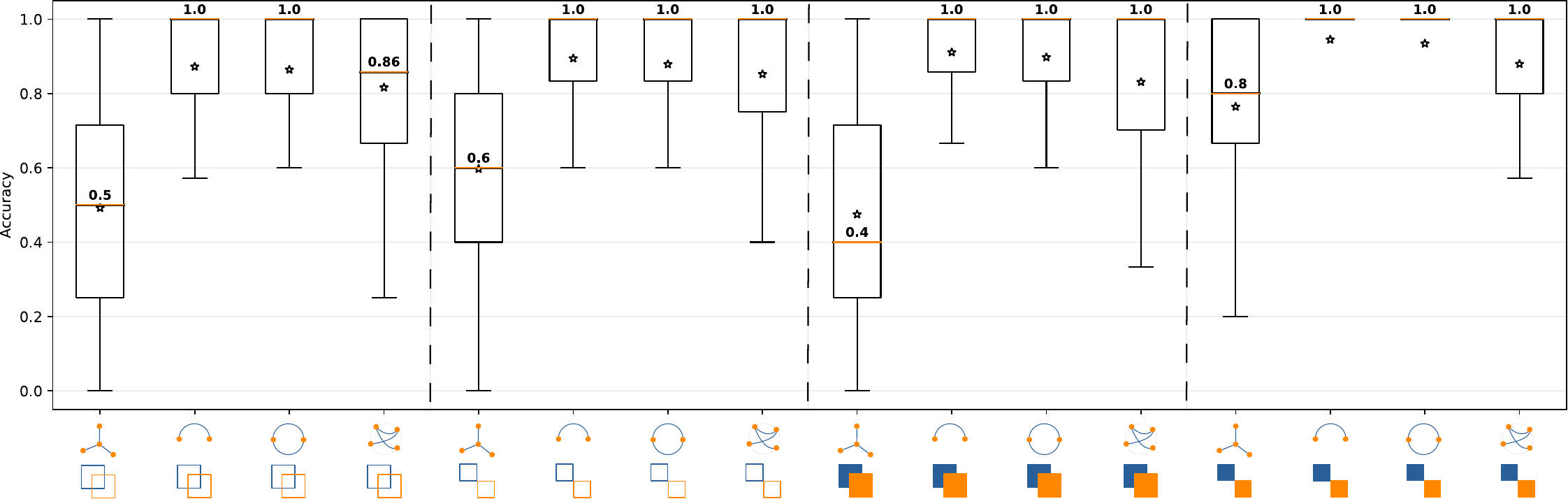}
    \hfill        
    \includegraphics[height=0.235\textwidth]{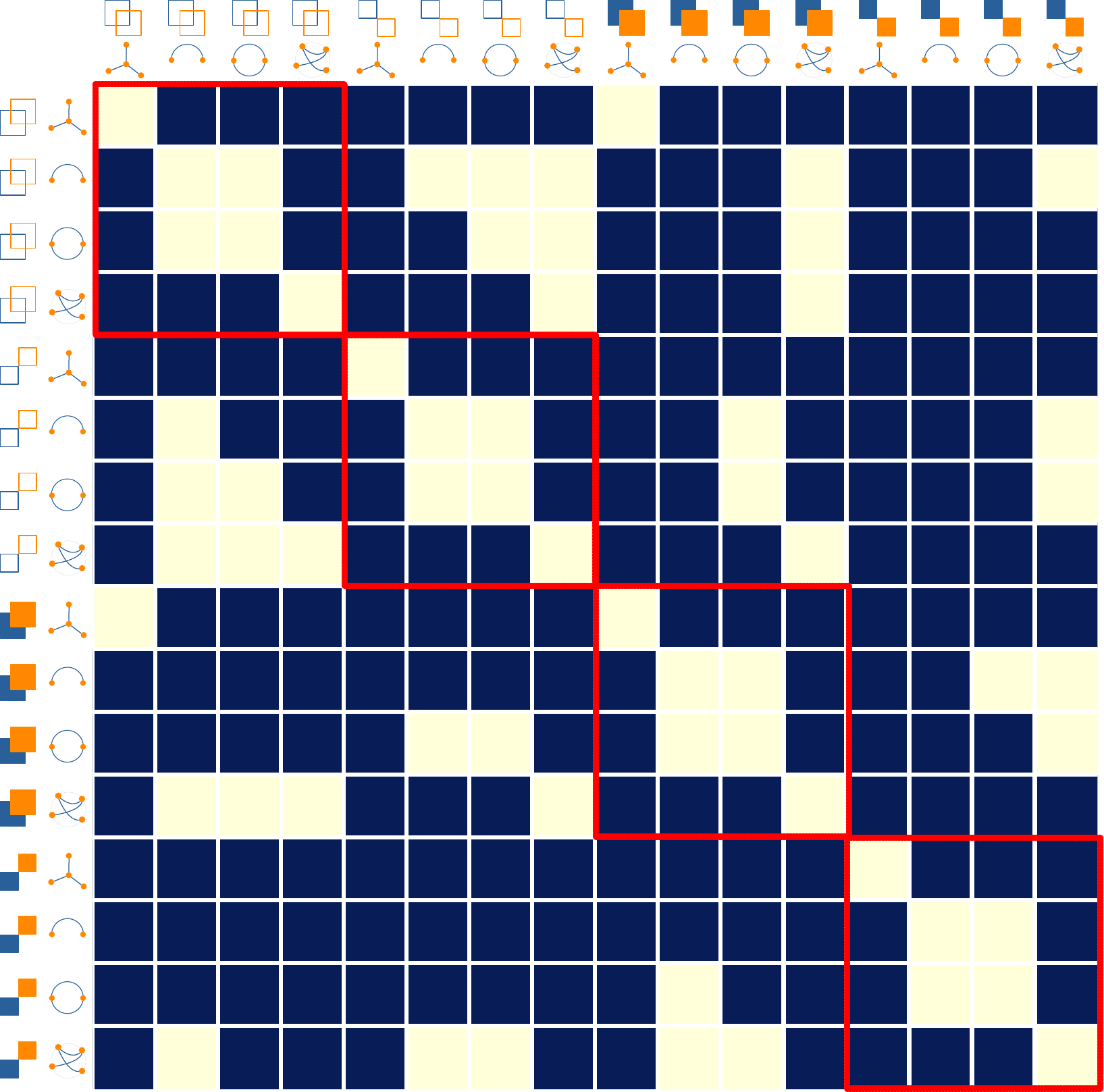}
\vspace{-0.5em}    
    \caption{Overall \textbf{accuracy} of cluster count judgments for sizes = all, clusters = all, \textbf{order = GEN}, baseline = sfdp. On the right, the dark blue matrix cells correspond to the Bonferroni corrected $p\leq0.05$ for all pairwise comparisons. sfdp is significantly (up to 60\%) slower than GEN-ordered orderable layouts for all cluster settings. All orderable layouts achieve 100\% median accuracy. Radial layouts have a greater variance than arc and symmetric arc layouts.}

   \label{fig:overall_sfdp_gen_acc}
\vspace{1.5em}

    \includegraphics[height=0.235\textwidth]{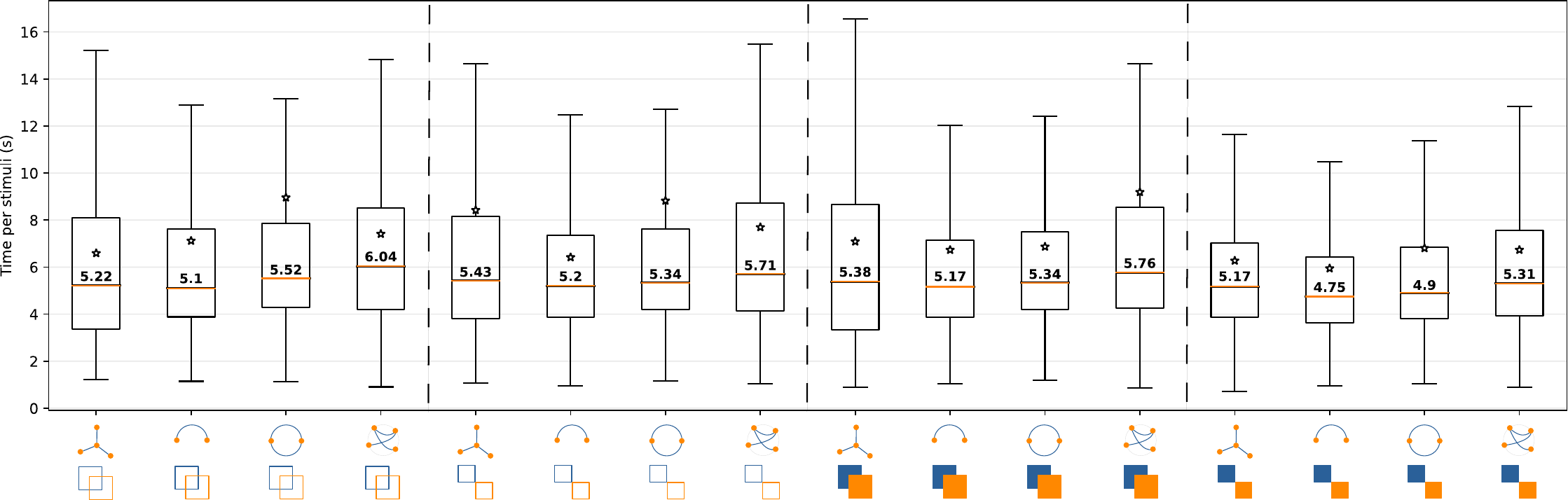}
    \hfill        
    \includegraphics[height=0.235\textwidth]{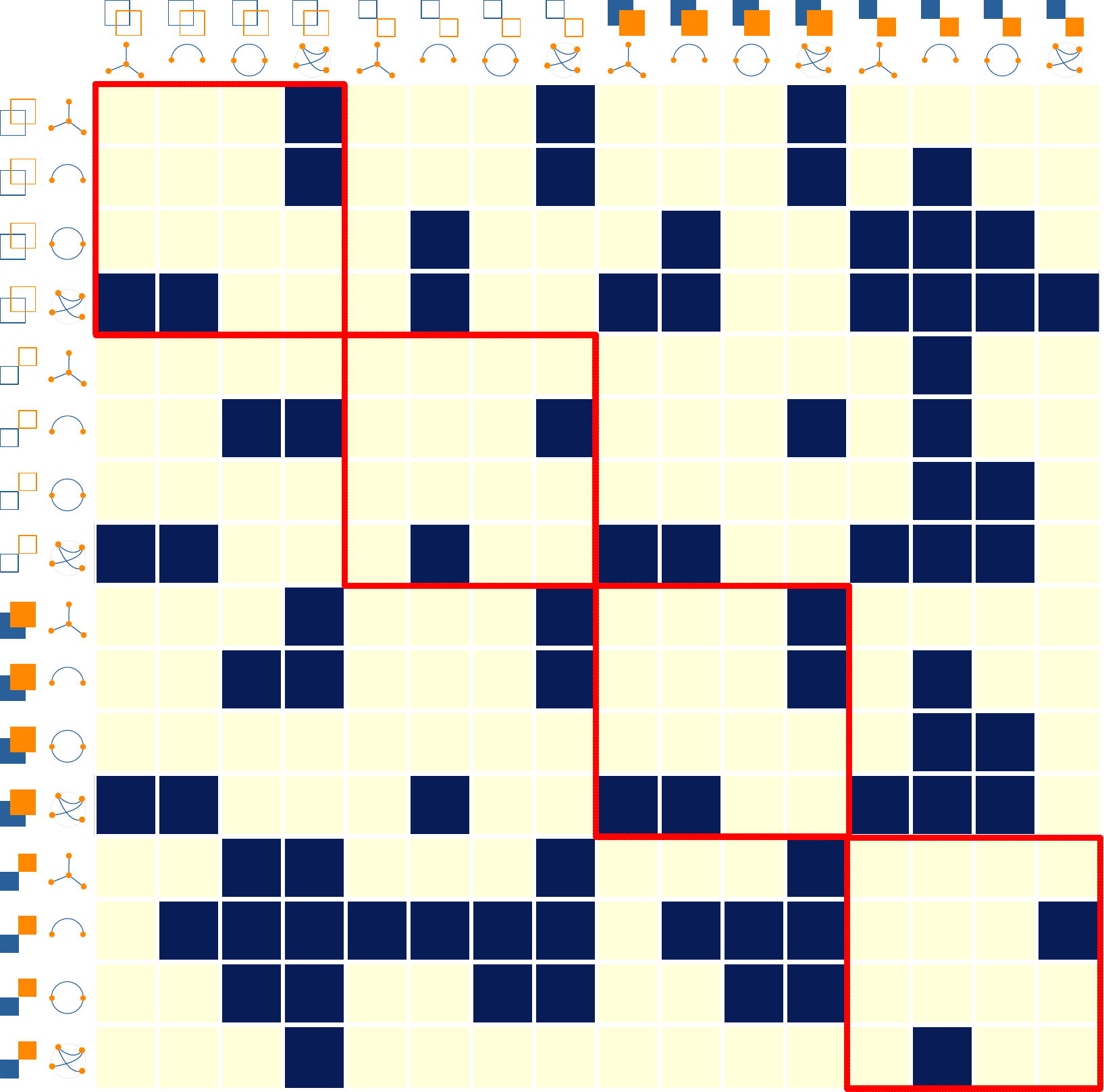}
\vspace{-0.5em}    
    \caption{Overall task \textbf{completion time} for sizes = all, clusters = all, \textbf{order = GEN}, baseline = sfdp. On the right, the dark blue matrix cells correspond to the Bonferroni corrected $p\leq0.05$ for all pairwise comparisons. The poorer accuracy of sfdp observed in \autoref{fig:overall_sfdp_gen_acc} comes with short completion time, indicating that the participants were discouraged by the relatively higher visual clutter in sfdp plots.}
   \label{fig:overall_sfdp_gen_time}
   \vspace{1.5em}

    \includegraphics[height=0.235\textwidth]{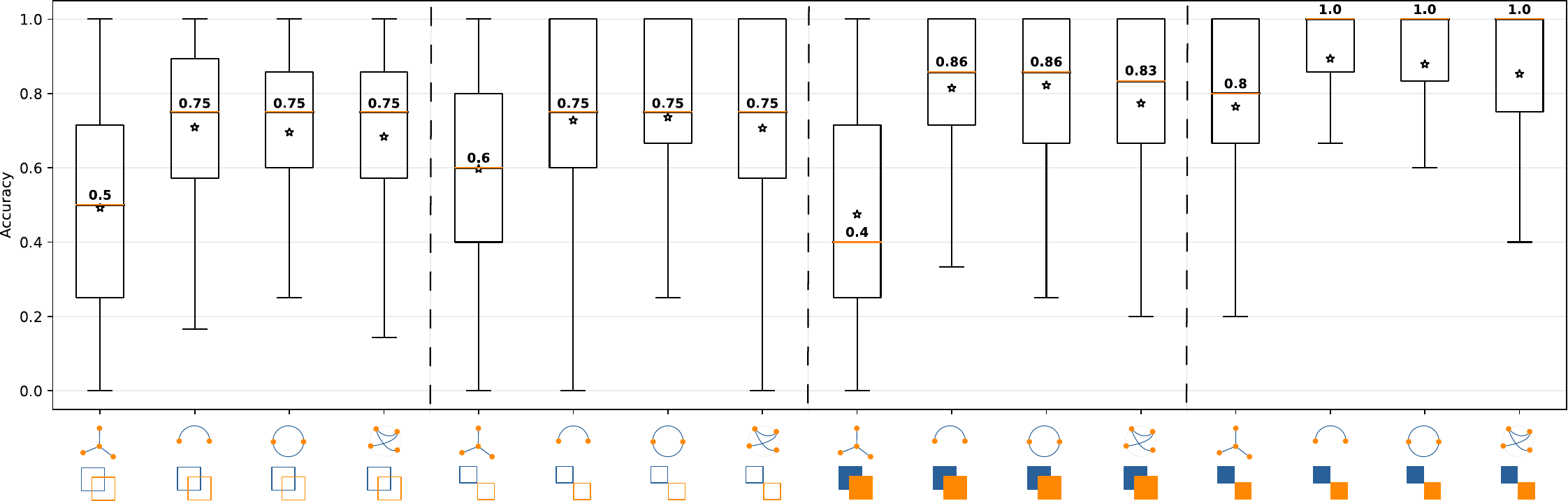}
    \hfill        
    \includegraphics[height=0.235\textwidth]{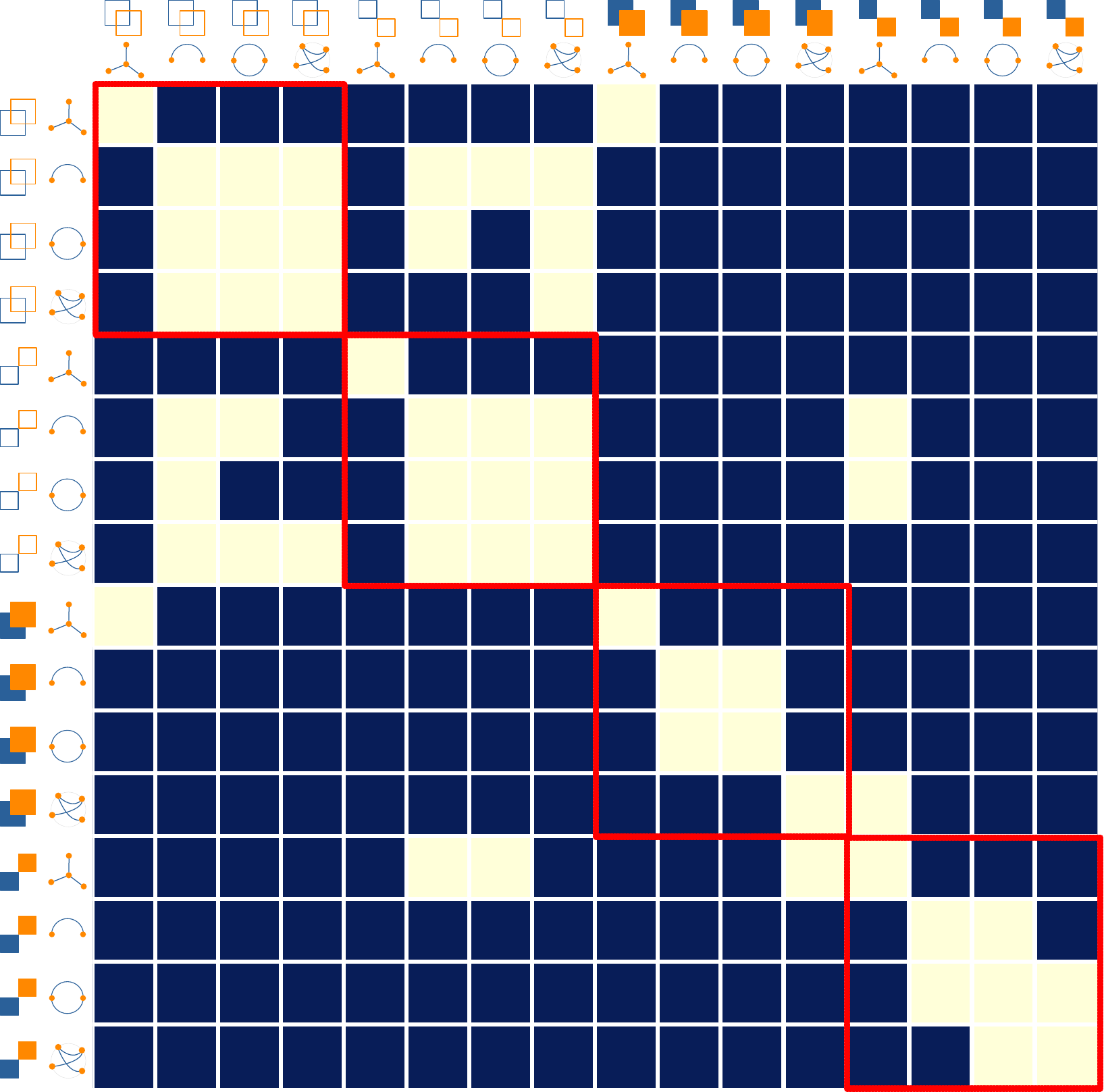}
\vspace{-0.5em}    
    \caption{Overall \textbf{accuracy} of cluster count judgments for sizes = all, clusters = all, \textbf{order = CR}, baseline = sfdp. On the right, the dark blue matrix cells correspond to the Bonferroni corrected $p\leq0.05$ for all pairwise comparisons. sfdp is significantly (up to 46\%) slower than CR-ordered orderable layouts for all cluster settings. All orderable layouts achieve high median accuracy, betwen 75\% and 100\%.}

   \label{fig:overall_sfdp_cr_acc}

\end{figure*}

\begin{figure*} [h]
\centering
    \includegraphics[height=0.235\textwidth]{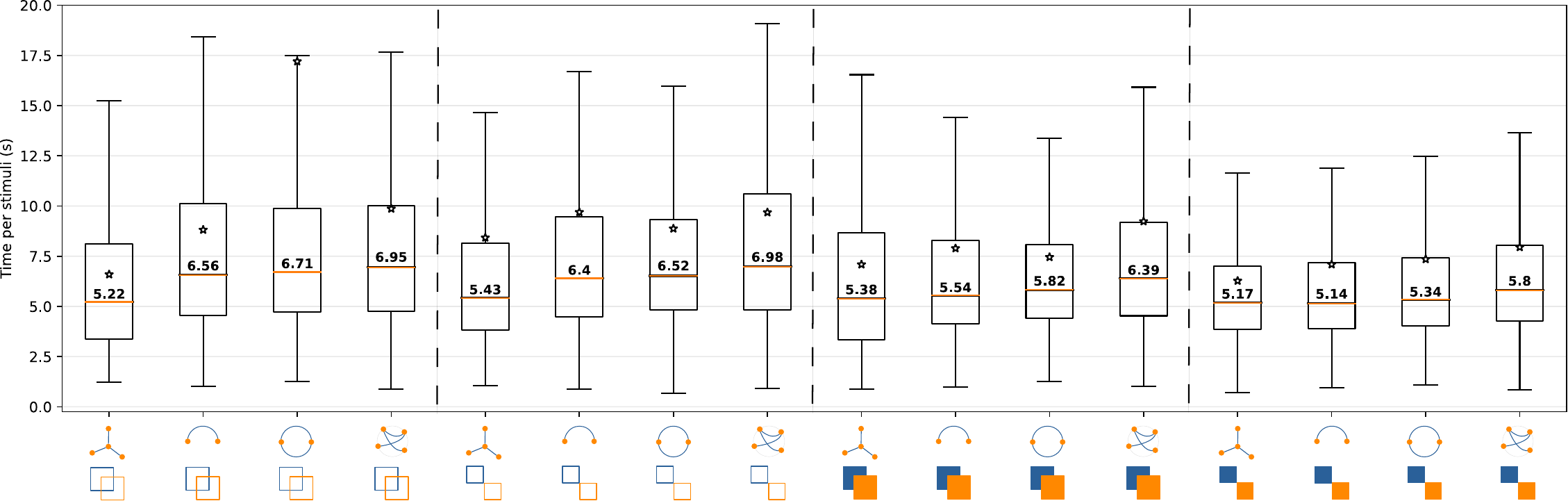}
    \hfill        
    \includegraphics[height=0.235\textwidth]{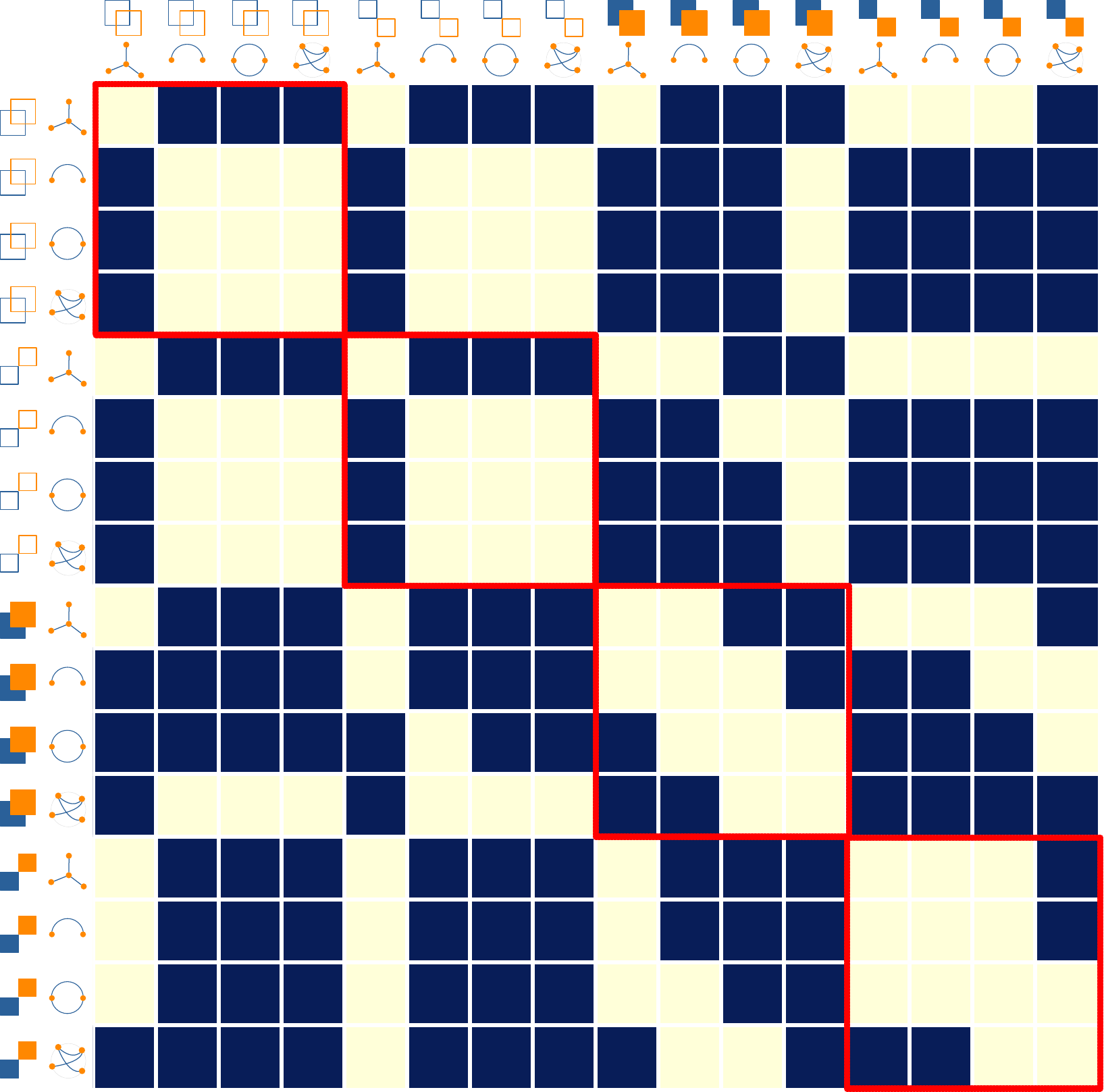}
\vspace{-0.5em}    
    \caption{Overall task \textbf{completion time} for sizes = all, clusters = all, \textbf{order = CR}, baseline = sfdp. On the right, the dark blue matrix cells correspond to the Bonferroni corrected $p\leq0.05$ for all pairwise comparisons. The poorer accuracy of sfdp observed in \autoref{fig:overall_sfdp_cr_acc} comes with short completion time, indicating that the participants were discouraged by the relatively higher visual clutter in sfdp plots.}
   \label{fig:overall_sfdp_cr_time}
   \vspace{1.5em}       
    \includegraphics[height=0.235\textwidth]{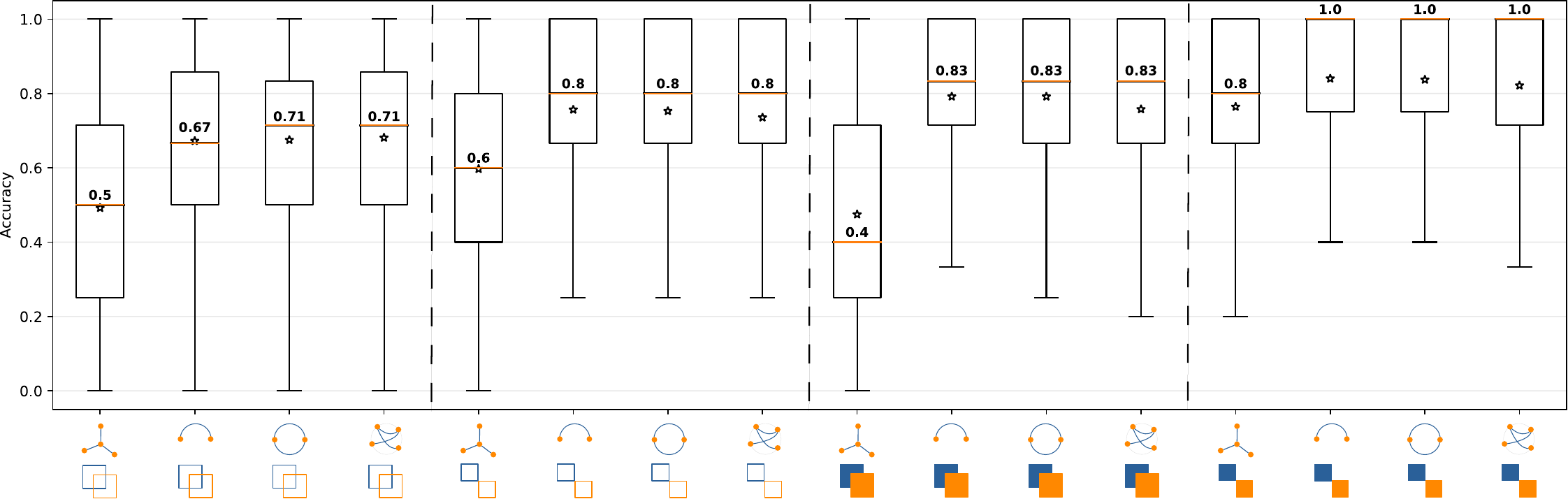}
    \hfill        
    \includegraphics[height=0.235\textwidth]{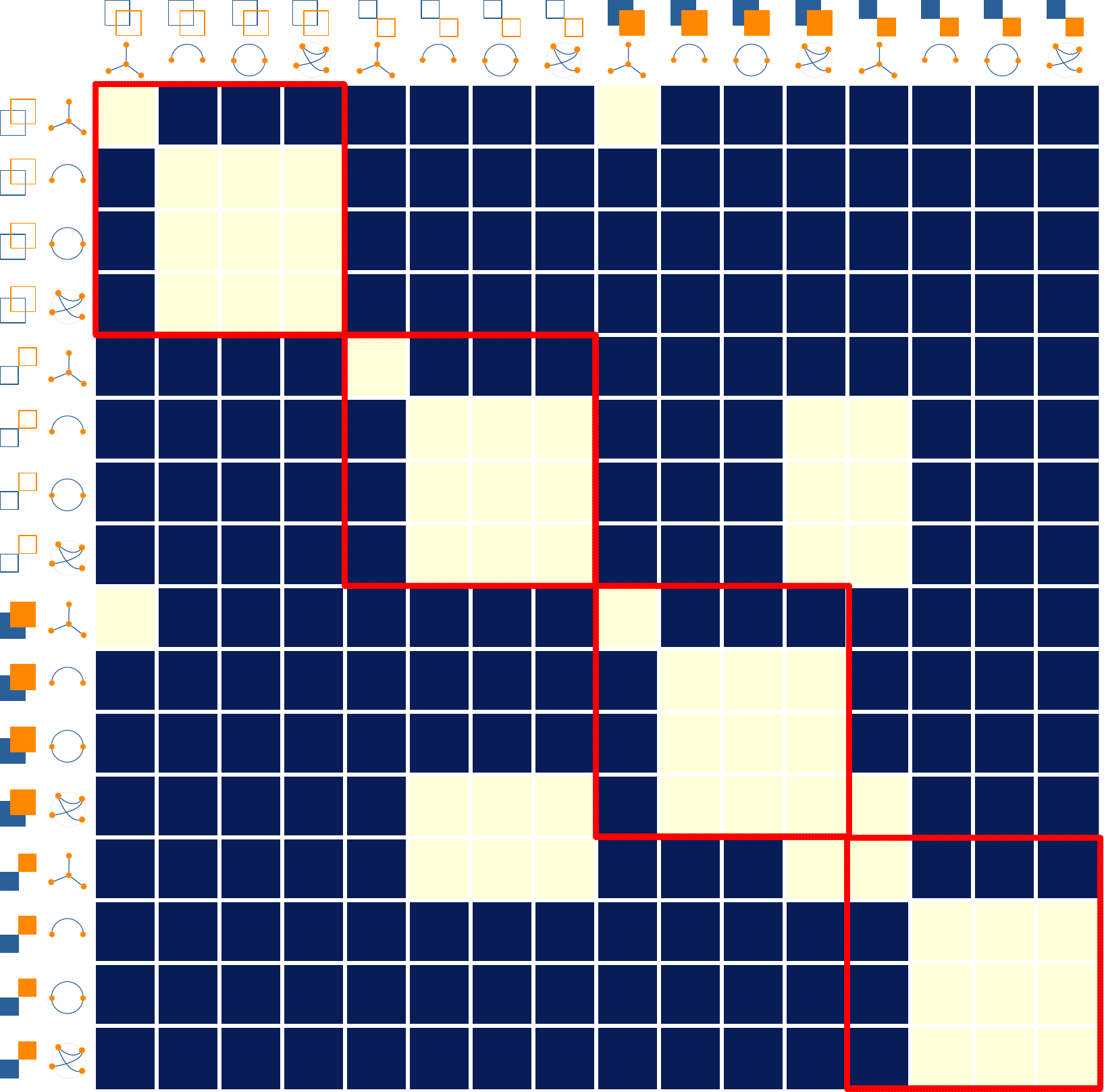}
\vspace{-0.5em}    
    \caption{Overall \textbf{accuracy} of cluster count judgments for sizes = all, clusters = all, \textbf{order = OLO}, baseline = sfdp. On the right, the dark blue matrix cells correspond to the Bonferroni corrected $p\leq0.05$ for all pairwise comparisons. sfdp is significantly (up to 43\%) slower than OLO-ordered orderable layouts for all cluster settings. All orderable layouts achieve high median accuracy, betwen 71\% and 100\%.}
   \label{fig:overall_sfdp_olo_acc}
   \vspace{1.5em}       

    \includegraphics[height=0.235\textwidth]{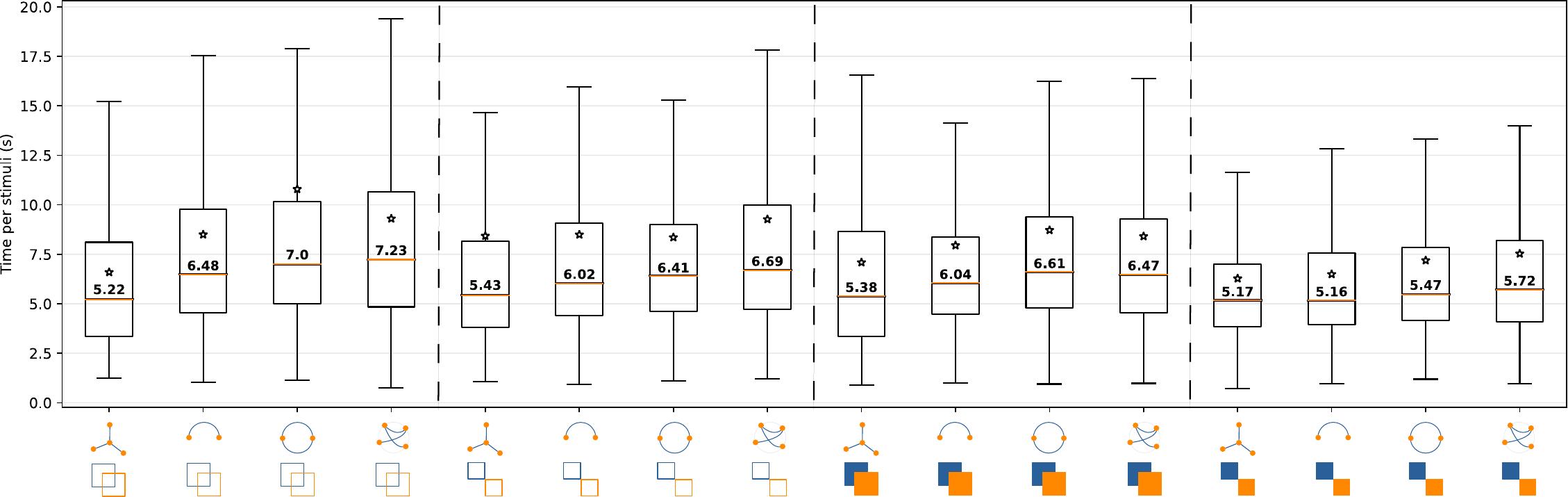}
    \hfill        
    \includegraphics[height=0.235\textwidth]{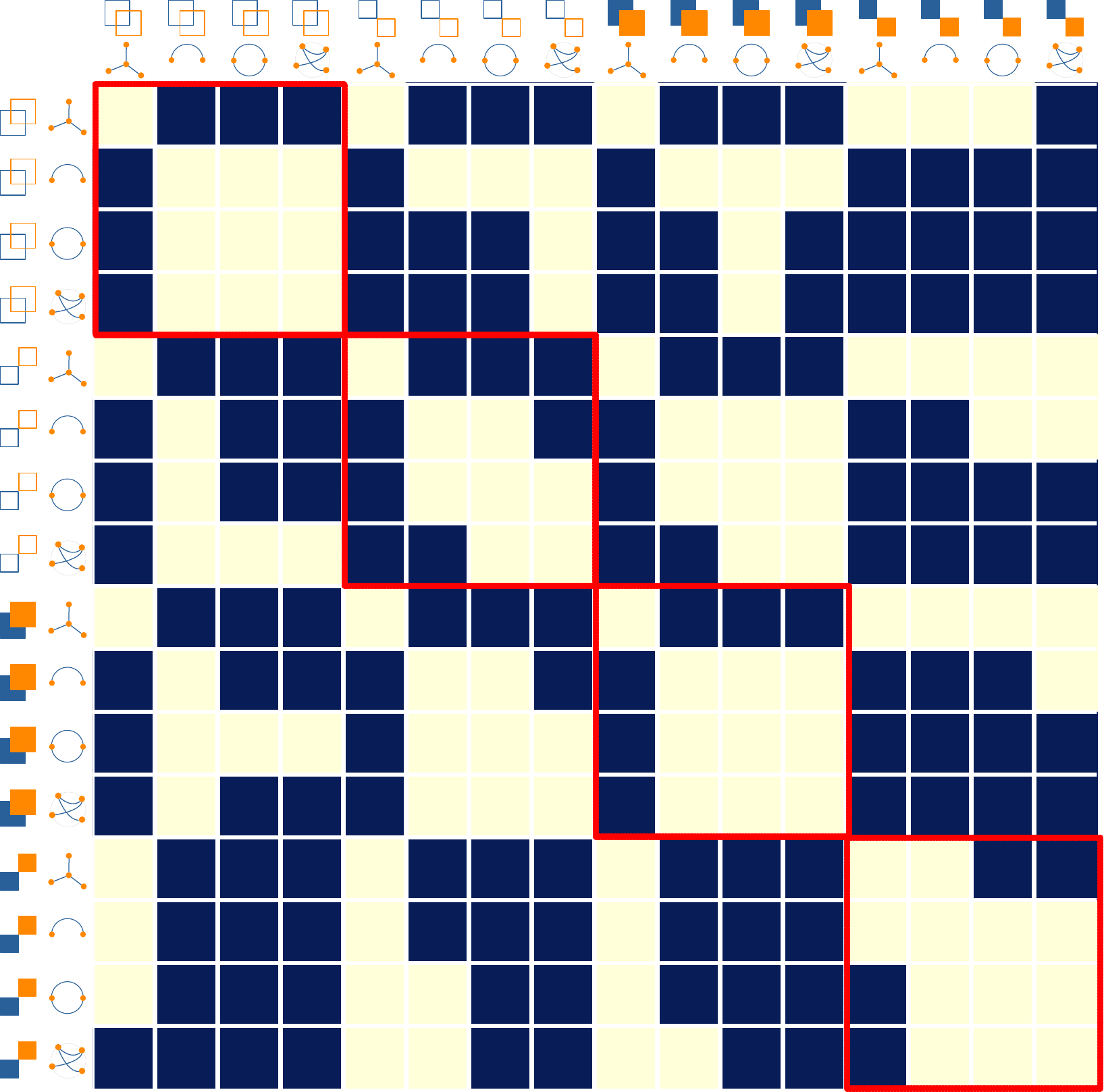}
\vspace{-0.5em}    
    \caption{Overall task \textbf{completion time} for sizes = all, clusters = all, \textbf{order = OLO}, baseline = sfdp. On the right, the dark blue matrix cells correspond to the Bonferroni corrected $p\leq0.05$ for all pairwise comparisons. The poorer accuracy of sfdp observed in \autoref{fig:overall_sfdp_olo_acc} comes with short completion time, indicating that the participants were discouraged by the relatively higher visual clutter in sfdp plots.}
   \label{fig:overall_sfdp_olo_time}
\end{figure*}

\input{inc-tables}

\subsection*{Detailed p-value Tables}
\input{inc-pvalues-tables}

%% file: inc-tables.tex
\clearpage
\subsection*{The Effect of Network Size on Cluster Count in Force-Directed Layouts}

\begin{table*}[h]
\centering
\noindent\begin{minipage}{\textwidth}
    \bgroup
    \def\arraystretch{1.1}
        \begin{adjustbox}{width =0.5\textwidth, center}
    \begin{tabular}{|c|c|c|c|}
\hline
    \textbf{Layout}     &  \textbf{Size} & \textbf{p-value} & \textbf{$\Delta$ Median accuracy}\\
\hline
    \multirow{3}{*}{
        \makecell[c]{\includegraphics[height=18pt]{img/icons/NL.pdf}\\ Linlog}}     
         & 50*-100 & $< 0.001$      &  \textbf{8.33\%}         \\
         & 50*-300     &  $< 0.001$     &   \textbf{8.33\%}        \\
         & 100*-300     &  $0.11$     &   0\%        \\
\hline
    \multirow{3}{*}{
        \makecell[c]{\includegraphics[height=18pt]{img/icons/NL.pdf}\\ Backbone}}     
         & 50*-100 &  $< 0.001$     &  \textbf{5\%}        \\
         & 50*-300     & $< 0.001$      &   \textbf{8.6\%}        \\
         & 100*-300     & $0.236$      &   3.6\%        \\
\hline
    \multirow{3}{*}{
        \makecell[c]{\includegraphics[height=18pt]{img/icons/NL.pdf}\\ SFDP}}     
         & 50*-100 & $< 0.001$      &   \textbf{15\%}        \\
         & 50*-300     & $< 0.001$      &   \textbf{32.1\%}        \\
         & 100*-300    &  $< 0.001$     &    \textbf{17.1\%}       \\
\hline
    \end{tabular}
    \end{adjustbox}
    \egroup
    \vspace{1em}
     \caption{The difference in median accuracies for cluster count judgements between pairs of network sizes. Bold-faced values are statistically significant. A star \mbox{*} marks the network size with higher accuracy. For Linlog, Backbone and SFDP, the accuracy is always better for the smaller network size.}
    \label{tab:FD_size}        
    \end{minipage}
\end{table*}

%% file: inc-pvalues-tables.tex
\begin{table*}[h]
\centering
\noindent\begin{minipage}{\textwidth}
\bgroup
\begin{adjustbox}{width =\textwidth, center}
\begin{tabular}{lrrrrrrrrrrrrrrr}
\toprule
 & \makecell[c]{\includegraphics[height=10pt]{img/icons/loose-inseparable.pdf}\\ \includegraphics[height=10pt]{img/icons/NL.pdf}} & \makecell[c]{\includegraphics[height=10pt]{img/icons/loose-inseparable.pdf}\\ \includegraphics[height=10pt]{img/icons/Arc.pdf}} & \makecell[c]{\includegraphics[height=10pt]{img/icons/loose-inseparable.pdf}\\ \includegraphics[height=10pt]{img/icons/ArcSym.pdf}} & \makecell[c]{\includegraphics[height=10pt]{img/icons/loose-inseparable.pdf}\\ \includegraphics[height=10pt]{img/icons/radial.pdf}} & \makecell[c]{\includegraphics[height=10pt]{img/icons/loose-separable.pdf}\\ \includegraphics[height=10pt]{img/icons/NL.pdf}} & \makecell[c]{\includegraphics[height=10pt]{img/icons/loose-separable.pdf}\\ \includegraphics[height=10pt]{img/icons/Arc.pdf}} & \makecell[c]{\includegraphics[height=10pt]{img/icons/loose-separable.pdf}\\ \includegraphics[height=10pt]{img/icons/ArcSym.pdf}} & \makecell[c]{\includegraphics[height=10pt]{img/icons/loose-separable.pdf}\\ \includegraphics[height=10pt]{img/icons/radial.pdf}} & \makecell[c]{\includegraphics[height=10pt]{img/icons/compact-inseparable.pdf}\\ \includegraphics[height=10pt]{img/icons/NL.pdf}} & \makecell[c]{\includegraphics[height=10pt]{img/icons/compact-inseparable.pdf}\\ \includegraphics[height=10pt]{img/icons/Arc.pdf}} & \makecell[c]{\includegraphics[height=10pt]{img/icons/compact-inseparable.pdf}\\ \includegraphics[height=10pt]{img/icons/ArcSym.pdf}} & \makecell[c]{\includegraphics[height=10pt]{img/icons/compact-inseparable.pdf}\\ \includegraphics[height=10pt]{img/icons/radial.pdf}} & \makecell[c]{\includegraphics[height=10pt]{img/icons/compact-separable.pdf}\\ \includegraphics[height=10pt]{img/icons/NL.pdf}} & \makecell[c]{\includegraphics[height=10pt]{img/icons/compact-separable.pdf}\\ \includegraphics[height=10pt]{img/icons/Arc.pdf}} & \makecell[c]{\includegraphics[height=10pt]{img/icons/compact-separable.pdf}\\ \includegraphics[height=10pt]{img/icons/ArcSym.pdf}} \\
\bottomrule
\end{tabular}

\end{adjustbox}
\egroup
\vspace{1em}
\caption{P-Values for cluster count accuracy for sizes = all, clusters = all, order = GEN, baseline = Linlog (see the corresponding matrix Fig.~\ref{fig:overall_gen_linlog_acc}).}
\label{tab:pVal_acc_linlog_gen}
\end{minipage}
\end{table*}

\begin{table*}[h]
\centering
\noindent\begin{minipage}{\textwidth}
\bgroup
\begin{adjustbox}{width =\textwidth, center}
\begin{tabular}{lrrrrrrrrrrrrrrr}
\toprule
 & \makecell[c]{\includegraphics[height=10pt]{img/icons/loose-inseparable.pdf}\\ \includegraphics[height=10pt]{img/icons/NL.pdf}} & \makecell[c]{\includegraphics[height=10pt]{img/icons/loose-inseparable.pdf}\\ \includegraphics[height=10pt]{img/icons/Arc.pdf}} & \makecell[c]{\includegraphics[height=10pt]{img/icons/loose-inseparable.pdf}\\ \includegraphics[height=10pt]{img/icons/ArcSym.pdf}} & \makecell[c]{\includegraphics[height=10pt]{img/icons/loose-inseparable.pdf}\\ \includegraphics[height=10pt]{img/icons/radial.pdf}} & \makecell[c]{\includegraphics[height=10pt]{img/icons/loose-separable.pdf}\\ \includegraphics[height=10pt]{img/icons/NL.pdf}} & \makecell[c]{\includegraphics[height=10pt]{img/icons/loose-separable.pdf}\\ \includegraphics[height=10pt]{img/icons/Arc.pdf}} & \makecell[c]{\includegraphics[height=10pt]{img/icons/loose-separable.pdf}\\ \includegraphics[height=10pt]{img/icons/ArcSym.pdf}} & \makecell[c]{\includegraphics[height=10pt]{img/icons/loose-separable.pdf}\\ \includegraphics[height=10pt]{img/icons/radial.pdf}} & \makecell[c]{\includegraphics[height=10pt]{img/icons/compact-inseparable.pdf}\\ \includegraphics[height=10pt]{img/icons/NL.pdf}} & \makecell[c]{\includegraphics[height=10pt]{img/icons/compact-inseparable.pdf}\\ \includegraphics[height=10pt]{img/icons/Arc.pdf}} & \makecell[c]{\includegraphics[height=10pt]{img/icons/compact-inseparable.pdf}\\ \includegraphics[height=10pt]{img/icons/ArcSym.pdf}} & \makecell[c]{\includegraphics[height=10pt]{img/icons/compact-inseparable.pdf}\\ \includegraphics[height=10pt]{img/icons/radial.pdf}} & \makecell[c]{\includegraphics[height=10pt]{img/icons/compact-separable.pdf}\\ \includegraphics[height=10pt]{img/icons/NL.pdf}} & \makecell[c]{\includegraphics[height=10pt]{img/icons/compact-separable.pdf}\\ \includegraphics[height=10pt]{img/icons/Arc.pdf}} & \makecell[c]{\includegraphics[height=10pt]{img/icons/compact-separable.pdf}\\ \includegraphics[height=10pt]{img/icons/ArcSym.pdf}} \\
\bottomrule
\end{tabular}

\end{adjustbox}
\egroup
\vspace{1em}
\caption{P-Values for cluster count accuracy for sizes = all, clusters = all, order = CR, baseline = Linlog (see the corresponding matrix Fig.~\ref{fig:overall_cr_linlog_acc}).}
\label{tab:pVal_acc_linlog_cr}
\end{minipage}
\end{table*}

\begin{table*}[h]
\centering
\noindent\begin{minipage}{\textwidth}
\bgroup
\begin{adjustbox}{width =\textwidth, center}
\begin{tabular}{lrrrrrrrrrrrrrrr}
\toprule
 & \makecell[c]{\includegraphics[height=10pt]{img/icons/loose-inseparable.pdf}\\ \includegraphics[height=10pt]{img/icons/NL.pdf}} & \makecell[c]{\includegraphics[height=10pt]{img/icons/loose-inseparable.pdf}\\ \includegraphics[height=10pt]{img/icons/Arc.pdf}} & \makecell[c]{\includegraphics[height=10pt]{img/icons/loose-inseparable.pdf}\\ \includegraphics[height=10pt]{img/icons/ArcSym.pdf}} & \makecell[c]{\includegraphics[height=10pt]{img/icons/loose-inseparable.pdf}\\ \includegraphics[height=10pt]{img/icons/radial.pdf}} & \makecell[c]{\includegraphics[height=10pt]{img/icons/loose-separable.pdf}\\ \includegraphics[height=10pt]{img/icons/NL.pdf}} & \makecell[c]{\includegraphics[height=10pt]{img/icons/loose-separable.pdf}\\ \includegraphics[height=10pt]{img/icons/Arc.pdf}} & \makecell[c]{\includegraphics[height=10pt]{img/icons/loose-separable.pdf}\\ \includegraphics[height=10pt]{img/icons/ArcSym.pdf}} & \makecell[c]{\includegraphics[height=10pt]{img/icons/loose-separable.pdf}\\ \includegraphics[height=10pt]{img/icons/radial.pdf}} & \makecell[c]{\includegraphics[height=10pt]{img/icons/compact-inseparable.pdf}\\ \includegraphics[height=10pt]{img/icons/NL.pdf}} & \makecell[c]{\includegraphics[height=10pt]{img/icons/compact-inseparable.pdf}\\ \includegraphics[height=10pt]{img/icons/Arc.pdf}} & \makecell[c]{\includegraphics[height=10pt]{img/icons/compact-inseparable.pdf}\\ \includegraphics[height=10pt]{img/icons/ArcSym.pdf}} & \makecell[c]{\includegraphics[height=10pt]{img/icons/compact-inseparable.pdf}\\ \includegraphics[height=10pt]{img/icons/radial.pdf}} & \makecell[c]{\includegraphics[height=10pt]{img/icons/compact-separable.pdf}\\ \includegraphics[height=10pt]{img/icons/NL.pdf}} & \makecell[c]{\includegraphics[height=10pt]{img/icons/compact-separable.pdf}\\ \includegraphics[height=10pt]{img/icons/Arc.pdf}} & \makecell[c]{\includegraphics[height=10pt]{img/icons/compact-separable.pdf}\\ \includegraphics[height=10pt]{img/icons/ArcSym.pdf}} \\
\bottomrule
\end{tabular}

\end{adjustbox}
\egroup
\vspace{1em}
\caption{P-Values for cluster count accuracy for sizes = all, clusters = all, order = OLO, baseline = Linlog (see the corresponding matrix Fig.~\ref{fig:overall_olo_linlog_acc}).}
\label{tab:pVal_acc_linlog_olo}
\end{minipage}
\end{table*}

\begin{table*}[h]
\centering
\noindent\begin{minipage}{\textwidth}
\bgroup
\begin{adjustbox}{width =\textwidth, center}
\begin{tabular}{lrrrrrrrrrrrrrrr}
\toprule
 & \makecell[c]{\includegraphics[height=10pt]{img/icons/loose-inseparable.pdf}\\ \includegraphics[height=10pt]{img/icons/NL.pdf}} & \makecell[c]{\includegraphics[height=10pt]{img/icons/loose-inseparable.pdf}\\ \includegraphics[height=10pt]{img/icons/Arc.pdf}} & \makecell[c]{\includegraphics[height=10pt]{img/icons/loose-inseparable.pdf}\\ \includegraphics[height=10pt]{img/icons/ArcSym.pdf}} & \makecell[c]{\includegraphics[height=10pt]{img/icons/loose-inseparable.pdf}\\ \includegraphics[height=10pt]{img/icons/radial.pdf}} & \makecell[c]{\includegraphics[height=10pt]{img/icons/loose-separable.pdf}\\ \includegraphics[height=10pt]{img/icons/NL.pdf}} & \makecell[c]{\includegraphics[height=10pt]{img/icons/loose-separable.pdf}\\ \includegraphics[height=10pt]{img/icons/Arc.pdf}} & \makecell[c]{\includegraphics[height=10pt]{img/icons/loose-separable.pdf}\\ \includegraphics[height=10pt]{img/icons/ArcSym.pdf}} & \makecell[c]{\includegraphics[height=10pt]{img/icons/loose-separable.pdf}\\ \includegraphics[height=10pt]{img/icons/radial.pdf}} & \makecell[c]{\includegraphics[height=10pt]{img/icons/compact-inseparable.pdf}\\ \includegraphics[height=10pt]{img/icons/NL.pdf}} & \makecell[c]{\includegraphics[height=10pt]{img/icons/compact-inseparable.pdf}\\ \includegraphics[height=10pt]{img/icons/Arc.pdf}} & \makecell[c]{\includegraphics[height=10pt]{img/icons/compact-inseparable.pdf}\\ \includegraphics[height=10pt]{img/icons/ArcSym.pdf}} & \makecell[c]{\includegraphics[height=10pt]{img/icons/compact-inseparable.pdf}\\ \includegraphics[height=10pt]{img/icons/radial.pdf}} & \makecell[c]{\includegraphics[height=10pt]{img/icons/compact-separable.pdf}\\ \includegraphics[height=10pt]{img/icons/NL.pdf}} & \makecell[c]{\includegraphics[height=10pt]{img/icons/compact-separable.pdf}\\ \includegraphics[height=10pt]{img/icons/Arc.pdf}} & \makecell[c]{\includegraphics[height=10pt]{img/icons/compact-separable.pdf}\\ \includegraphics[height=10pt]{img/icons/ArcSym.pdf}} \\
\bottomrule
\end{tabular}

\end{adjustbox}
\egroup
\vspace{1em}
\caption{P-Values for overall task completion time for sizes = all, clusters = all, order = GEN, baseline = Linlog (see the corresponding matrix Fig.~\ref{fig:overall_gen_linlog_time}).}
\label{tab:pVal_time_linlog_gen}
\end{minipage}
\end{table*}

\begin{table*}[h]
\centering
\noindent\begin{minipage}{\textwidth}
\bgroup
\begin{adjustbox}{width =\textwidth, center}
\begin{tabular}{lrrrrrrrrrrrrrrr}
\toprule
 & \makecell[c]{\includegraphics[height=10pt]{img/icons/loose-inseparable.pdf}\\ \includegraphics[height=10pt]{img/icons/NL.pdf}} & \makecell[c]{\includegraphics[height=10pt]{img/icons/loose-inseparable.pdf}\\ \includegraphics[height=10pt]{img/icons/Arc.pdf}} & \makecell[c]{\includegraphics[height=10pt]{img/icons/loose-inseparable.pdf}\\ \includegraphics[height=10pt]{img/icons/ArcSym.pdf}} & \makecell[c]{\includegraphics[height=10pt]{img/icons/loose-inseparable.pdf}\\ \includegraphics[height=10pt]{img/icons/radial.pdf}} & \makecell[c]{\includegraphics[height=10pt]{img/icons/loose-separable.pdf}\\ \includegraphics[height=10pt]{img/icons/NL.pdf}} & \makecell[c]{\includegraphics[height=10pt]{img/icons/loose-separable.pdf}\\ \includegraphics[height=10pt]{img/icons/Arc.pdf}} & \makecell[c]{\includegraphics[height=10pt]{img/icons/loose-separable.pdf}\\ \includegraphics[height=10pt]{img/icons/ArcSym.pdf}} & \makecell[c]{\includegraphics[height=10pt]{img/icons/loose-separable.pdf}\\ \includegraphics[height=10pt]{img/icons/radial.pdf}} & \makecell[c]{\includegraphics[height=10pt]{img/icons/compact-inseparable.pdf}\\ \includegraphics[height=10pt]{img/icons/NL.pdf}} & \makecell[c]{\includegraphics[height=10pt]{img/icons/compact-inseparable.pdf}\\ \includegraphics[height=10pt]{img/icons/Arc.pdf}} & \makecell[c]{\includegraphics[height=10pt]{img/icons/compact-inseparable.pdf}\\ \includegraphics[height=10pt]{img/icons/ArcSym.pdf}} & \makecell[c]{\includegraphics[height=10pt]{img/icons/compact-inseparable.pdf}\\ \includegraphics[height=10pt]{img/icons/radial.pdf}} & \makecell[c]{\includegraphics[height=10pt]{img/icons/compact-separable.pdf}\\ \includegraphics[height=10pt]{img/icons/NL.pdf}} & \makecell[c]{\includegraphics[height=10pt]{img/icons/compact-separable.pdf}\\ \includegraphics[height=10pt]{img/icons/Arc.pdf}} & \makecell[c]{\includegraphics[height=10pt]{img/icons/compact-separable.pdf}\\ \includegraphics[height=10pt]{img/icons/ArcSym.pdf}} \\
\bottomrule
\end{tabular}

\end{adjustbox}
\egroup
\vspace{1em}
\caption{P-Values for overall task completion time for sizes = all, clusters = all, order = CR, baseline = Linlog (see the corresponding matrix Fig.~\ref{fig:overall_cr_linlog_time}).}
\label{tab:pVal_time_linlog_cr}
\end{minipage}
\end{table*}

\begin{table*}[h]
\centering
\noindent\begin{minipage}{\textwidth}
\bgroup
\begin{adjustbox}{width =\textwidth, center}
\begin{tabular}{lrrrrrrrrrrrrrrr}
\toprule
 & \makecell[c]{\includegraphics[height=10pt]{img/icons/loose-inseparable.pdf}\\ \includegraphics[height=10pt]{img/icons/NL.pdf}} & \makecell[c]{\includegraphics[height=10pt]{img/icons/loose-inseparable.pdf}\\ \includegraphics[height=10pt]{img/icons/Arc.pdf}} & \makecell[c]{\includegraphics[height=10pt]{img/icons/loose-inseparable.pdf}\\ \includegraphics[height=10pt]{img/icons/ArcSym.pdf}} & \makecell[c]{\includegraphics[height=10pt]{img/icons/loose-inseparable.pdf}\\ \includegraphics[height=10pt]{img/icons/radial.pdf}} & \makecell[c]{\includegraphics[height=10pt]{img/icons/loose-separable.pdf}\\ \includegraphics[height=10pt]{img/icons/NL.pdf}} & \makecell[c]{\includegraphics[height=10pt]{img/icons/loose-separable.pdf}\\ \includegraphics[height=10pt]{img/icons/Arc.pdf}} & \makecell[c]{\includegraphics[height=10pt]{img/icons/loose-separable.pdf}\\ \includegraphics[height=10pt]{img/icons/ArcSym.pdf}} & \makecell[c]{\includegraphics[height=10pt]{img/icons/loose-separable.pdf}\\ \includegraphics[height=10pt]{img/icons/radial.pdf}} & \makecell[c]{\includegraphics[height=10pt]{img/icons/compact-inseparable.pdf}\\ \includegraphics[height=10pt]{img/icons/NL.pdf}} & \makecell[c]{\includegraphics[height=10pt]{img/icons/compact-inseparable.pdf}\\ \includegraphics[height=10pt]{img/icons/Arc.pdf}} & \makecell[c]{\includegraphics[height=10pt]{img/icons/compact-inseparable.pdf}\\ \includegraphics[height=10pt]{img/icons/ArcSym.pdf}} & \makecell[c]{\includegraphics[height=10pt]{img/icons/compact-inseparable.pdf}\\ \includegraphics[height=10pt]{img/icons/radial.pdf}} & \makecell[c]{\includegraphics[height=10pt]{img/icons/compact-separable.pdf}\\ \includegraphics[height=10pt]{img/icons/NL.pdf}} & \makecell[c]{\includegraphics[height=10pt]{img/icons/compact-separable.pdf}\\ \includegraphics[height=10pt]{img/icons/Arc.pdf}} & \makecell[c]{\includegraphics[height=10pt]{img/icons/compact-separable.pdf}\\ \includegraphics[height=10pt]{img/icons/ArcSym.pdf}} \\
\bottomrule
\end{tabular}

\end{adjustbox}
\egroup
\vspace{1em}
\caption{P-Values for overall task completion time for sizes = all, clusters = all, order = OLO, baseline = Linlog (see the corresponding matrix Fig.~\ref{fig:overall_olo_linlog_time}).}
\label{tab:pVal_time_linlog_olo}
\end{minipage}
\end{table*}

\begin{table*}[h]
\centering
\noindent\begin{minipage}{\textwidth}
\bgroup
\begin{adjustbox}{width =\textwidth, center}
\begin{tabular}{lrrrrrrrrrrrrrrr}
\toprule
 & \makecell[c]{\includegraphics[height=10pt]{img/icons/loose-inseparable.pdf}\\ \includegraphics[height=10pt]{img/icons/NL.pdf}} & \makecell[c]{\includegraphics[height=10pt]{img/icons/loose-inseparable.pdf}\\ \includegraphics[height=10pt]{img/icons/Arc.pdf}} & \makecell[c]{\includegraphics[height=10pt]{img/icons/loose-inseparable.pdf}\\ \includegraphics[height=10pt]{img/icons/ArcSym.pdf}} & \makecell[c]{\includegraphics[height=10pt]{img/icons/loose-inseparable.pdf}\\ \includegraphics[height=10pt]{img/icons/radial.pdf}} & \makecell[c]{\includegraphics[height=10pt]{img/icons/loose-separable.pdf}\\ \includegraphics[height=10pt]{img/icons/NL.pdf}} & \makecell[c]{\includegraphics[height=10pt]{img/icons/loose-separable.pdf}\\ \includegraphics[height=10pt]{img/icons/Arc.pdf}} & \makecell[c]{\includegraphics[height=10pt]{img/icons/loose-separable.pdf}\\ \includegraphics[height=10pt]{img/icons/ArcSym.pdf}} & \makecell[c]{\includegraphics[height=10pt]{img/icons/loose-separable.pdf}\\ \includegraphics[height=10pt]{img/icons/radial.pdf}} & \makecell[c]{\includegraphics[height=10pt]{img/icons/compact-inseparable.pdf}\\ \includegraphics[height=10pt]{img/icons/NL.pdf}} & \makecell[c]{\includegraphics[height=10pt]{img/icons/compact-inseparable.pdf}\\ \includegraphics[height=10pt]{img/icons/Arc.pdf}} & \makecell[c]{\includegraphics[height=10pt]{img/icons/compact-inseparable.pdf}\\ \includegraphics[height=10pt]{img/icons/ArcSym.pdf}} & \makecell[c]{\includegraphics[height=10pt]{img/icons/compact-inseparable.pdf}\\ \includegraphics[height=10pt]{img/icons/radial.pdf}} & \makecell[c]{\includegraphics[height=10pt]{img/icons/compact-separable.pdf}\\ \includegraphics[height=10pt]{img/icons/NL.pdf}} & \makecell[c]{\includegraphics[height=10pt]{img/icons/compact-separable.pdf}\\ \includegraphics[height=10pt]{img/icons/Arc.pdf}} & \makecell[c]{\includegraphics[height=10pt]{img/icons/compact-separable.pdf}\\ \includegraphics[height=10pt]{img/icons/ArcSym.pdf}} \\
\bottomrule
\end{tabular}

\end{adjustbox}
\egroup
\vspace{1em}
\caption{P-Values for overall accuracy of cluster count judgments for sizes = all, clusters = all, order = GEN, baseline = backbone (see the corresponding matrix Fig.~\ref{fig:overall_backbone_gen_acc}).}
\label{tab:pVal_acc_backbone_gen}
\end{minipage}
\end{table*}

\begin{table*}[h]
\centering
\noindent\begin{minipage}{\textwidth}
\bgroup
\begin{adjustbox}{width =\textwidth, center}
\begin{tabular}{lrrrrrrrrrrrrrrr}
\toprule
 & \makecell[c]{\includegraphics[height=10pt]{img/icons/loose-inseparable.pdf}\\ \includegraphics[height=10pt]{img/icons/NL.pdf}} & \makecell[c]{\includegraphics[height=10pt]{img/icons/loose-inseparable.pdf}\\ \includegraphics[height=10pt]{img/icons/Arc.pdf}} & \makecell[c]{\includegraphics[height=10pt]{img/icons/loose-inseparable.pdf}\\ \includegraphics[height=10pt]{img/icons/ArcSym.pdf}} & \makecell[c]{\includegraphics[height=10pt]{img/icons/loose-inseparable.pdf}\\ \includegraphics[height=10pt]{img/icons/radial.pdf}} & \makecell[c]{\includegraphics[height=10pt]{img/icons/loose-separable.pdf}\\ \includegraphics[height=10pt]{img/icons/NL.pdf}} & \makecell[c]{\includegraphics[height=10pt]{img/icons/loose-separable.pdf}\\ \includegraphics[height=10pt]{img/icons/Arc.pdf}} & \makecell[c]{\includegraphics[height=10pt]{img/icons/loose-separable.pdf}\\ \includegraphics[height=10pt]{img/icons/ArcSym.pdf}} & \makecell[c]{\includegraphics[height=10pt]{img/icons/loose-separable.pdf}\\ \includegraphics[height=10pt]{img/icons/radial.pdf}} & \makecell[c]{\includegraphics[height=10pt]{img/icons/compact-inseparable.pdf}\\ \includegraphics[height=10pt]{img/icons/NL.pdf}} & \makecell[c]{\includegraphics[height=10pt]{img/icons/compact-inseparable.pdf}\\ \includegraphics[height=10pt]{img/icons/Arc.pdf}} & \makecell[c]{\includegraphics[height=10pt]{img/icons/compact-inseparable.pdf}\\ \includegraphics[height=10pt]{img/icons/ArcSym.pdf}} & \makecell[c]{\includegraphics[height=10pt]{img/icons/compact-inseparable.pdf}\\ \includegraphics[height=10pt]{img/icons/radial.pdf}} & \makecell[c]{\includegraphics[height=10pt]{img/icons/compact-separable.pdf}\\ \includegraphics[height=10pt]{img/icons/NL.pdf}} & \makecell[c]{\includegraphics[height=10pt]{img/icons/compact-separable.pdf}\\ \includegraphics[height=10pt]{img/icons/Arc.pdf}} & \makecell[c]{\includegraphics[height=10pt]{img/icons/compact-separable.pdf}\\ \includegraphics[height=10pt]{img/icons/ArcSym.pdf}} \\
\bottomrule
\end{tabular}

\end{adjustbox}
\egroup
\vspace{1em}
\caption{P-Values for overall completion time for network sizes = all, clusters = all, order = GEN, baseline = backbone (see the corresponding matrix Fig.~\ref{fig:overall_backbone_gen_time}).}
\label{tab:pVal_time_backbone_gen}
\end{minipage}
\end{table*}

\begin{table*}[h]
\centering
\noindent\begin{minipage}{\textwidth}
\bgroup
\begin{adjustbox}{width =\textwidth, center}
\begin{tabular}{lrrrrrrrrrrrrrrr}
\toprule
 & \makecell[c]{\includegraphics[height=10pt]{img/icons/loose-inseparable.pdf}\\ \includegraphics[height=10pt]{img/icons/NL.pdf}} & \makecell[c]{\includegraphics[height=10pt]{img/icons/loose-inseparable.pdf}\\ \includegraphics[height=10pt]{img/icons/Arc.pdf}} & \makecell[c]{\includegraphics[height=10pt]{img/icons/loose-inseparable.pdf}\\ \includegraphics[height=10pt]{img/icons/ArcSym.pdf}} & \makecell[c]{\includegraphics[height=10pt]{img/icons/loose-inseparable.pdf}\\ \includegraphics[height=10pt]{img/icons/radial.pdf}} & \makecell[c]{\includegraphics[height=10pt]{img/icons/loose-separable.pdf}\\ \includegraphics[height=10pt]{img/icons/NL.pdf}} & \makecell[c]{\includegraphics[height=10pt]{img/icons/loose-separable.pdf}\\ \includegraphics[height=10pt]{img/icons/Arc.pdf}} & \makecell[c]{\includegraphics[height=10pt]{img/icons/loose-separable.pdf}\\ \includegraphics[height=10pt]{img/icons/ArcSym.pdf}} & \makecell[c]{\includegraphics[height=10pt]{img/icons/loose-separable.pdf}\\ \includegraphics[height=10pt]{img/icons/radial.pdf}} & \makecell[c]{\includegraphics[height=10pt]{img/icons/compact-inseparable.pdf}\\ \includegraphics[height=10pt]{img/icons/NL.pdf}} & \makecell[c]{\includegraphics[height=10pt]{img/icons/compact-inseparable.pdf}\\ \includegraphics[height=10pt]{img/icons/Arc.pdf}} & \makecell[c]{\includegraphics[height=10pt]{img/icons/compact-inseparable.pdf}\\ \includegraphics[height=10pt]{img/icons/ArcSym.pdf}} & \makecell[c]{\includegraphics[height=10pt]{img/icons/compact-inseparable.pdf}\\ \includegraphics[height=10pt]{img/icons/radial.pdf}} & \makecell[c]{\includegraphics[height=10pt]{img/icons/compact-separable.pdf}\\ \includegraphics[height=10pt]{img/icons/NL.pdf}} & \makecell[c]{\includegraphics[height=10pt]{img/icons/compact-separable.pdf}\\ \includegraphics[height=10pt]{img/icons/Arc.pdf}} & \makecell[c]{\includegraphics[height=10pt]{img/icons/compact-separable.pdf}\\ \includegraphics[height=10pt]{img/icons/ArcSym.pdf}} \\
\bottomrule
\end{tabular}

\end{adjustbox}
\egroup
\vspace{1em}
\caption{P-Values for overall accuracy of cluster count judgments for sizes = all, clusters = all, order = CR, baseline = backbone (see the corresponding matrix Fig.~\ref{fig:overall_backbone_cr_acc}).}
\label{tab:pVal_acc_backbone_cr}
\end{minipage}
\end{table*}

\begin{table*}[h]
\centering
\noindent\begin{minipage}{\textwidth}
\bgroup
\begin{adjustbox}{width =\textwidth, center}
\begin{tabular}{lrrrrrrrrrrrrrrr}
\toprule
 & \makecell[c]{\includegraphics[height=10pt]{img/icons/loose-inseparable.pdf}\\ \includegraphics[height=10pt]{img/icons/NL.pdf}} & \makecell[c]{\includegraphics[height=10pt]{img/icons/loose-inseparable.pdf}\\ \includegraphics[height=10pt]{img/icons/Arc.pdf}} & \makecell[c]{\includegraphics[height=10pt]{img/icons/loose-inseparable.pdf}\\ \includegraphics[height=10pt]{img/icons/ArcSym.pdf}} & \makecell[c]{\includegraphics[height=10pt]{img/icons/loose-inseparable.pdf}\\ \includegraphics[height=10pt]{img/icons/radial.pdf}} & \makecell[c]{\includegraphics[height=10pt]{img/icons/loose-separable.pdf}\\ \includegraphics[height=10pt]{img/icons/NL.pdf}} & \makecell[c]{\includegraphics[height=10pt]{img/icons/loose-separable.pdf}\\ \includegraphics[height=10pt]{img/icons/Arc.pdf}} & \makecell[c]{\includegraphics[height=10pt]{img/icons/loose-separable.pdf}\\ \includegraphics[height=10pt]{img/icons/ArcSym.pdf}} & \makecell[c]{\includegraphics[height=10pt]{img/icons/loose-separable.pdf}\\ \includegraphics[height=10pt]{img/icons/radial.pdf}} & \makecell[c]{\includegraphics[height=10pt]{img/icons/compact-inseparable.pdf}\\ \includegraphics[height=10pt]{img/icons/NL.pdf}} & \makecell[c]{\includegraphics[height=10pt]{img/icons/compact-inseparable.pdf}\\ \includegraphics[height=10pt]{img/icons/Arc.pdf}} & \makecell[c]{\includegraphics[height=10pt]{img/icons/compact-inseparable.pdf}\\ \includegraphics[height=10pt]{img/icons/ArcSym.pdf}} & \makecell[c]{\includegraphics[height=10pt]{img/icons/compact-inseparable.pdf}\\ \includegraphics[height=10pt]{img/icons/radial.pdf}} & \makecell[c]{\includegraphics[height=10pt]{img/icons/compact-separable.pdf}\\ \includegraphics[height=10pt]{img/icons/NL.pdf}} & \makecell[c]{\includegraphics[height=10pt]{img/icons/compact-separable.pdf}\\ \includegraphics[height=10pt]{img/icons/Arc.pdf}} & \makecell[c]{\includegraphics[height=10pt]{img/icons/compact-separable.pdf}\\ \includegraphics[height=10pt]{img/icons/ArcSym.pdf}} \\
\bottomrule
\end{tabular}

\end{adjustbox}
\egroup
\vspace{1em}
\caption{P-Values for overall completion time for network sizes = all, clusters = all, order = CR, baseline = backbone (see the corresponding matrix Fig.~\ref{fig:overall_backbone_cr_time}).}
\label{tab:pVal_time_backbone_cr}
\end{minipage}
\end{table*}

\begin{table*}[h]
\centering
\noindent\begin{minipage}{\textwidth}
\bgroup
\begin{adjustbox}{width =\textwidth, center}
\begin{tabular}{lrrrrrrrrrrrrrrr}
\toprule
 & \makecell[c]{\includegraphics[height=10pt]{img/icons/loose-inseparable.pdf}\\ \includegraphics[height=10pt]{img/icons/NL.pdf}} & \makecell[c]{\includegraphics[height=10pt]{img/icons/loose-inseparable.pdf}\\ \includegraphics[height=10pt]{img/icons/Arc.pdf}} & \makecell[c]{\includegraphics[height=10pt]{img/icons/loose-inseparable.pdf}\\ \includegraphics[height=10pt]{img/icons/ArcSym.pdf}} & \makecell[c]{\includegraphics[height=10pt]{img/icons/loose-inseparable.pdf}\\ \includegraphics[height=10pt]{img/icons/radial.pdf}} & \makecell[c]{\includegraphics[height=10pt]{img/icons/loose-separable.pdf}\\ \includegraphics[height=10pt]{img/icons/NL.pdf}} & \makecell[c]{\includegraphics[height=10pt]{img/icons/loose-separable.pdf}\\ \includegraphics[height=10pt]{img/icons/Arc.pdf}} & \makecell[c]{\includegraphics[height=10pt]{img/icons/loose-separable.pdf}\\ \includegraphics[height=10pt]{img/icons/ArcSym.pdf}} & \makecell[c]{\includegraphics[height=10pt]{img/icons/loose-separable.pdf}\\ \includegraphics[height=10pt]{img/icons/radial.pdf}} & \makecell[c]{\includegraphics[height=10pt]{img/icons/compact-inseparable.pdf}\\ \includegraphics[height=10pt]{img/icons/NL.pdf}} & \makecell[c]{\includegraphics[height=10pt]{img/icons/compact-inseparable.pdf}\\ \includegraphics[height=10pt]{img/icons/Arc.pdf}} & \makecell[c]{\includegraphics[height=10pt]{img/icons/compact-inseparable.pdf}\\ \includegraphics[height=10pt]{img/icons/ArcSym.pdf}} & \makecell[c]{\includegraphics[height=10pt]{img/icons/compact-inseparable.pdf}\\ \includegraphics[height=10pt]{img/icons/radial.pdf}} & \makecell[c]{\includegraphics[height=10pt]{img/icons/compact-separable.pdf}\\ \includegraphics[height=10pt]{img/icons/NL.pdf}} & \makecell[c]{\includegraphics[height=10pt]{img/icons/compact-separable.pdf}\\ \includegraphics[height=10pt]{img/icons/Arc.pdf}} & \makecell[c]{\includegraphics[height=10pt]{img/icons/compact-separable.pdf}\\ \includegraphics[height=10pt]{img/icons/ArcSym.pdf}} \\
\bottomrule
\end{tabular}

\end{adjustbox}
\egroup
\vspace{1em}
\caption{P-Values for overall accuracy of cluster count judgments for sizes = all, clusters = all, order = OLO, baseline = backbone (see the corresponding matrix Fig.~\ref{fig:overall_backbone_olo_acc}).}
\label{tab:pVal_acc_backbone_olo}
\end{minipage}
\end{table*}

\begin{table*}[h]
\centering
\noindent\begin{minipage}{\textwidth}
\bgroup
\begin{adjustbox}{width =\textwidth, center}
\begin{tabular}{lrrrrrrrrrrrrrrr}
\toprule
 & \makecell[c]{\includegraphics[height=10pt]{img/icons/loose-inseparable.pdf}\\ \includegraphics[height=10pt]{img/icons/NL.pdf}} & \makecell[c]{\includegraphics[height=10pt]{img/icons/loose-inseparable.pdf}\\ \includegraphics[height=10pt]{img/icons/Arc.pdf}} & \makecell[c]{\includegraphics[height=10pt]{img/icons/loose-inseparable.pdf}\\ \includegraphics[height=10pt]{img/icons/ArcSym.pdf}} & \makecell[c]{\includegraphics[height=10pt]{img/icons/loose-inseparable.pdf}\\ \includegraphics[height=10pt]{img/icons/radial.pdf}} & \makecell[c]{\includegraphics[height=10pt]{img/icons/loose-separable.pdf}\\ \includegraphics[height=10pt]{img/icons/NL.pdf}} & \makecell[c]{\includegraphics[height=10pt]{img/icons/loose-separable.pdf}\\ \includegraphics[height=10pt]{img/icons/Arc.pdf}} & \makecell[c]{\includegraphics[height=10pt]{img/icons/loose-separable.pdf}\\ \includegraphics[height=10pt]{img/icons/ArcSym.pdf}} & \makecell[c]{\includegraphics[height=10pt]{img/icons/loose-separable.pdf}\\ \includegraphics[height=10pt]{img/icons/radial.pdf}} & \makecell[c]{\includegraphics[height=10pt]{img/icons/compact-inseparable.pdf}\\ \includegraphics[height=10pt]{img/icons/NL.pdf}} & \makecell[c]{\includegraphics[height=10pt]{img/icons/compact-inseparable.pdf}\\ \includegraphics[height=10pt]{img/icons/Arc.pdf}} & \makecell[c]{\includegraphics[height=10pt]{img/icons/compact-inseparable.pdf}\\ \includegraphics[height=10pt]{img/icons/ArcSym.pdf}} & \makecell[c]{\includegraphics[height=10pt]{img/icons/compact-inseparable.pdf}\\ \includegraphics[height=10pt]{img/icons/radial.pdf}} & \makecell[c]{\includegraphics[height=10pt]{img/icons/compact-separable.pdf}\\ \includegraphics[height=10pt]{img/icons/NL.pdf}} & \makecell[c]{\includegraphics[height=10pt]{img/icons/compact-separable.pdf}\\ \includegraphics[height=10pt]{img/icons/Arc.pdf}} & \makecell[c]{\includegraphics[height=10pt]{img/icons/compact-separable.pdf}\\ \includegraphics[height=10pt]{img/icons/ArcSym.pdf}} \\
\bottomrule
\end{tabular}

\end{adjustbox}
\egroup
\vspace{1em}
\caption{P-Values for overall completion time for network sizes = all, clusters = all, order = OLO, baseline = backbone (see the corresponding matrix Fig.~\ref{fig:overall_backbone_olo_time}).}
\label{tab:pVal_time_backbone_olo}
\end{minipage}
\end{table*}

\begin{table*}[h]
\centering
\noindent\begin{minipage}{\textwidth}
\bgroup
\begin{adjustbox}{width =\textwidth, center}
\begin{tabular}{lrrrrrrrrrrrrrrr}
\toprule
 & \makecell[c]{\includegraphics[height=10pt]{img/icons/loose-inseparable.pdf}\\ \includegraphics[height=10pt]{img/icons/NL.pdf}} & \makecell[c]{\includegraphics[height=10pt]{img/icons/loose-inseparable.pdf}\\ \includegraphics[height=10pt]{img/icons/Arc.pdf}} & \makecell[c]{\includegraphics[height=10pt]{img/icons/loose-inseparable.pdf}\\ \includegraphics[height=10pt]{img/icons/ArcSym.pdf}} & \makecell[c]{\includegraphics[height=10pt]{img/icons/loose-inseparable.pdf}\\ \includegraphics[height=10pt]{img/icons/radial.pdf}} & \makecell[c]{\includegraphics[height=10pt]{img/icons/loose-separable.pdf}\\ \includegraphics[height=10pt]{img/icons/NL.pdf}} & \makecell[c]{\includegraphics[height=10pt]{img/icons/loose-separable.pdf}\\ \includegraphics[height=10pt]{img/icons/Arc.pdf}} & \makecell[c]{\includegraphics[height=10pt]{img/icons/loose-separable.pdf}\\ \includegraphics[height=10pt]{img/icons/ArcSym.pdf}} & \makecell[c]{\includegraphics[height=10pt]{img/icons/loose-separable.pdf}\\ \includegraphics[height=10pt]{img/icons/radial.pdf}} & \makecell[c]{\includegraphics[height=10pt]{img/icons/compact-inseparable.pdf}\\ \includegraphics[height=10pt]{img/icons/NL.pdf}} & \makecell[c]{\includegraphics[height=10pt]{img/icons/compact-inseparable.pdf}\\ \includegraphics[height=10pt]{img/icons/Arc.pdf}} & \makecell[c]{\includegraphics[height=10pt]{img/icons/compact-inseparable.pdf}\\ \includegraphics[height=10pt]{img/icons/ArcSym.pdf}} & \makecell[c]{\includegraphics[height=10pt]{img/icons/compact-inseparable.pdf}\\ \includegraphics[height=10pt]{img/icons/radial.pdf}} & \makecell[c]{\includegraphics[height=10pt]{img/icons/compact-separable.pdf}\\ \includegraphics[height=10pt]{img/icons/NL.pdf}} & \makecell[c]{\includegraphics[height=10pt]{img/icons/compact-separable.pdf}\\ \includegraphics[height=10pt]{img/icons/Arc.pdf}} & \makecell[c]{\includegraphics[height=10pt]{img/icons/compact-separable.pdf}\\ \includegraphics[height=10pt]{img/icons/ArcSym.pdf}} \\
\bottomrule
\end{tabular}

\end{adjustbox}
\egroup
\vspace{1em}
\caption{P-Values for overall accuracy of cluster count judgments for sizes = all, clusters = all, order = GEN, baseline = sfdp (see the corresponding matrix Fig.~\ref{fig:overall_sfdp_gen_acc}).}
\label{tab:pVal_acc_sfdp_gen}
\end{minipage}
\end{table*}

\begin{table*}[h]
\centering
\noindent\begin{minipage}{\textwidth}
\bgroup
\begin{adjustbox}{width =\textwidth, center}
\begin{tabular}{lrrrrrrrrrrrrrrr}
\toprule
 & \makecell[c]{\includegraphics[height=10pt]{img/icons/loose-inseparable.pdf}\\ \includegraphics[height=10pt]{img/icons/NL.pdf}} & \makecell[c]{\includegraphics[height=10pt]{img/icons/loose-inseparable.pdf}\\ \includegraphics[height=10pt]{img/icons/Arc.pdf}} & \makecell[c]{\includegraphics[height=10pt]{img/icons/loose-inseparable.pdf}\\ \includegraphics[height=10pt]{img/icons/ArcSym.pdf}} & \makecell[c]{\includegraphics[height=10pt]{img/icons/loose-inseparable.pdf}\\ \includegraphics[height=10pt]{img/icons/radial.pdf}} & \makecell[c]{\includegraphics[height=10pt]{img/icons/loose-separable.pdf}\\ \includegraphics[height=10pt]{img/icons/NL.pdf}} & \makecell[c]{\includegraphics[height=10pt]{img/icons/loose-separable.pdf}\\ \includegraphics[height=10pt]{img/icons/Arc.pdf}} & \makecell[c]{\includegraphics[height=10pt]{img/icons/loose-separable.pdf}\\ \includegraphics[height=10pt]{img/icons/ArcSym.pdf}} & \makecell[c]{\includegraphics[height=10pt]{img/icons/loose-separable.pdf}\\ \includegraphics[height=10pt]{img/icons/radial.pdf}} & \makecell[c]{\includegraphics[height=10pt]{img/icons/compact-inseparable.pdf}\\ \includegraphics[height=10pt]{img/icons/NL.pdf}} & \makecell[c]{\includegraphics[height=10pt]{img/icons/compact-inseparable.pdf}\\ \includegraphics[height=10pt]{img/icons/Arc.pdf}} & \makecell[c]{\includegraphics[height=10pt]{img/icons/compact-inseparable.pdf}\\ \includegraphics[height=10pt]{img/icons/ArcSym.pdf}} & \makecell[c]{\includegraphics[height=10pt]{img/icons/compact-inseparable.pdf}\\ \includegraphics[height=10pt]{img/icons/radial.pdf}} & \makecell[c]{\includegraphics[height=10pt]{img/icons/compact-separable.pdf}\\ \includegraphics[height=10pt]{img/icons/NL.pdf}} & \makecell[c]{\includegraphics[height=10pt]{img/icons/compact-separable.pdf}\\ \includegraphics[height=10pt]{img/icons/Arc.pdf}} & \makecell[c]{\includegraphics[height=10pt]{img/icons/compact-separable.pdf}\\ \includegraphics[height=10pt]{img/icons/ArcSym.pdf}} \\
\bottomrule
\end{tabular}

\end{adjustbox}
\egroup
\vspace{1em}
\caption{P-Values for overall completion time for network sizes = all, clusters = all, order = GEN, baseline = sfdp (see the corresponding matrix Fig.~\ref{fig:overall_sfdp_gen_time}).}
\label{tab:pVal_time_sfdp_gen}
\end{minipage}
\end{table*}

\begin{table*}[h]
\centering
\noindent\begin{minipage}{\textwidth}
\bgroup
\begin{adjustbox}{width =\textwidth, center}
\begin{tabular}{lrrrrrrrrrrrrrrr}
\toprule
 & \makecell[c]{\includegraphics[height=10pt]{img/icons/loose-inseparable.pdf}\\ \includegraphics[height=10pt]{img/icons/NL.pdf}} & \makecell[c]{\includegraphics[height=10pt]{img/icons/loose-inseparable.pdf}\\ \includegraphics[height=10pt]{img/icons/Arc.pdf}} & \makecell[c]{\includegraphics[height=10pt]{img/icons/loose-inseparable.pdf}\\ \includegraphics[height=10pt]{img/icons/ArcSym.pdf}} & \makecell[c]{\includegraphics[height=10pt]{img/icons/loose-inseparable.pdf}\\ \includegraphics[height=10pt]{img/icons/radial.pdf}} & \makecell[c]{\includegraphics[height=10pt]{img/icons/loose-separable.pdf}\\ \includegraphics[height=10pt]{img/icons/NL.pdf}} & \makecell[c]{\includegraphics[height=10pt]{img/icons/loose-separable.pdf}\\ \includegraphics[height=10pt]{img/icons/Arc.pdf}} & \makecell[c]{\includegraphics[height=10pt]{img/icons/loose-separable.pdf}\\ \includegraphics[height=10pt]{img/icons/ArcSym.pdf}} & \makecell[c]{\includegraphics[height=10pt]{img/icons/loose-separable.pdf}\\ \includegraphics[height=10pt]{img/icons/radial.pdf}} & \makecell[c]{\includegraphics[height=10pt]{img/icons/compact-inseparable.pdf}\\ \includegraphics[height=10pt]{img/icons/NL.pdf}} & \makecell[c]{\includegraphics[height=10pt]{img/icons/compact-inseparable.pdf}\\ \includegraphics[height=10pt]{img/icons/Arc.pdf}} & \makecell[c]{\includegraphics[height=10pt]{img/icons/compact-inseparable.pdf}\\ \includegraphics[height=10pt]{img/icons/ArcSym.pdf}} & \makecell[c]{\includegraphics[height=10pt]{img/icons/compact-inseparable.pdf}\\ \includegraphics[height=10pt]{img/icons/radial.pdf}} & \makecell[c]{\includegraphics[height=10pt]{img/icons/compact-separable.pdf}\\ \includegraphics[height=10pt]{img/icons/NL.pdf}} & \makecell[c]{\includegraphics[height=10pt]{img/icons/compact-separable.pdf}\\ \includegraphics[height=10pt]{img/icons/Arc.pdf}} & \makecell[c]{\includegraphics[height=10pt]{img/icons/compact-separable.pdf}\\ \includegraphics[height=10pt]{img/icons/ArcSym.pdf}} \\
\bottomrule
\end{tabular}

\end{adjustbox}
\egroup
\vspace{1em}
\caption{P-Values for overall accuracy of cluster count judgments for sizes = all, clusters = all, order = CR, baseline = sfdp (see the corresponding matrix Fig.~\ref{fig:overall_sfdp_cr_acc}).}
\label{tab:pVal_acc_sfdp_cr}
\end{minipage}
\end{table*}

\begin{table*}[h]
\centering
\noindent\begin{minipage}{\textwidth}
\bgroup
\begin{adjustbox}{width =\textwidth, center}
\begin{tabular}{lrrrrrrrrrrrrrrr}
\toprule
 & \makecell[c]{\includegraphics[height=10pt]{img/icons/loose-inseparable.pdf}\\ \includegraphics[height=10pt]{img/icons/NL.pdf}} & \makecell[c]{\includegraphics[height=10pt]{img/icons/loose-inseparable.pdf}\\ \includegraphics[height=10pt]{img/icons/Arc.pdf}} & \makecell[c]{\includegraphics[height=10pt]{img/icons/loose-inseparable.pdf}\\ \includegraphics[height=10pt]{img/icons/ArcSym.pdf}} & \makecell[c]{\includegraphics[height=10pt]{img/icons/loose-inseparable.pdf}\\ \includegraphics[height=10pt]{img/icons/radial.pdf}} & \makecell[c]{\includegraphics[height=10pt]{img/icons/loose-separable.pdf}\\ \includegraphics[height=10pt]{img/icons/NL.pdf}} & \makecell[c]{\includegraphics[height=10pt]{img/icons/loose-separable.pdf}\\ \includegraphics[height=10pt]{img/icons/Arc.pdf}} & \makecell[c]{\includegraphics[height=10pt]{img/icons/loose-separable.pdf}\\ \includegraphics[height=10pt]{img/icons/ArcSym.pdf}} & \makecell[c]{\includegraphics[height=10pt]{img/icons/loose-separable.pdf}\\ \includegraphics[height=10pt]{img/icons/radial.pdf}} & \makecell[c]{\includegraphics[height=10pt]{img/icons/compact-inseparable.pdf}\\ \includegraphics[height=10pt]{img/icons/NL.pdf}} & \makecell[c]{\includegraphics[height=10pt]{img/icons/compact-inseparable.pdf}\\ \includegraphics[height=10pt]{img/icons/Arc.pdf}} & \makecell[c]{\includegraphics[height=10pt]{img/icons/compact-inseparable.pdf}\\ \includegraphics[height=10pt]{img/icons/ArcSym.pdf}} & \makecell[c]{\includegraphics[height=10pt]{img/icons/compact-inseparable.pdf}\\ \includegraphics[height=10pt]{img/icons/radial.pdf}} & \makecell[c]{\includegraphics[height=10pt]{img/icons/compact-separable.pdf}\\ \includegraphics[height=10pt]{img/icons/NL.pdf}} & \makecell[c]{\includegraphics[height=10pt]{img/icons/compact-separable.pdf}\\ \includegraphics[height=10pt]{img/icons/Arc.pdf}} & \makecell[c]{\includegraphics[height=10pt]{img/icons/compact-separable.pdf}\\ \includegraphics[height=10pt]{img/icons/ArcSym.pdf}} \\
\bottomrule
\end{tabular}

\end{adjustbox}
\egroup
\vspace{1em}
\caption{P-Values for overall completion time for network sizes = all, clusters = all, order = CR, baseline = sfdp (see the corresponding matrix Fig.~\ref{fig:overall_sfdp_cr_time}).}
\label{tab:pVal_time_sfdp_cr}
\end{minipage}
\end{table*}

\begin{table*}[h]
\centering
\noindent\begin{minipage}{\textwidth}
\bgroup
\begin{adjustbox}{width =\textwidth, center}
\begin{tabular}{lrrrrrrrrrrrrrrr}
\toprule
 & \makecell[c]{\includegraphics[height=10pt]{img/icons/loose-inseparable.pdf}\\ \includegraphics[height=10pt]{img/icons/NL.pdf}} & \makecell[c]{\includegraphics[height=10pt]{img/icons/loose-inseparable.pdf}\\ \includegraphics[height=10pt]{img/icons/Arc.pdf}} & \makecell[c]{\includegraphics[height=10pt]{img/icons/loose-inseparable.pdf}\\ \includegraphics[height=10pt]{img/icons/ArcSym.pdf}} & \makecell[c]{\includegraphics[height=10pt]{img/icons/loose-inseparable.pdf}\\ \includegraphics[height=10pt]{img/icons/radial.pdf}} & \makecell[c]{\includegraphics[height=10pt]{img/icons/loose-separable.pdf}\\ \includegraphics[height=10pt]{img/icons/NL.pdf}} & \makecell[c]{\includegraphics[height=10pt]{img/icons/loose-separable.pdf}\\ \includegraphics[height=10pt]{img/icons/Arc.pdf}} & \makecell[c]{\includegraphics[height=10pt]{img/icons/loose-separable.pdf}\\ \includegraphics[height=10pt]{img/icons/ArcSym.pdf}} & \makecell[c]{\includegraphics[height=10pt]{img/icons/loose-separable.pdf}\\ \includegraphics[height=10pt]{img/icons/radial.pdf}} & \makecell[c]{\includegraphics[height=10pt]{img/icons/compact-inseparable.pdf}\\ \includegraphics[height=10pt]{img/icons/NL.pdf}} & \makecell[c]{\includegraphics[height=10pt]{img/icons/compact-inseparable.pdf}\\ \includegraphics[height=10pt]{img/icons/Arc.pdf}} & \makecell[c]{\includegraphics[height=10pt]{img/icons/compact-inseparable.pdf}\\ \includegraphics[height=10pt]{img/icons/ArcSym.pdf}} & \makecell[c]{\includegraphics[height=10pt]{img/icons/compact-inseparable.pdf}\\ \includegraphics[height=10pt]{img/icons/radial.pdf}} & \makecell[c]{\includegraphics[height=10pt]{img/icons/compact-separable.pdf}\\ \includegraphics[height=10pt]{img/icons/NL.pdf}} & \makecell[c]{\includegraphics[height=10pt]{img/icons/compact-separable.pdf}\\ \includegraphics[height=10pt]{img/icons/Arc.pdf}} & \makecell[c]{\includegraphics[height=10pt]{img/icons/compact-separable.pdf}\\ \includegraphics[height=10pt]{img/icons/ArcSym.pdf}} \\
\bottomrule
\end{tabular}

\end{adjustbox}
\egroup
\vspace{1em}
\caption{P-Values for overall accuracy of cluster count judgments for sizes = all, clusters = all, order = OLO, baseline = sfdp (see the corresponding matrix Fig.~\ref{fig:overall_sfdp_olo_acc}).}
\label{tab:pVal_acc_sfdp_olo_3}
\end{minipage}
\end{table*}

\begin{table*}[h]
\centering
\noindent\begin{minipage}{\textwidth}
\bgroup
\begin{adjustbox}{width =\textwidth, center}
\begin{tabular}{lrrrrrrrrrrrrrrr}
\toprule
 & \makecell[c]{\includegraphics[height=10pt]{img/icons/loose-inseparable.pdf}\\ \includegraphics[height=10pt]{img/icons/NL.pdf}} & \makecell[c]{\includegraphics[height=10pt]{img/icons/loose-inseparable.pdf}\\ \includegraphics[height=10pt]{img/icons/Arc.pdf}} & \makecell[c]{\includegraphics[height=10pt]{img/icons/loose-inseparable.pdf}\\ \includegraphics[height=10pt]{img/icons/ArcSym.pdf}} & \makecell[c]{\includegraphics[height=10pt]{img/icons/loose-inseparable.pdf}\\ \includegraphics[height=10pt]{img/icons/radial.pdf}} & \makecell[c]{\includegraphics[height=10pt]{img/icons/loose-separable.pdf}\\ \includegraphics[height=10pt]{img/icons/NL.pdf}} & \makecell[c]{\includegraphics[height=10pt]{img/icons/loose-separable.pdf}\\ \includegraphics[height=10pt]{img/icons/Arc.pdf}} & \makecell[c]{\includegraphics[height=10pt]{img/icons/loose-separable.pdf}\\ \includegraphics[height=10pt]{img/icons/ArcSym.pdf}} & \makecell[c]{\includegraphics[height=10pt]{img/icons/loose-separable.pdf}\\ \includegraphics[height=10pt]{img/icons/radial.pdf}} & \makecell[c]{\includegraphics[height=10pt]{img/icons/compact-inseparable.pdf}\\ \includegraphics[height=10pt]{img/icons/NL.pdf}} & \makecell[c]{\includegraphics[height=10pt]{img/icons/compact-inseparable.pdf}\\ \includegraphics[height=10pt]{img/icons/Arc.pdf}} & \makecell[c]{\includegraphics[height=10pt]{img/icons/compact-inseparable.pdf}\\ \includegraphics[height=10pt]{img/icons/ArcSym.pdf}} & \makecell[c]{\includegraphics[height=10pt]{img/icons/compact-inseparable.pdf}\\ \includegraphics[height=10pt]{img/icons/radial.pdf}} & \makecell[c]{\includegraphics[height=10pt]{img/icons/compact-separable.pdf}\\ \includegraphics[height=10pt]{img/icons/NL.pdf}} & \makecell[c]{\includegraphics[height=10pt]{img/icons/compact-separable.pdf}\\ \includegraphics[height=10pt]{img/icons/Arc.pdf}} & \makecell[c]{\includegraphics[height=10pt]{img/icons/compact-separable.pdf}\\ \includegraphics[height=10pt]{img/icons/ArcSym.pdf}} \\
\bottomrule
\end{tabular}

\end{adjustbox}
\egroup
\vspace{1em}
\caption{P-Values for overall completion time for network sizes = all, clusters = all, order = OLO, baseline = sfdp (see the corresponding matrix Fig.~\ref{fig:overall_sfdp_olo_time}).}
\label{tab:pVal_time_sfdp_olo}
\end{minipage}
\end{table*}